\documentclass[print]{aa}

\usepackage{txfonts}
\usepackage{graphicx}
\usepackage{xspace}\usepackage{natbib}
\usepackage{longtable}
\usepackage[center]{subfigure}
\usepackage{lscape}
\usepackage{amssymb}
\usepackage{color}
\definecolor{lightgrey}{rgb}{.8, .8, .8}
\definecolor{darkgrey}{rgb}{.5, .5, .5}

\def\aap{A\&A}

\def\apjl{ApJL}

\def\um{$\mu$m\xspace}

\def\12c{$^{12}$C\xspace}
\def\h2o{H$_2$O\xspace}
\def\so2{SO$_2$\xspace}
\def\co2{CO$_2$\xspace}
\def\hcn{HCN\xspace}
\def\c2h2{C$_2$H$_2$\xspace}
\def\sio{SiO\xspace}
\def\sis{SiS\xspace}

\begin{document}

   \title{The Spitzer Spectroscopic Survey of S-type Stars}

   \author{K. Smolders\inst{1} \fnmsep \thanks{Aspirant Fellow of the Fund for Scientific Research, Flanders}
           \and
          P.~Neyskens\inst{2} \fnmsep \thanks{Fellowship ``boursier F.R.I.A'', Belgium}
           \and
          J.A.D.L.~Blommaert\inst{1}
           \and
          S.~Hony\inst{3}
           \and
          H.~Van Winckel\inst{1}
           \and
          L.~Decin\inst{1} \fnmsep \thanks{Postdoctoral Fellows of the Fund for Scientific Research, Flanders}
           \and
          S.~Van Eck\inst{2}
           \and
          G.~C.~Sloan\inst{4}
           \and
          J.~Cami\inst{5}
           \and
          S.~Uttenthaler\inst{1,6}
           \and
          P.~Degroote\inst{1}
           \and
          D.~Barry\inst{4}
           \and
          M.~Feast\inst{7,8}
           \and
          M.A.T.~Groenewegen\inst{9}
           \and
          M.~Matsuura\inst{10}
           \and
          J.~Menzies\inst{8}
           \and
          R.~Sahai\inst{11}
           \and
          J.~Th.~van Loon\inst{12}
           \and
          A.A.~Zijlstra\inst{13}
           \and
          B.~Acke\inst{1} \fnmsep $^{\star\star\star}$
           \and
          S.~Bloemen\inst{1}
           \and
          N.~Cox\inst{1}
           \and
          P.~de Cat\inst{9}
           \and
          M.~Desmet\inst{1}
           \and
          K.~Exter\inst{1}
           \and
          D.~Ladjal\inst{1}
		   \and
          R.~\O stensen\inst{1}
           \and
          S.~Saesen\inst{1, 14}
           \and
          F.~van~Wyk\inst{8}
 		   \and
          T.~Verhoelst\inst{1}\fnmsep$^{\star\star\star}$
 		   \and
          W.~Zima\inst{1}
}

   \offprints{K. Smolders}

   \institute{
  			Instituut voor Sterrenkunde (IvS),
             Katholieke Universiteit Leuven,
             Celestijnenlaan 200 D,
             B-3001 Leuven, Belgium \\
             \email{kristof.smolders@ster.kuleuven.be}
		\and
             Institut d'Astronomie et d'Astrophysique (IAA), 
             Universit\'e Libre de Bruxelles, 
             C.P.226, Boulevard du Triomphe,
             B-1050 Bruxelles, Belgium
		\and
             Service d'Astrophysique,
             CEA Saclay,
             91191 Gif-sur-Yvette, France \\
             \email{sacha.hony@cea.fr}
         \and
             Astronomy Department,
			Cornell University,
			Ithaca, NY 14853-6801, USA
	     \and
		     Department of Physics and Astronomy,
  		     University of Western Ontario, 
		     London, Ontario N6A 3K7, Canada
	     \and
			Department of Astronomy,
			University of Vienna,
			T\"urkenschanzstra\ss e 17, 1180 Vienna, Austria
		\and
			Astrophysics, Cosmology and Gravity Centre, Astronomy dept.,
			University of Cape Town, 7701 South Africa
		\and 
			South African Astronomical Observatory,
			PO Box 9, Observatory, 
			7935, South Africa
		\and
			Koninklijke Sterrenwacht van Belgi\"e, 
			Ringlaan 3, 
			B-1180 Brussel, Belgium
		\and
	         Department of Physics and Astronomy,
	         University College London,
	         Gower Street, London WC1E 6BT, UK
        \and
			Jet Propulsion Laboratory,
			California Institute of Technology,
			Pasadena, CA 91109, USA
		\and
			Lennard-Jones Laboratories, 
			Keele University, 
			Staffordshire ST5 5BG, UK
        \and
             Jodrell Bank Centre for Astrophysics, 
             The University of Manchester, 
             School of Physics \& Astronomy, Manchester M13 9PL, UK
		\and
		    Observatoire de Gen\`{e}ve, 
		    Universit\'{e} de Gen\`{e}ve, 
		    Chemin des Maillettes 51, 
		    1290 Sauverny, Switzerland
}

\date{\today}

 \abstract
 {S-type AGB stars are thought to be in the transitional phase between M-type and C-type AGB stars. Because the composition of the circumstellar environment reflects the photospheric abundances, one may expect a strong influence of the stellar C/O ratio on the molecular chemistry and the mineralogy of the circumstellar dust.}{In this paper, we present a large sample of 87 intrinsic galactic S-type AGB stars, observed at infrared wavelengths with the Spitzer Space Telescope, and supplemented with ground-based optical data.}{On the one hand, we derive the stellar parameters from the optical spectroscopy and photometry, using a grid of model atmospheres. On the other, we decompose the infrared spectra to quantify the flux-contributions from the different dust species. Finally, we compare the independently determined stellar parameters and dust properties.}{For the stars without significant dust emission features, we detect a strict relation between the presence of SiS absorption in the Spitzer spectra and the C/O ratio of the stellar atmosphere. These absorption bands can thus be used as an additional diagnostic for the C/O ratio. For stars with significant dust emission, we define three distinct groups, based on the relative contribution of certain dust species to the infrared flux. We find a strong link between group-membership and C/O ratio. Furthermore, we show that these groups can be explained by assuming that the dust-condensation can be cut short before silicates are produced, while the remaining free atoms and molecules can then be used to form the observed magnesium sulfides or the carriers of the unidentified 13\,\um and 20\,\um features. Finally, we present the detection of emission features attributed to molecules and dust characteristic to C-type stars, such as molecular SiS, hydrocarbons and magnesium sulfide grains. We show that we often detect magnesium sulfides together with molecular SiS and we propose that it is formed by a reaction of SiS molecules with Mg.}{}

 \maketitle

\section{Introduction}
The dredge-up of carbon in asymptotic giant branch (AGB) stars plays a key role in the chemical evolution from oxygen-rich M-type stars to carbon-rich C-type stars \citep{Iben1983, Herwig2005, Habing2004}. S-type stars are often considered to be an intermediate phase of AGB evolution between M- and C-type stars in which the C/O ratio of stars changes from the solar value to ratios larger than unity\footnote{It is often said that S-type stars have C/O ratio close to one, but a star with strong s-element enrichment can be classified as an S-type star while the C/O ratio is as low as 0.5 \citep{VanEck2010}.}. In the spectral classification scheme, S-type stars are defined by the presence of absorption bands of ZrO and LaO molecules in the optical spectra. The transition from oxygen-rich to carbon-rich marks an important chemical transition in the AGB evolution.

Following the chemical pathways relevant in the circumstellar environment, CO is the first molecule to form, locking up all of the available carbon or oxygen, whichever is less abundant. Subsequently, in M-type stars (C/O~$<$~1) almost all C atoms will be consumed by CO and oxygen-rich molecules such as TiO, SiO, and H$_2$O will be present. In carbon-rich stars (C/O~$>$~1) the dominant molecules are CO, CH, C$_2$, and C$_2$H$_2$. However, it is important to realize that this is a simplified picture of the circumstellar chemistry and only valid under the assumption of local thermal equilibrium (LTE). Due to non-equilibrium chemistry effects, oxygen-rich stars could display molecules that are typical for carbon-rich stars and vice versa \citep{Cherchneff2006} .

The chemical and physical properties of circumstellar dust grains are linked to the chemical composition of the stellar atmosphere. In an oxygen-rich environment, silicates and oxides are formed, while graphite, amorphous carbon, and silicon carbide are formed in carbon-rich environments. When the C/O ratio is very close to unity, both O and C are almost completely consumed in CO, in which case the dust condensation sequence is not well studied nor understood. Theoretical models predict the presence of FeSi and metallic Fe, but this still needs observational confirmation \citep{Ferrarotti2002}. Furthermore,  some molecules and dust species that contain neither C nor O are also limited to certain chemical environments, for example magnesium sulfide dust and SiS molecules are typical for carbon-rich environments. This indicates that the C/O ratio does not only influence the carbon- and oxygen-abundance, but indirectly effects the entire chemical network, of which only a couple of molecular products are detected. 

The chemistry of the circumstellar material is key to understanding the driving mechanism of mass-loss in AGB stars, one of the major open questions in stellar evolution. It is generally accepted that the mass loss of AGB stars is dust-driven \citep{Lamers1999}. Pulsations bring the envelope far from the stellar equilibrium radius, where the temperatures are low enough for dust-condensation. This dust is easily accelerated by radiative pressure and it drags the gas along. For carbon-rich stars, the opacity of amorphous carbon is high enough in the near-infrared to drive mass loss. For oxygen-rich stars however, theoretical studies suggest that silicates are not able to drive the wind \citep{Woitke2006, Hofner2007, Hofner2008}. It is clear that an exact understanding of the process is lacking.

A recent study of the  infrared spectra of nine S-type AGB stars from the Short Wavelength Spectrometer onboard of the Infrared Space Observatory (ISO/SWS), presented by \citet{Hony2009} showed that the infrared spectra of S-type AGB stars can be significantly different from those of oxygen-rich AGB stars and already reported on the detection of dust emission at 28\,\um attributed to magnesium sulfides. Although this study shows that S-type AGB stars exhibit several unique dust characteristics, the study is limited because ISO/SWS is only sensitive to the brightest (at infrared wavelengths) AGB stars and thus to the S-type stars with higher mass-loss rates, while a large and homogeneous sample is necessary to draw conclusions about the dust condensation in S-type stars.

In this paper we present a combination of optical spectroscopy, infrared photometry and infrared spectroscopy for a sample of 87 \emph{intrinsic}\footnote{Intrinsic S-type stars are genuine thermally pulsating AGB stars, contrary to extrinsic S-type stars which are enriched in C or s-process elements by accretion from an evolved companion \citep{Groenewegen1993, VanEck1999}.} S-type stars. This sample allows us to study the interaction between the chemistry and dynamics of the stellar atmosphere and the dusty circumstellar environment. In \S \ref{sct:sample} and \S \ref{sct:data} we discuss the sample selection procedure and the data reduction. In \S \ref{sct:stelparam} we derive the chemical and variability characteristics of the stars in our sample. In \S \ref{sct:nakedstelatm} we give an overview of the ``naked'' stellar atmospheres and we show that their spectral appearance of depends strongly on the C/O ratio of the star. In \S \ref{sct:dustemis} we present a tool to decompose infrared (IR) spectra into the contribution of the stellar atmosphere and the different dust species, allowing us to classify the IR spectra of dust-rich stars into three distinct groups. We show that there is a relation between the C/O ratio and those groups. 

\section{The sample of S-type AGB stars}\label{sct:sample}
In order to study the interaction between the stellar atmosphere and the infrared properties of S-type AGB stars, we selected a large sample of S-type stars designed to include all intrinsic S-type AGB stars bright enough to be detectable with the Spitzer IRS spectrograph. This sample spans the entire range of C/O ratios, temperatures and s-process abundances found in S-type AGB stars, but is slightly biased towards stars with lower mass-loss rates (see Sect. \ref{sct:bias}).

\subsection{Sample selection}
A first sample was drawn from the second edition of the General Catalog of S stars \citep[GCSS2,][]{Stephenson2}. This catalog classifies 1346 stars as S-type stars, based on the presence of ZrO or LaO bands in the red part of the optical spectrum. This method only selects stars based on an enrichment of s-process elements in their lower envelope, which not only includes stars that have undergone a recent third dredge-up event, the \emph{intrinsic S stars}, but also stars that are enhanced with s-process elements by binary interaction, called \emph{extrinsic S stars}~\citep{Brown1990}. In order to study the dust-sequence on S-type AGB stars, we need to exclude the \emph{extrinsic} stars from our sample.

Several properties distinguish both groups, such as luminosity, the $K-[12]$ color, the presence of technetium in the atmosphere, and variability \citep{VanEck1999}. The best distinction method is based on the presence of technetium (Tc), which is brought to the surface during the third dredge-up events. With a laboratory half-life time of approximately $2\times10^5$ years, the detection of Tc in the spectrum points to a recent third dredge-up event. Although the detection of Tc is the most secure way to include only \emph{intrinsic} stars in the sample, it requires high resolution spectroscopy in the blue region where these cool stars are faint. A criterion based on the luminosity would introduce strong and unwanted biases on the distance or intrinsic brightness of the stars, while putting a lower limit on the infrared color $K-[12]$ would favor stars with a dusty circumstellar environment. Considering all this, source variability turns out to be the most efficient way to discriminate between intrinsic and extrinsic S-type stars.

The ASAS light curves of the sample presented by \citet{VanEck1999b} were examined in the light of the intrinsic/extrinsic classification of these authors. For the extrinsic stars (excluding the symbiotic binaries), we found an upper limit of 0.12\,mag for the standard deviations of the $V$-band light curves. Hence applying a lower limit of 0.12\,mag on the standard deviations of the $V$-band light curves should result in the exclusion of the extrinsic S-type AGB stars. The remaining sample consisted of 459 presumably \emph{intrinsic} S-type stars with variability amplitudes larger than 0.12 mag.

To estimate the source variability of the S-type stars, we have cross-identified all stars in the GCSS2 with the two major available variability surveys: the All Sky Automated Survey (ASAS) located at the Las Campanas Observatory \citep{ASAS} and the Northern Sky Variability Survey\footnote{See http://skydot.lanl.gov.} (NSVS) operated at Los Alamos National Laboratory \citep{NSVS}. This resulted in 1154 S-type stars with V-band light curves. 

In the last step, the sample was reduced to contain only stars that could be observed by Spitzer within a reasonable integration time. Therefore we used a cross-identification of the remaining sources with the catalogues of the Infrared Astronomical Satellite (IRAS) and the Two Micron All Sky Survey (2MASS). From the relation between the $K$ magnitude and the 12~\um IRAS flux\footnote{Actually, the IRAS fluxes are given in Jy and are thus \emph{monochromatic flux densities}, but for ease of use, we will continue to use the term \emph{flux}.} we estimate that S stars with $4.5\,\mathrm{mag} < K < 5.5\,\mathrm{mag}$ exhibit maximum fluxes between 0.3 and 3 Jy in the Spitzer wavelength range, not so bright as to saturate the detector but sufficiently bright to allow short integration times. The final target list contained 90 sources (3 of which will be rejected based on the ground-based optical spectra).

\subsection{Extrinsic intruders in our sample}
Because the cut-off criterion on the standard deviations of the V-band light curve is not strict, we can expect a small number of extrinsic S-type stars in our sample. As said in the previous section, the best way to (a posteriori) confirm that a star is an S-type AGB star is to check for the presence of the 4238\,\AA~, 4262\,\AA~ and 4297\,\AA~Tc lines in high-resolution spectra \citep{Uttenthaler2007}.

As a filler program, we obtained high-resolution spectra for some of the brightest stars in our sample. The target selection was inhomogeneous, focussing on (i) stars with low-amplitude light curves and (ii) stars with an unusual appearance in the low-resolution optical spectra. The first group of stars was targeted because the chance of finding an extrinsic S-type star amongst the low-amplitude pulsators is larger since true AGB stars are expected to have larger pulsation amplitudes. The stars with an unusual low-resolution optical spectrum could be mis-classified as S-type stars in the GCSS2 \citep{Stephenson2} (e.g. \object{BD$+$02~4571}). For more information on the low- and high-resolution spectra and the lightcurves, see Sect. \ref{sct:data}.

As shown by \citet{Uttenthaler2007}, we can make the distinction between Tc-rich and Tc-poor stars by comparing the flux ratios in the Tc line to a pseudo-continuum flux. Using this method a reliable distinction of this kind can be made, even for low SNR spectra.

Three stars did not show significant Tc lines and were removed from our sample (\object{CPD$-$19~1672}, \object{BD$+$02~4571} and \object{CSS~718}). Because of the inhomogeneous subset of stars with high-resolution spectra, this ratio of 3 stars without Tc lines in a subset of 13 stars is not representative for the entire sample. Discarding these stars, we are left with a total sample of 87 S-type AGB stars.

\subsection{Bias towards low mass-loss rates}\label{sct:bias}
During the target selection phase, we put a limit on the $K$-band magnitude and cross-identified the sample with the ASAS and NSVS catalogs. Because both surveys are limited to $V$-band magnitudes of approximately 14 and the $K$-band magnitude is limited to the 4.5--5.5\,mag range, we have effectively limited the sample to stars with $V-K < 10\,\mathrm{mag}$. 

Based on the {\sc marcs} model atmospheres for S-type AGB stars without detectable circumstellar material (see Section \ref{sct:stelparam}), we can expect $V-K$ values up to 10\,mag or more for stars with $T_\mathrm{eff}$ lower than 2700\,K, even without circumstellar material. Stars \emph{with} circumstellar dust are reddened and thus have higher values for the $V-K$ color. The limitation on this color will exclude the stars with the highest mass-loss rates because they appear redder \citep{Guandalini2010}.

\begin{figure}
  \resizebox{\hsize}{!}{\includegraphics{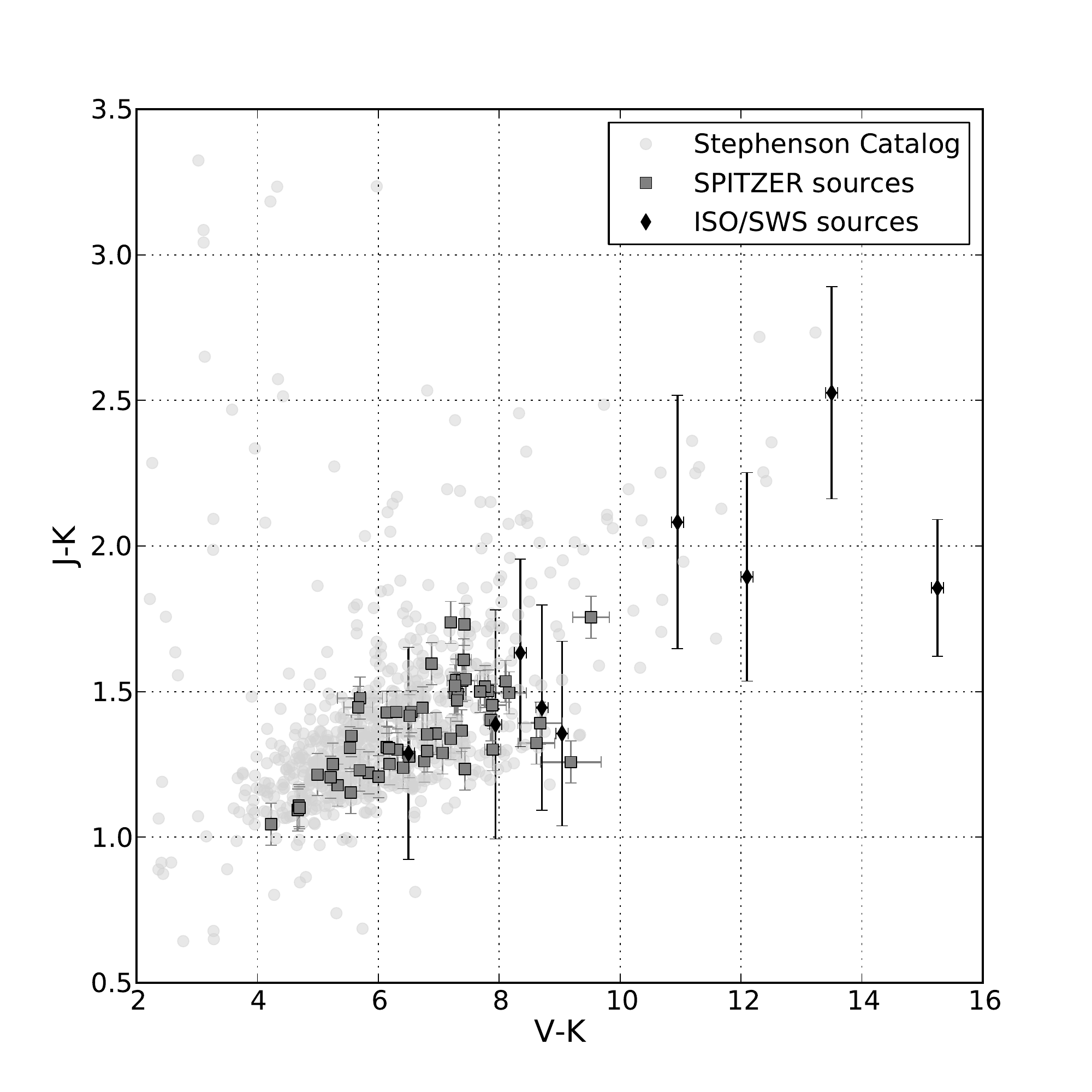}}
  \caption{The $V-K$ vs.\ $J-K$ diagram of the S-type stars in the Stephenson catalog (light grey dots), the ISO/SWS targets presented in \citet{Hony2009} (black diamonds) and the Spitzer sample presented in this paper (dark grey squares).
  }
  \label{fig:vkjk_ss}
\end{figure}

In Fig. \ref{fig:vkjk_ss}, we show the $J-K$ and $V-K$ colors for our stars, the sample of S stars observed with ISO/SWS \citep{Hony2009} and all S-type stars found in the Stephenson Catalog for comparison. For our stars, the phase-calibrated $V$-band magnitudes are derived from the light curves in the ASAS or AAVSO database and new measurements with the MkII photometer, discussed in more detail in Sect.~\ref{sct:SAAOphot}. For the ISO/SWS sources we use the AAVSO database to derive the phase-calibrated $V$ magnitudes. The relations presented in \citet{Carpenter2001} were used to convert the $K_\mathrm{2MASS}$ magnitudes to $K_\mathrm{SAAO}$ magnitudes. For the S-type stars found in the Stephenson catalog that are not observed with Spitzer or ISO/SWS, the $V-K$ and $J-K$ colors are not phase-calibrated and not corrected for the interstellar reddening because the distances and pulsation characteristics are very uncertain.

From this figure, we can see that although we have indirectly removed stars with $V-K > 10\,\mathrm{mag}$ from our sample, this effect is only small and we excluded only the coolest stars which have high mass-loss rates, less than 5\% of all S-type stars. Furthermore, the figure shows that the sample of S-type AGB presented in \citet{Hony2009} is biased towards higher mass-loss rates, because the limited sensitivity of the ISO/SWS spectrograph favors stars with a strong infrared excess. We have to keep both biases in mind when we compare our results with literature.

\section{Data}\label{sct:data}
\subsection{Spitzer Space Telescope}
The IRS data of all 90 S-type stars were obtained using the IRS Standard Staring mode. The targets were observed in two positions in each aperture (SL1, SL2, LL1 and LL2) to obtain low-resolution spectra over the entire wavelength range (5--38\,$\mu$m). All data were taken during a period from September 2006 to October 2007 (Program ID 30737, P.I. S. Hony).

For the data reduction we used the FEPS pipeline, developed for the Spitzer legacy program ``Formation and evolution of planetary systems''. A detailed description can be found in \citet{Hines2005}. For the extraction we start from the intermediate droop data products as delivered by the Spitzer Science Center, together with the SMART reduction package tools described in detail in \citet{Higdon2004}. The spectra were extracted using a fixed width aperture, so that 99\% of the source flux falls within the aperture. After the extraction of the spectrum for each nod position and cycle, a mean spectrum over all slit positions and cycles is computed for each individual order. Thereafter, the orders are combined. In regions where there is order overlap, the fluxes are replaced by the mean flux value at each wavelength point. The IRSFRINGE package was used for the de-fringing of the spectra. For the calibration, spectral response functions were calculated from standard stars and corresponding stellar models. The estimated average uncertainty on the spectral response function  and uncertainty on the extracted spectra are incorporated in the final estimated error.  Finally, for each target a one-dimensional spectrum is computed by stitching together overlapping orders. This is done by multiplying the LL orders with a correction factor to match the mean flux with the SL orders in the overlapping wavelength range. No wavelength shifting or spectral tilting is performed.

\subsection{Additional ground-based data}
\subsubsection{SAAO}\label{sct:SAAOphot}
For 69 of the 87 stars in our sample, we have obtained JHKL magnitudes (1.25, 1.65, 2.2 and 3.5 \um) at the South African Astronomical Observatory (SAAO). The observations were made with the MkII infrared photometer on the 0.75-m telescope. The accuracy is typically better than 0.03 mag in the J, H, and K band, and better than 0.05 mag in the L band. All JHKL magnitudes derived for these stars are given in Table \ref{table:data}.

\subsubsection{Low-resolution optical spectra}\label{sct:lowres}
Low resolution optical spectra were obtained for 81 of the 87 stars in our sample. The southern objects were observed with the 1.9-m telescope at SAAO, whereas the more northern objects were observed with the ISIS spectrograph mounted on the 4.2-m William Herschel Telescope (WHT) at the Observatory de Roque de los Muchachos (ORM) situated at La Palma. 

The spectra were wavelength calibrated with a CuNe+CuAr lamp. The correction for atmospheric extinction was performed using the average curve for the local, continuous atmospheric extinction at the site. The resulting SAAO spectra cover the 4800-7700$\,\AA$ range with a resolving power of $\mathrm{R} = \frac{\lambda}{\Delta \lambda} \approx 3800$. The spectra of WHT consist of a blue arm, spanning 3700-5350$\,\AA$ with a resolving power of $\mathrm{R} \approx 5000$ and a red arm, spanning 5100-8100$\,\AA$ with a resolving power of $\mathrm{R} \approx 3800$.

\subsubsection{High-resolution optical spectra}\label{sct:highres}
We obtained high-resolution spectra for some of the brightest targets in our sample. Eight southern sources were observed with the Coralie \'{e}chelle spectrograph at the Euler telescope in La Silla, spaning 3900-6800$\,\AA$, with a resolving power of $\mathrm{R} \approx 50\,000$. Thirteen northern sources were observed with the HERMES \'{e}chelle spectrograph at the Mercator telescope in La Palma \citep{Raskin2011}, covering 3770-9000$\,\AA$ with a resolving power of $\mathrm{R} \approx 85\,000$. We use the end-products of the automatic data reduction pipelines for both instruments.

\subsubsection{Dereddening}\label{sct:dereddening}
To remove the effects of interstellar, wavelength dependent extinction from the ground-based and Spitzer data, we start from the 3-dimensional reddening models of \citet{Drimmel2003} and \citet{Marshall2006}. The Marshall model returns the K-band extinction for over 64 000 lines of sight, each separated by $15^\prime$, based on the 2MASS point source catalogue. When possible, we use this model to estimate the extinction, but the Marshall model only covers the inner Galaxy ($|l| \le 100^\circ, |b| \le 10^\circ$). The model presented in \citet{Drimmel2003} returns V-band extinctions for the entire sky. The data are dereddened, using the extinction laws of \citet{Cardelli1989} for the optical and NIR data and \citet{Chiar2006} at longer wavelengths.

\section{Stellar parameters}\label{sct:stelparam}
Because the spectral appearance of S-type AGB stars depends strongly on the stellar parameters such as ${\rm T}_{\rm eff}$, C/O ratio, s-process element abundance and iron abundance, it is possible to constrain these stellar parameters by comparing model atmospheres to observational data. These stellar parameters characterize the \emph{chemical} conditions in the stellar atmosphere. Especially the C/O ratio is expected to have a very profound impact. 

Because the sample selection is based on the ASAS and NSVS catalogs, we are able to constrain the pulsation characteristics of most stars. The period and amplitude of the V-band variations characterize the \emph{variability} of the stellar atmosphere.

\subsection{Chemical characterization}
\subsubsection{Comparison with MARCS model atmospheres}
The stellar parameters were derived using a grid of model atmospheres specially computed for S-type stars with the {\sc marcs} code \citep{Gustafsson2008}. 
This model grid has been described in \citet{VanEck2010} and \citet{Neyskens2010}. The full detailed description will be published in a forthcoming paper (Plez et al., in preperation).
To summarize briefly, the grid covers the following parameter space: \begin{itemize}
\item $2700 \le T_{\rm eff} $ (K) $ \le 4000$ in steps of 100\,K
\item C/O = 0.5, 0.750, 0.899, 0.925, 0.951, 0.971, 0.991
\item $[\rm{s}/\rm{Fe}]$ = 0, +1, +2 dex
\item $[\rm{Fe}/\rm{H}]$ = -0.5, 0 dex.
\end{itemize}
The unequal C/O spacing has been designed to cover in an optimal way the change of the absorption features affecting the optical spectra when C/O increases from 0.5 to unity. All models were computed for a stellar mass of $M = 1\,M_\odot$ and with $[\alpha/\rm{Fe}]~=-0.4~\times~[\rm{Fe}/\rm{H}]$. It is essential to take into account the non-solar C/O ratio of S-type stars since it has an important impact on the partial pressures of H$_2$O and TiO, two major opacity contributors. Coupled with s-process overabundances, it affects the depth of the prominent ZrO bands observed in S-type stars. The effect of the gravity was found to be too degenerate to be constrained solely with the observational data available for the present study. Therefore it was fixed in the following way, devised to approximately follow the portion of the AGB covered by the S stars under consideration: $\log{g} = 0$ if $T_{\rm eff}  < 3400$\,K, $\log{g} = 1$ otherwise. Similarly, low-resolution spectra do not allow to lift the degeneracy between metallicity and s-process overabundance.  

A set of well-chosen photometric and narrow-band indices were then computed on the basis of low-resolution synthetic and observed spectra (see \citet{VanEck2010} for more details). The low-resolution spectra used here consist of Boller \& Chivens spectra ($\Delta \lambda = 3$\,\AA , $4400-8200$\,\AA), SAAO spectra ($\Delta \lambda = 1.8$\,\AA , $4800-7700$\,\AA) and WHT spectra ($\Delta \lambda=1.8$\,\AA , $5100-8100$\,\AA). Two colour indices $V-K$ and $J-K$ as well as three spectroscopic indices (strengths of the carefully-selected ZrO and TiO bands, as well as the strength of the Na D lines) were used to define a total of five indices.

Chi-square minimization between observed and synthetic indices then led to a set of plausible physical parameters (effective temperature, gravity, metallicity, C/O ratio, and [s/Fe]). The agreement between selected synthetic spectra and the observed optical and infrared low-resolution spectra is in most cases very good. However, it is sometimes difficult to select one out of the first few ``best-matching'' synthetic spectra. In fact, the uncertainties inherent to the whole procedure (e.g., non-simultaneous photometry and spectroscopy, stellar variability, or approximate values of the reddening) have to be taken into account in estimating the best model and the uncertainty on the stellar parameters. The resulting value and uncertainties on the stellar parameters $T_{\rm eff}$, C/O, and [s/Fe] that provide a reasonably good fit between the low-resolution observed and synthetic spectra are given in Table~\ref{table:1}.

\subsubsection{Stellar properties}
\begin{figure}
  \resizebox{\hsize}{!}{\includegraphics{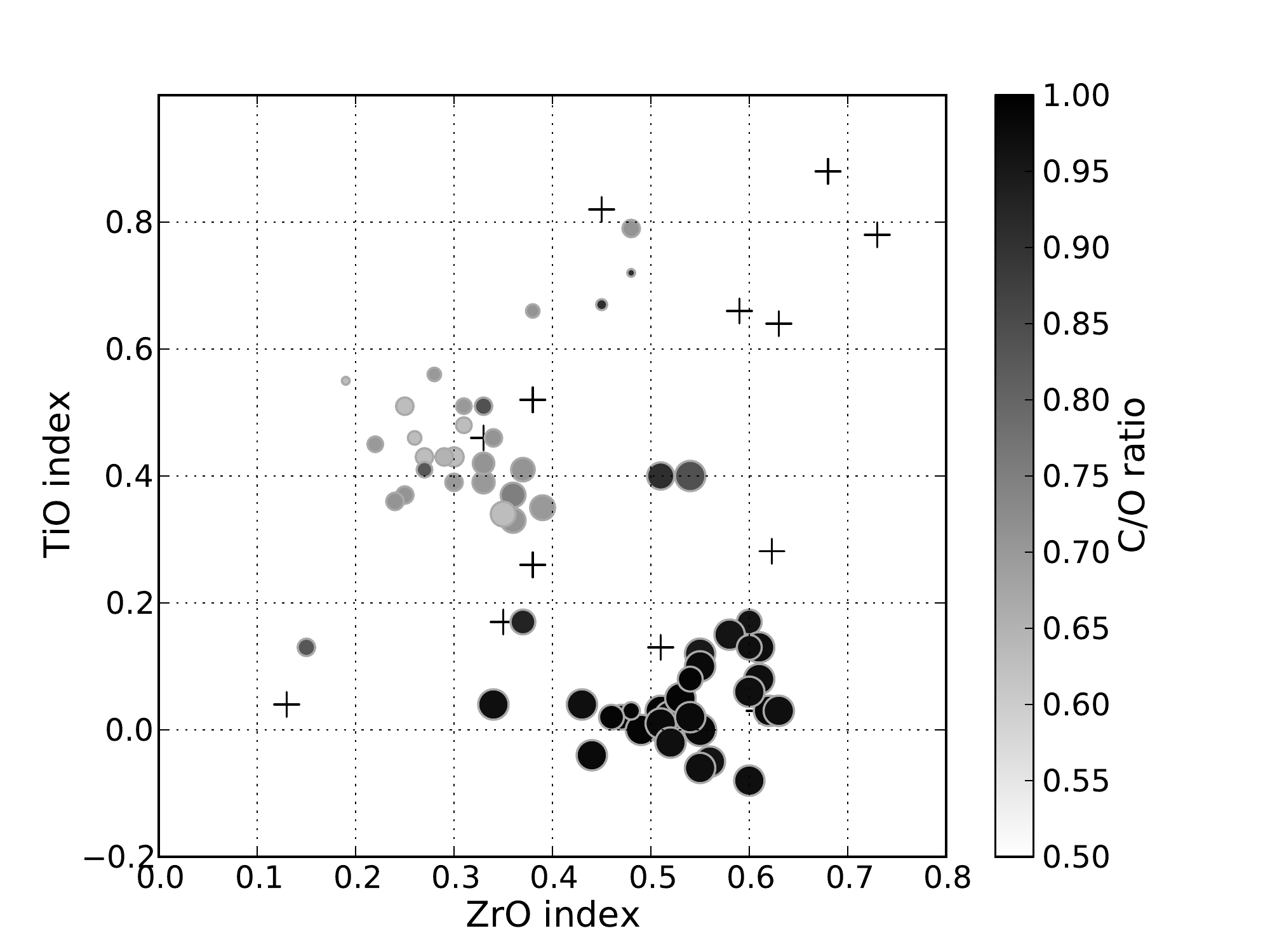}}
  \caption{The average line-to-continuum-ratios of the ZrO and TiO absorption bands  within the spectral range of WHT and SAAO low-resolution spectra (i.e. ZrO and TiO indices) for all stars in the sample. The size of the symbols increases with [s/Fe] and the color varies with C/O. Crosses indicate stars for which $\mathrm{T_{eff}}$, C/O and [s/Fe] are unknown.
  }
  \label{fig:zrotiosfeco}
\end{figure}

From the results in Table~\ref{table:1} we can get a general overview of the chemical properties of the stars in our sample. The effective temperature spans a range from 2900\,K for the coolest S-type star to 3650\,K for the hottest one, with most stars with $T_{\rm eff}$ around 3300\,K. There is a strong correlation between the place in the ZrO--TiO plot and the C/O ratio. The ZrO and TiO indices can thus be used as indirect indicators for the C/O ratio (see Fig.~\ref{fig:zrotiosfeco}). We will make use of this relation throughout the entire paper. The figure shows that one can separate two groups, stars with a low C/O ratio are located at the upper left corner and the stars with a high C/O ratio towards the lower right corner. The position in this diagram can thus be used as a decent estimate for the C/O ratio.

From this figure, we can also see that there is a significant correlation between the [s/Fe] values and the C/O ratios. This obvious correlation was predicted and observationally confirmed in AGB stars \citep{Smith1990, Mowlavi1997} and S-type stars in particular \citep{VanEck2010}. Because all combinations of C/O and [s/Fe] were allowed, the presence of the relation is not inherent to the method to derive the stellar parameters, but rather shows that the method works well and can reproduce this relationship.

\subsection{Variability characterization}\label{sct:periodanalysis}
To derive the pulsation parameters, we use the databases of NSVS and ASAS. A complete description of the instruments and reduction processes can be found in \citet{Wozniak2004} and \citet{Pojmanski2004}. For some stars, additional $V$-band light curves are available from the American Association of Variable Star Observers (AAVSO).

Because for most stars only a few data-points are available, we use the \emph{generalized least-squares} periodogram to find the pulsation periods of all stars \citep{Zechmeister2009}. This periodogram takes into account that for small datasets the sample average can differ significantly from the true average. Some semi-regular variables are known to exhibit long secondary periods (LSP), typically 5 to 15 times longer than the primary period. When detected, such variations are removed (prewhitened) from the light curve and the period analysis is repeated. The results of this analysis can be found in Table \ref{table:periods}.

The period analysis resulted in detected periods for 54 of the 87 stars in our sample. For most cases where no significant periodicity was found, this can be explained by the short timespan of the NSVS light curves. These short timespans allow an identification of the amplitude of the variations, but a period analysis needs longer timespans. For eight stars, the lack of periodicity is due to intrinsic irregular behavior.

In addition to these periods, the General Catalog of Variable Stars (GCVS) gives the period to 18 stars. For 11 of these stars, we also derived a period, based on the NSVS, ASAS and AAVSO light curves. As shown in Table \ref{table:periods}, the derived period for 10 of these 11 stars is in close agreement with the periods shown in the GCVS. However, the recent data shows that \object{V899~Aql} has a period of 374 days and is most likely a Mira, this star is classified as a semi-regular variable in the GCVS, with a period of 100 days. In further calculations, we will use the periods and pulsation types derived from the NSVS, ASAS, and AAVSO light curves if available. The sample contains 6 irregular pulsators, 38 semi-regular variables and 21 Miras.

\section{Naked stellar atmospheres}\label{sct:nakedstelatm}
The 87 S-type stars in our sample show a wide variety of features in the infrared spectra. Some stars show clear dust/molecular emission features, while others only show the molecular absorption features typical for ``naked'' stellar atmospheres. Because stars with and without emission features have to be treated separately, we classify the targets into naked stellar atmospheres and stars with additional emission features. Out of the 87 stars, 47 are classified as naked stellar atmospheres. The other stars show relatively strong emission features on top of these stellar atmospheres which are due to (i) molecular emission from extended atmospheres (3 stars), (ii) the typical emission features from hydrocarbons (4 stars) or (iii) dust emission (32 stars). Here, we focus on the \emph{naked} stellar atmospheres. The dust emission will be discussed in more detail in Sect. \ref{sct:dustemis} and the molecular emission in Sect. \ref{sct:molemis}.

\begin{figure*}
  \resizebox{\hsize}{!}{\includegraphics{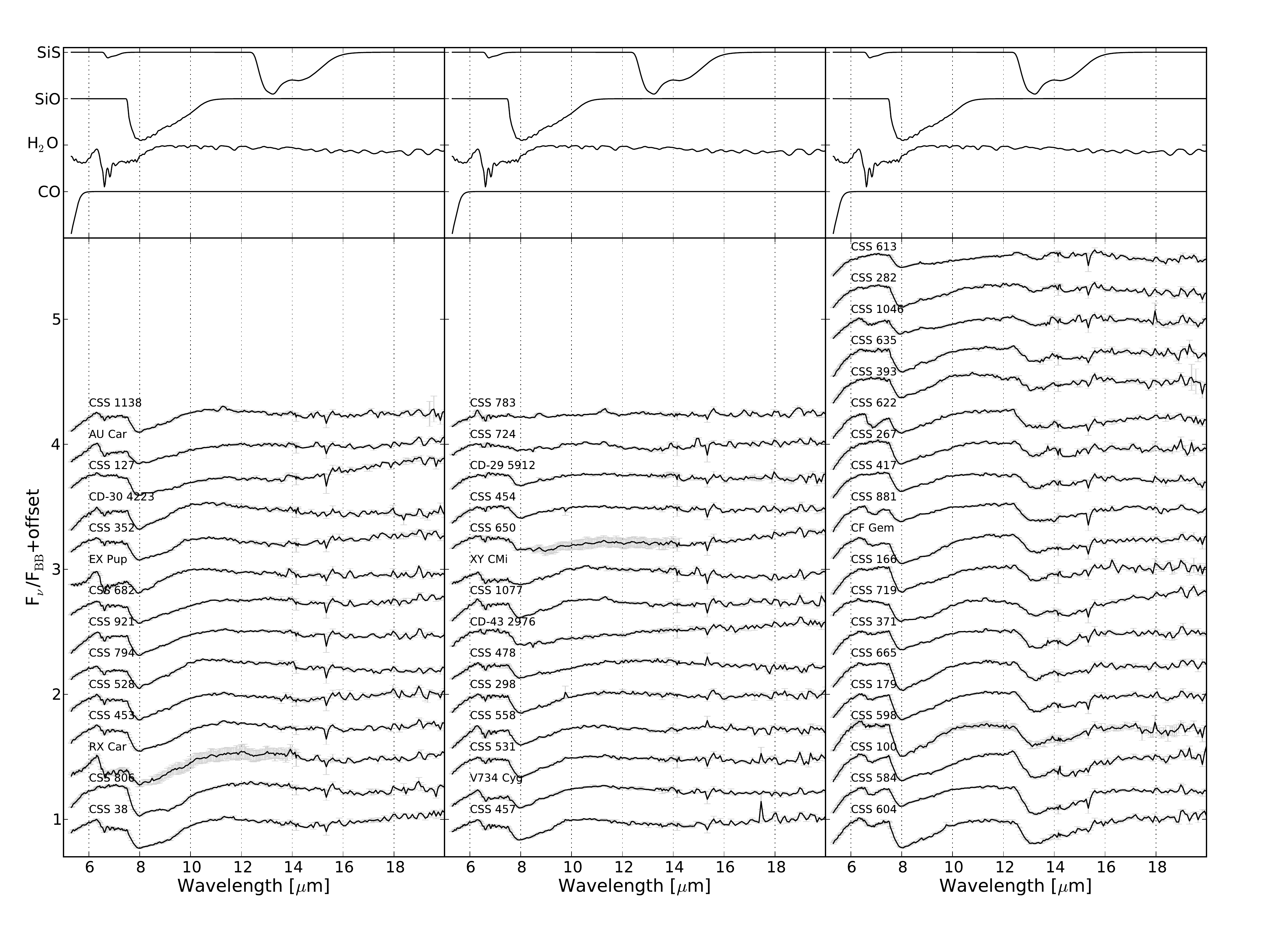}}
  \caption{Top panel: the template spectra of SiS, SiO, \h2o, and CO. Bottom panel: the spectra of all naked stars from 5\,\um to 20\,\um, normalized with a blackbody spectrum. The the two panels on the left show the stars without SiS, the right-most panel shows all stars with significant SiS absorption. The sharp feature at 15.3\,\um seen in \emph{all} stars is most likely an artifact of the data reduction process.}
  \label{fig:nakedstars}
\end{figure*}

\subsection{Overview of the observed absorption bands}
In the bottom panel of Fig. \ref{fig:nakedstars} we show the normalized infrared spectra of all 47 naked stellar atmospheres of our sample; the top panel shows the normalized templates used for comparison. These templates have column densities and temperatures that are representative for AGB stars \citep[CO at $T=2000\,\mathrm{K}$ and $N = 10^{21}\,\mathrm{cm}^{-2}$; H$_2$O at $T=1500\,\mathrm{K}$ and $N = 10^{20}\,\mathrm{cm}^{-2}$; SiO at $T=2000\,\mathrm{K}$ and $N = 10^{21}\,\mathrm{cm}^{-2}$; SiS at $T=1500\,\mathrm{K}$ and $N = 10^{19}\,\mathrm{cm}^{-2}$;][]{Cami2002, Cami2009}.

From this figure it is clear that \emph{all} stars in our sample show absorption bands of oxygen-rich molecules such as H$_2$O and SiO. The right panel shows the 19 stars with a significant SiS absorption band at 13\,\um. The remaining panels show the 28 stars without SiS absorption. We did not detect the strong absorption features at 7.2 (SO$_2$), 7.6 (H$_2$O$_2$), 13.8 (C$_2$H$_2$), 14 (HCN) or 15~\um (CO$_2$) typical for of SO$_2$, H$_2$O$_2$, \c2h2, \hcn and \co2.

\subsection{The relation between SiS absorption and the C/O ratio}
\begin{figure}
  \resizebox{\hsize}{!}{\includegraphics{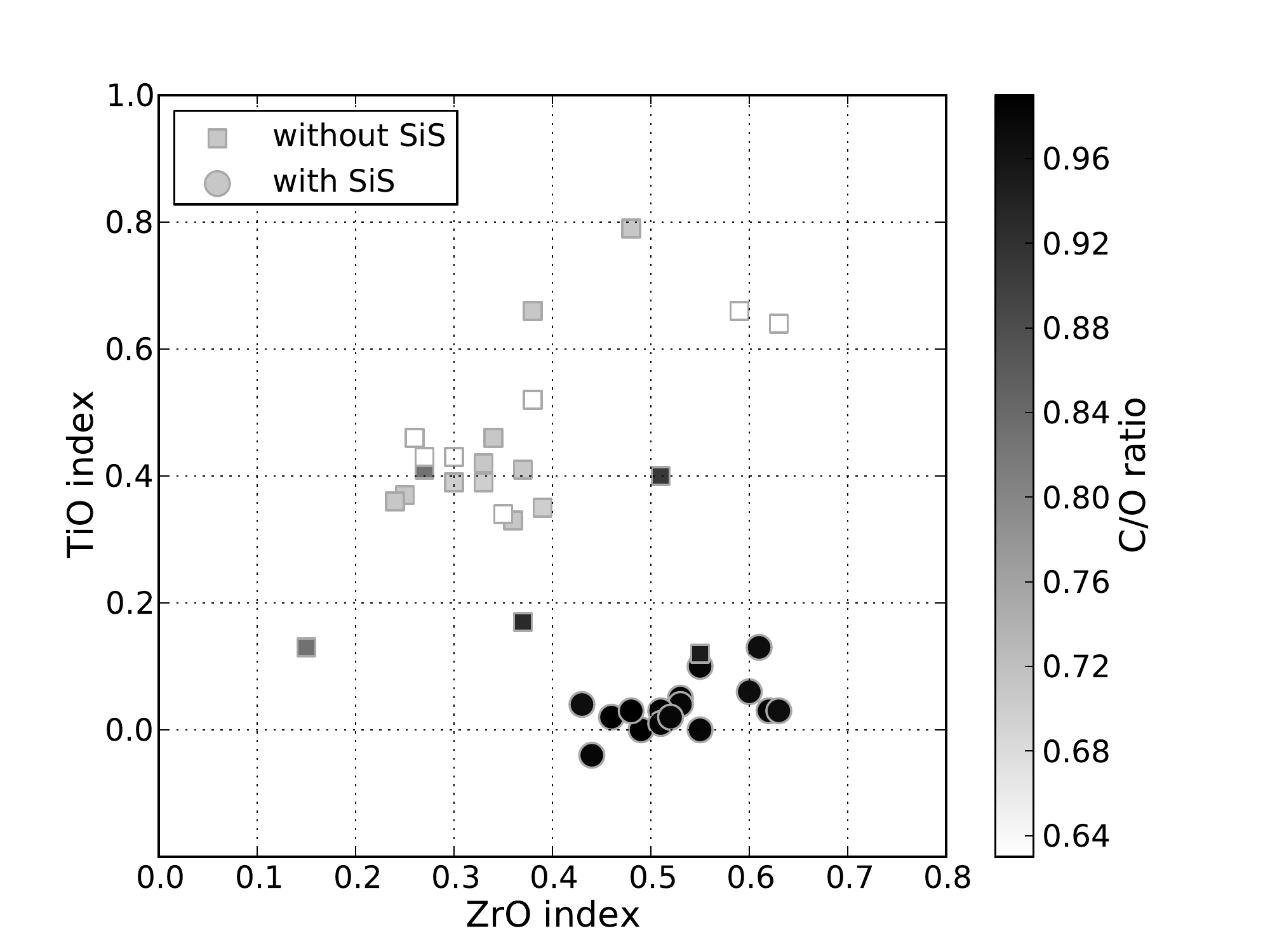}}
  \caption{The ZrO and TiO indices for the naked stellar atmospheres in the sample. Squares and circles indicate stars with and without significant SiS absorption with a colorscale that indicates the C/O ratio. Stars for which we have no estimates for $\mathrm{T_{eff}}$, C/O, and [s/Fe] are indicated with open symbols. The figure shows a one-one relation between C/O ratio and presence of SiS absorption in the infrared spectrum.}
  \label{fig:sio_sis_co}
\end{figure}

Considering the partial pressures, \citet{Cami2009} predicted that SiS absorption will only be visible in the S-type AGB stars with C/O ratios very close to unity. Fig. \ref{fig:sio_sis_co} shows the ZrO and TiO index for all naked stars, with a color scale indicating the C/O ratio for stars with and without significant SiS absorption. From this figure it is clear that there is a strong relation between the estimated C/O ratio and the presence of SiS absorption: \emph{all} stars with $\mathrm{C/O} > 0.96$ show SiS absorption and \emph{all} stars with $\mathrm{C/O} < 0.96$ have no significant SiS absorption bands. It should be noted that the C/O and SiS determinations are completely independent. The C/O ratio is determined, based on the optical spectroscopy and optical and NIR photometry, while the SiS bands are only observed in the Spitzer spectra.

Two conclusions can be drawn from this: (i) the C/O ratio is key to understanding and modeling the IR spectra of S-type AGB stars and (ii) the presence of the SiS absorption bands can help constrain the C/O ratio in stars, which can be a useful tool when the C/O ratio cannot be derived otherwise.

\section{Dust identification}\label{sct:dustemis}
In this section we focus on the 32 stars with significant dust emission features in the IR spectra. We first give an overview of the dust formation sequence in M- and C-type stars and then we compare this with what we find for S-type AGB stars.

\subsection{Dust formation in AGB stars}
\begin{figure*}
  \resizebox{\hsize}{!}{\includegraphics{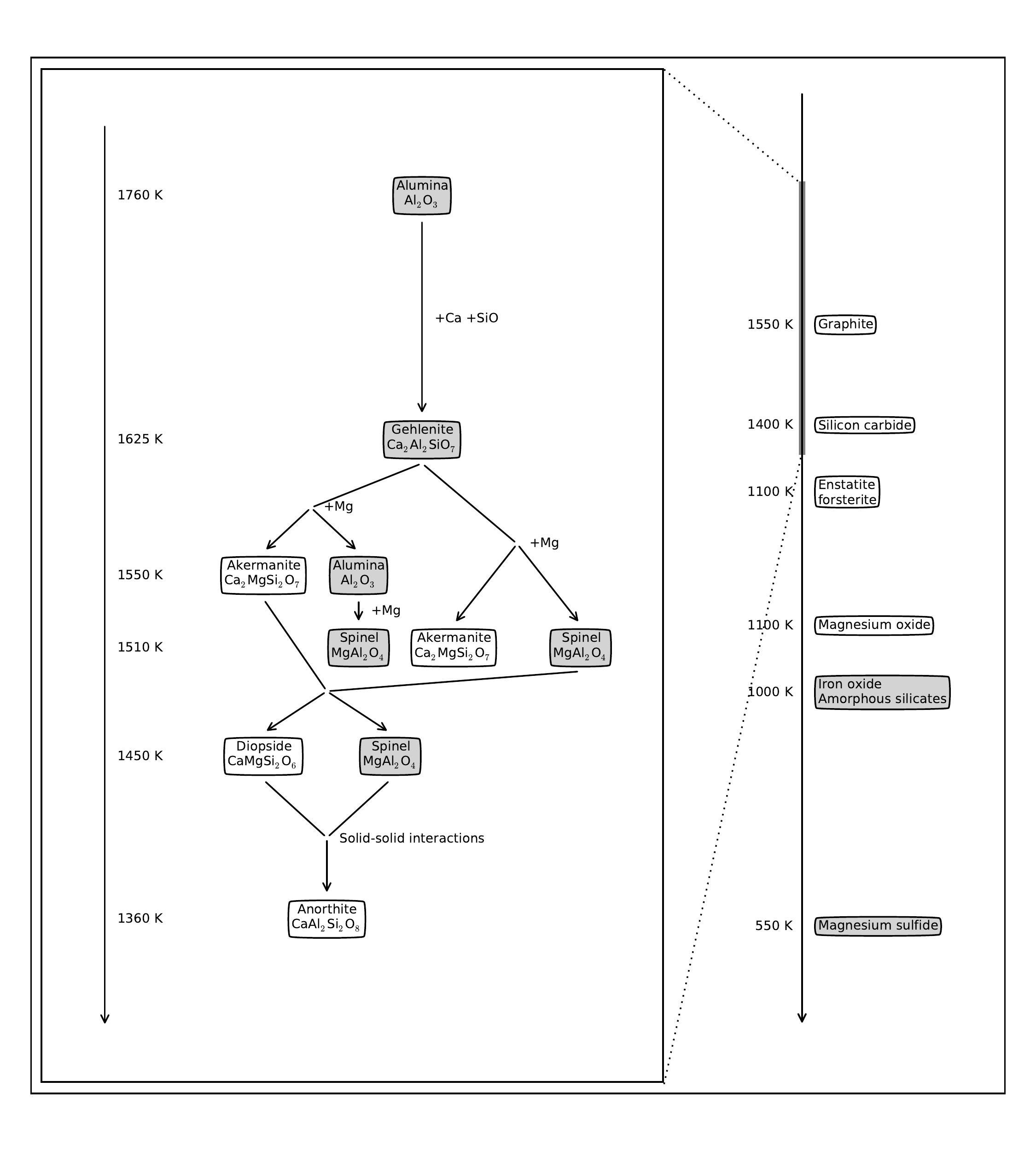}}
  \caption{The estimated dust condensation temperatures for different dust species found in AGB stars. The large inset panel zooms in on the dust condensation sequence for oxygen-rich stars, starting from alumina grains. We highlighted the species which were included in our dust decomposition tool in grey.}
  \label{fig:dustcondensation}
\end{figure*}

The study of the dust condensation chemistry in AGB stars has mostly focused on the formation of oxygen-rich dust species in M-type stars or carbon-rich dust species in C-stars. In Fig. \ref {fig:dustcondensation} we show the formation pathways for several dust species that are known to be important in oxygen-rich and/or carbon-rich stellar outflows.

Although it is not yet observationally confirmed, a consistent theory for the dust formation in \emph{M-type AGB stars} exists. Theoretically, the first condensate to form is alumina grains ($\mathrm{Al_2O_3}$), which can condense at temperatures as high as 1760$\,$K. When these grains interact with the circumstellar SiO and Ca, melilite is formed. This melilite is first pure gehlenite, but can later on become akermanite due to additional interactions with the circumstellar Mg. Due to the formation of akermanite from gehlenite grains, alumina grains are released from the gehlenite grains and can go on to form additional alumina grains (if the local temperature is above 1510$\,$K) or spinel grains (if the temperature is below 1510$\,$K). The melilite reacts at 1450$\,$K to produce diopside and more spinel. The reaction of diopside and spinel can form anorthite. Further on in the wind, at lower temperatures, these grains can be the seeds for the formation of enstatite, forsterite and the amorphous silicates we observe in many AGB stars. \citep[this scenario was proposed and adapted by][]{Grossman1974, Onaka1989, Tielens1990M, Cami2002}. Additionally, several species, that are not part of the formation path that eventually ends with silicates, can condense at certain temperatures. For example, due to the reaction of \h2o with Fe, iron oxides form at a temperatue of approximately 900\,K, while the magnesium-rich oxides can form at higher temperatures (up to 1200\,K) due to the reaction of \h2o with atomic magnesium \citep{Ferrarotti2003}.

The broad picture for \emph{C-type AGB stars} is simpler, but some key details have yet to be filled in. The main dust species detected in carbon-rich stars are amorphous carbon, silicon carbide and magnesium sulfide \citep{Treffers1974, Goebel1985, Martin1987, Andersen1999, Chan1990, Hony2002, Zijlstra2006, Lagadec2007, Leisenring2008}. One of the key questions in the carbon-species formation is the importance of mantle chemistry, for example the nucleation of amorphous carbon and magnesium sulfides as mantles around silicon carbide grains have been proposed as important dust formation mechanisms \citep{Lorenz-Martins2001, Zhukovska2008}.

Assuming a certain dust condensation sequence and an expanding stellar wind, one expects a process called \emph{freeze-out}, where the sequence is not completed but stops at an intermediate product when the wind density drops below the density required for the next step. This freeze-out has been tentatively observed in AGB stars \cite[e.g.][]{McCabe1979, Heras2005}. An alternative, but qualitatively similar effect could be the exhaustion of oxygen in these S-type environments. If the C/O ratio is close to unity, the formation of alumina grains can use up most of the oxygen atoms available after the formation of CO. If this happens, the density of free oxygen atoms is too low to continue the dust condensation sequence along the path presented in Fig. \ref{fig:dustcondensation}, analogous to the freeze-out \citep[as predicted by][]{SloanPrice1998}. We will refer to this process as \emph{oxygen depletion}.

\subsection{Dust decomposition tool}
The first crucial step towards understanding the dust condensation sequence of these S-type stars in the present sample is the identification of the dust emission features in the infrared spectra. We developed a simple tool to decompose the infrared spectra into a stellar continuum and the dust emission features of each individual dust species.

If we assume an optically thin circumstellar environment, where the dust condenses out instantaneously once the temperature gets below a certain threshold $T_0$ and where the dust grains are in thermal equilibrium, we can write the temperature profile in the circumstellar layers as:
\begin{equation}\label{eq:tempprofile}
T_\mathrm{dust}(r) = T_0\, \left(\frac{r}{r_0}\right)^{-1/2}
\end{equation}
where $r_0$ corresponds to the radius where the temperature in the outflow drops below $T_0$.

The flux coming from a single dust species (using spherical grains and Mie theory) in an optically thin stellar wind with a constant mass-loss rate ${\dot{M}_\mathrm{d}}$ and outflow velocity $v$ can be written as:
\begin{equation}\label{eq:shellemission}
\mathrm{F}_{\nu,\,\mathrm{dust}} = \frac{3 \dot{M}_\mathrm{d}}{4\,v\,\rho_d}\,\frac{Q_\nu}{a} \frac{r_0}{D^2} \int_1^\infty {B_\nu(T_\mathrm{d}(x))\,\mathrm{d}x}
\end{equation}
where $\rho_d$ is the density, $Q_\nu$ the absorption efficiency, $a$ the radius of the spherical dust grains, $r_0$ the inner radius of the dust shell, $D$ the distance to the object and $T_\mathrm{d}(x)$ the temperature profile as in Eq. \ref{eq:tempprofile} at a distance $x = r/r_0$ \citep{Schutte1989}. Finally, we can combine all constants and unknown parameters into a single parameter $\beta_i$ for each dust species $i$. The total spectrum can then be written as:
\begin{equation}\label{eq:dustemission}
\mathrm{F}_\nu = F_{\nu,\,\star} + \sum_i \beta_i \int_1^\infty {Q_{\nu,\,i}\,B_\nu\,(T_\mathrm{d}(x))\,\mathrm{d}x}
\end{equation}
where $F_{\nu,\,\star}$ is the flux of the central star at frequency $\nu$ and $\beta_{i}$ is the scaling constant that controls the relative strength of the emission of a specific type of dust.

To arrive at the best fitting model, we use the Lawson and Hanson Non-Negative Least-Squares algorithm (NNLS), which gives the best model under the condition that $\beta_{i} \geq 0$. To estimate the uncertainties on the $\beta_{i}$ parameters, we multiply the uncertainties on the data with a factor until the $\chi^2$ value for the best fit is 1. Using these new uncertainties, we add normally distributed noise to the spectrum and calculate the scaling factors $\beta_{i}$ for 1000 randomly disturbed spectra. If the mean value is significantly non-zero (i.e. outside the 99.9\% confidence interval), we consider this a positive detection of that dust species in the infrared spectrum.

Because the MARCS model atmospheres still need validation at infrared wavelengths, we prefer to fit the stellar continuum and the emission simultaneously. Therefore, we allow the dust decomposition tool to make any linear combination of the \emph{naked} stars presented in Sect.~\ref{sct:nakedstelatm} and Planck functions of different temperatures, ranging from 500\,K to 4000\,K, as an estimate for the stellar continuum, $F_{\nu,\,\star}$. The big advantage of using the \emph{naked} stars is that we can change the relative depth of the CO, \sio, \h2o, and \sis bands to give the optimal fit, while we can change the slope of each individual star with the Planck functions. The drawback, however, is that we will not be able to distinguish between the stellar continuum and additional featureless continuum emission from e.g. metallic iron or amorphous carbon.

Finally, we have to decide which dust species to include in the dust decomposition tool and which to leave out. This is of vital importance because too few dust species will leave us unable to explain some emission features, while too many species would make the interpretation difficult and would lead to degeneracy. We started from a wide range of dust-species, including titanium oxides, titanium carbides, silica and all dust species presented in Fig. \ref{fig:dustcondensation}. We rejected those species that did not improve the goodness-of-fit to the IRS spectrum. The final selection of dust species is:

\begin{description}
\item[\textbf{Amorphous alumina}] Amorphous alumina has a high condensation temperature and is thus expected to be at the start of the dust condensation sequence of oxygen-rich stars (see Fig. \ref{fig:dustcondensation}). The most prominent feature of amorphous alumina is at 11.2\,\um with a broad red wing, reaching up to 15.6\,\um. It is a common component of the circumstellar wind of M-type AGB stars. Optical constants have been taken from \citet{Begemann1997}.

\item[\textbf{Amorphous silicates}] The strongest emission features due to oxygen-rich dust are the two emission features at 9.8 and 18\,\um due to the stretching and bending of Si-O bonds in amorphous silicates. These features dominate the infrared spectra of most M-type stars. Amorphous silicates form at relatively low temperatures and are thus expected to be at the end of the dust condensation sequence. Here, we have chosen olivines as representative of the typical silicates that are observed in AGB stars. Optical constants have been taken from \citet{Jaeger1994}.

\item[\textbf{Gehlenite}] We know from \cite{Hony2009} that the peak of the silicate features in S stars is not at 9.8\,\um as in M-type stars, but shifted towards redder wavelengths, up to 10.6\,\um. Since this cannot be explained with only amorphous silicates, we need to include gehlenite, which is an alumino-silicate (Ca$_2$Al$_2$SiO$_7$). A combination of amorphous silicates, gehlenite and amorphous alumina can explain the broad and shifted 10.6 micron feature. Optical constants have been taken from \citet{Mutschke1998}.

\item[\textbf{19.8\,\um feature}] Although the identification of this distinct emission feature at 19.8\,\um is often attributed to iron oxides \citep{Cami2002, Posch2002}, other authors have argued that the 19.8\,\um emission, together with additional emission features at 28 and 32\,\um is due to crystalline silicates \citep{Sloan2003}. These features are visible in many M-type stars and most clearly in stars with low mass-loss rates. We were unable to reproduce the 19.8\,\um emission using only crystalline silicates without introducing strong additional emission features in the 9--11\,\um region, which we do not observe. Therefore, we prefer iron oxides (more specifically Mg$_{\mathbf{0.1}}$Fe$_{\mathbf{0.9}}$O). Optical constants have been taken from \citet{Henning1995}.

\item[\textbf{The 13\,\um feature}] Since the first detection of the 13\,\um feature by \cite{Vardya1986}, it has been attributed to different carriers: corundum \citep[crystalline $\mathrm{Al_2O_3}$,][]{Glaccum1995}, silica \citep[SiO$_2$,][]{Speck1998} and spinel \citep[MgAl$_2$O$_4$,][]{Posch1999}. The debate on the exact carrier of the 13\,\um feature is still ongoing \citep{Depew2006}. In our dust decomposition tool, we tried all three carriers and found that both corundum and spinel can reproduce the emission, but the best results are obtained with spinel. However, detailed modeling is required to attribute the 13\,\um unambiguously to any one carrier, which is beyond the scope of this paper. Optical constants for spinel have been taken from \citet{Fabian2001}.

\item[\textbf{Magnesium sulfides}] The most remarkable feature in the spectra of these S-type stars is the broad 26--30\,\um emission feature, attributed to magnesium sulfides. A similar feature has already been observed in two S-type stars, W~Aql and $\pi^1$~Gru \citep{Hony2009}. Because of the lower spectral resolution of our set of {\sc spitzer} spectra, the two separate emission peaks at 26 and 28\,\um visible in the ISO/SWS spectra are blended to a single, broad emission feature. In our dust decomposition tool we use the mass absorption coefficients of magnesium sulfide. Optical constants have been taken from \citet{Begemann1994}.
\end{description}

Although we could expect the presence of the narrow 11.3\,\um emission feature of \textbf{amorphous SiC grains} \citep[optical constants have been taken from][]{Mutschke1999} in stars with C/O ratios close to one, replacing one of the aforementioned dust species with amorphous SiC grains resulted in higher $\chi^2$ values for all spectra. We can confidently state that amorphous silicates, gehlenite and amorphous alumina are necessary to explain the data. Furthermore, the inclusion of SiC, together with the other dust species, did not significantly lower the $\chi^2$ for our sample of S-type stars. Therefore we did not include amorphous SiC grains in the simple fitting routine. More detailed modeling is necessary to make a statement on the absence or presence of SiC grains, but we can conclude that the SiC emission is not strong in any of our stars.

\subsection{Overview of the detected dust species}
The results of the dust decomposition tool can be found in the online Appendix. Although S-type stars can have C/O ratios very close to unity, the majority of the stars only show dust emission features similar to those found in oxygen-rich stars.

\subsubsection{Classification of the dust spectra} 
Based on the results of the dust decomposition tool, we can subdivide the infrared spectra into 3 groups of spectra with: (i) strong emission features of silicates and gehlenite at 10 and 18\,\um (16 stars), (ii) emission of alumina at 11.2\,\um and of magnesium sulfide at 26--30\,\um (10 stars), or (iii) a 13\,\um and 19.8\,\um emission feature (6 stars).

\begin{figure}
  \resizebox{\hsize}{!}{\includegraphics{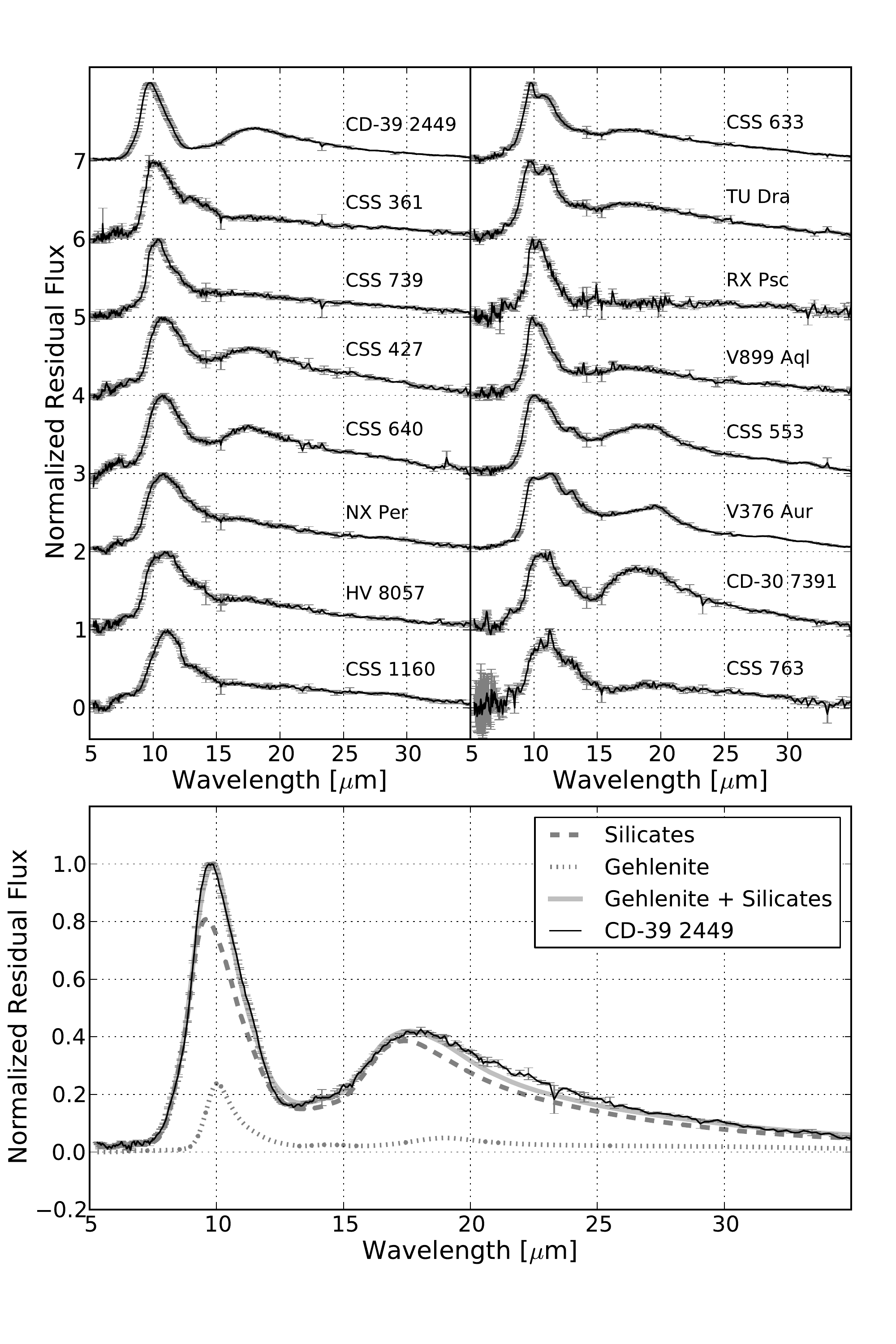}}
  \caption{The spectra of all stars in our sample with an infrared spectrum dominated by the 10 and 18\,\um features of silicates and gehlenite (Group~I). The top left panel shows the smoothest features at 10\,\um, while the features in the top right panel show substructure at 9.8, 11.2 and/or 13\,\um. The bottom panel shows \object{CD-39~2449}, with the relative flux contribution of silicates and gehlenite.}
  \label{fig:group1}
\end{figure}

\begin{figure}
  \resizebox{\hsize}{!}{\includegraphics{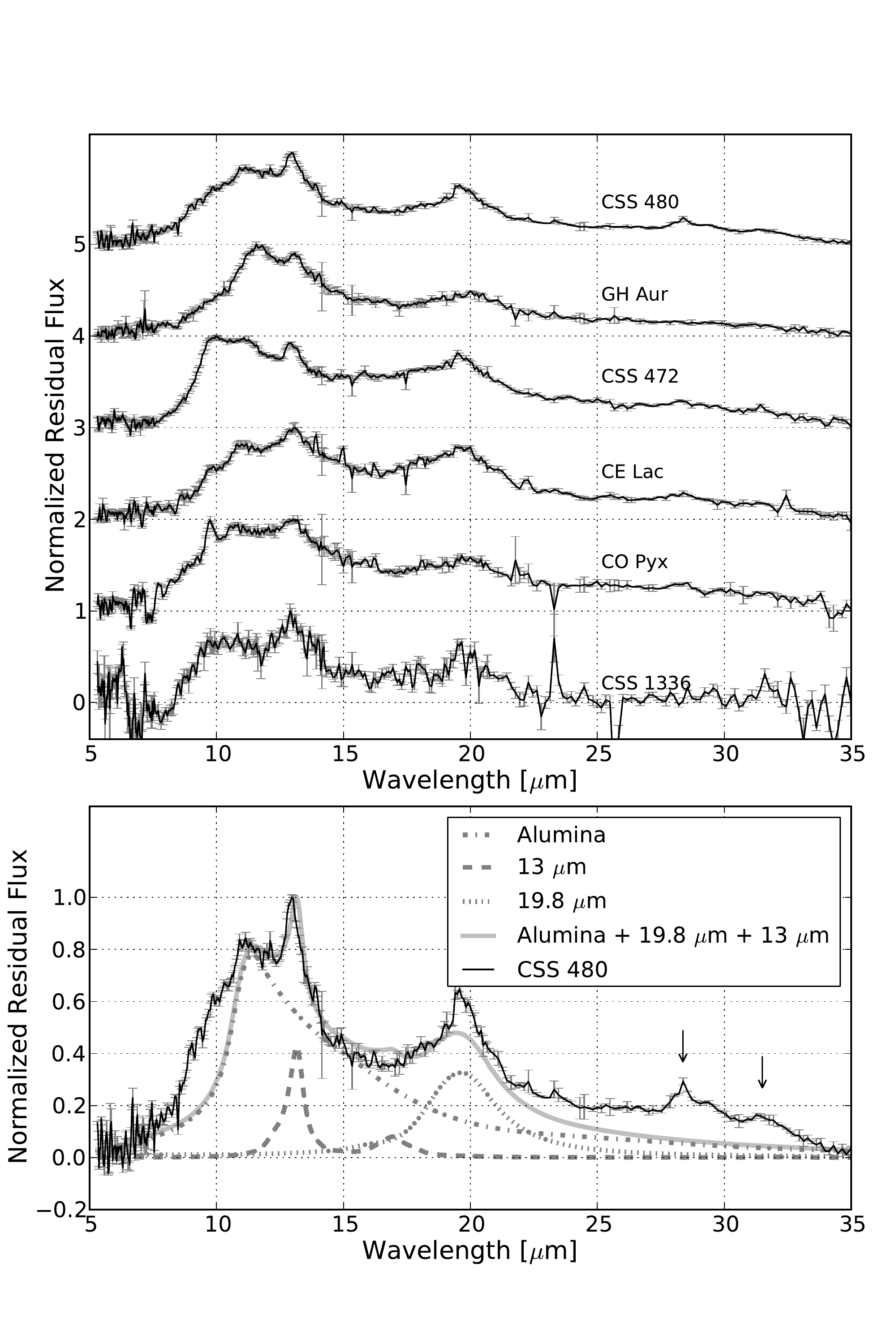}}
  \caption{The spectra of all stars in our sample with an infrared spectrum dominated by the emission of amorphous alumina at 11.2\,\um, the 13\,\um and the 19.8\,\um feature (Group~II). The bottom panel shows CSS~480, with the relative flux contribution each of these emission features. The previously detected, unidentified emission features at 28 and 32\,\um are indicated with black arrows.}
  \label{fig:group2}
\end{figure}

\begin{figure}
  \resizebox{\hsize}{!}{\includegraphics{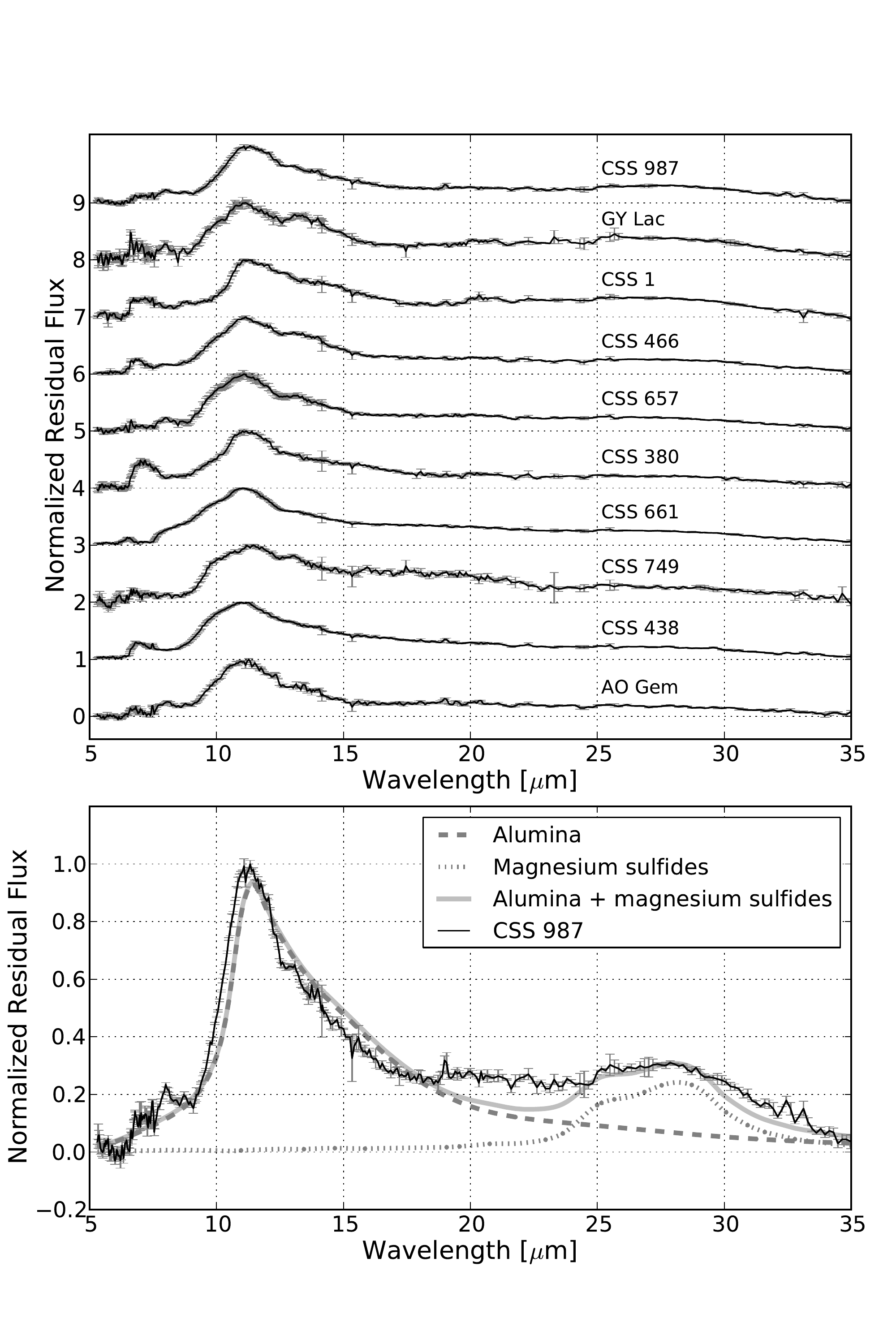}}
  \caption{The spectra of all stars in our sample with an infrared spectrum dominated by the emission of amorphous alumina at 11.2\,\um and of magnesium sulfides at 26--30\,\um (Group~III). The bottom panel shows CSS~987, with the relative flux contribution of alumina and magnesium sulfides.}
  \label{fig:group3}
\end{figure}

\begin{description}
\item[\textbf{Group~I, stars with strong silicate/gehlenite features:}] Approximately half of the stars with dust features in our sample, show a dominant emission feature around 10\,\um, due to the stretching of Si-O bonds of silicate or gehlenite dust grains. In Fig. \ref{fig:group1} we show all stars with a strong 10\,\um feature. All spectra in the right panel show substructure at 9.8 and 11\,\um, sometimes accompanied by a sharp peak at 13\,\um, the spectra in the left panel show a single broad emission feature that peaks at wavelengths up to 10.9\,\um. Although the dust decomposition tool can reproduce the emission features with substructure, it is impossible to reproduce the broad and smooth emission features, peaking at 10.5\,\um, visible in the left panel, without showing substructure.

\citet{Hony2009} found 11 stars with a single smooth emission feature at $\sim$\,$10.5$\,\um. This feature was only found in S-type AGB stars, while there were no oxygen-rich AGB stars that exhibit the same emission feature. The authors argue that the broad and smooth emission found in these S-type stars is fundamentally different from the substructured emission in oxygen-rich AGB stars and the shift towards longer wavelengths is due to an increase in the magnesium-to-iron ratio in S-type stars.

\item[\textbf{Group~II, stars with alumina, 13\,\um and 20\,\um features:}] Fig.~\ref{fig:group2} shows the six stars in Group II. Although some stars do show some indication for the presence of silicates, the dominant emission features are the 11.2 \,\um emission due to amorphous alumina, a strong and sharp 13\,\um feature and the broad 19.8\,\um emission. A similar combination is also common in oxygen-rich stars with low mass-loss rates \citep{Sloan2003}.

Many stars with a strong 13\,\um feature also show the previously detected, unidentified emission features at 28 and 32\,\um indicated with arrows in Fig.~\ref{fig:group2}. Because the 28 and 32\,\um features can blend into a broad plateau with two narrow emission features at 28 and 32\,\um, the dust decomposition tool sometimes \emph{wrongly} models this broad plateau with emission of magnesium sulfides. This error can easily be identified because the residuals still show the narrow 28 and 32\,\um emission peaks. 

Several carriers have been suggested for these features, the most important candidates being crystalline silicates. However, the dust decomposition tool cannot explain the 28 and 32\,\um emission features with these carriers without introducing strong additional emission features that we do not observe.

\item[\textbf{Group~III, stars with magnesium sulfide and alumina}] \item[\textbf{features:}] The ten stars in Fig.~\ref{fig:group3} show two dominant emission features, one attributed to amorphous alumina and the other to magnesium sulfide. Because of the absence of strong silicate features or a strong 19.8\,\um feature, the 26--30\,\um emission of magnesium sulfides is clearly visible on top of the 11.2\,\um of amorphous alumina and can easily be separated from the 28 and 32\,\um emission of Group II stars.

The substructure that can be seen in some of these stars at 6.6 and 13.6\,\um is most likely due to emission of SiS gas near the star, as explained in Sect. \ref{sct:sisemission}.
\end{description}

\subsubsection{C/O dichotomy}\label{sct:codichotomy}
\begin{figure}
  \resizebox{\hsize}{!}{\includegraphics{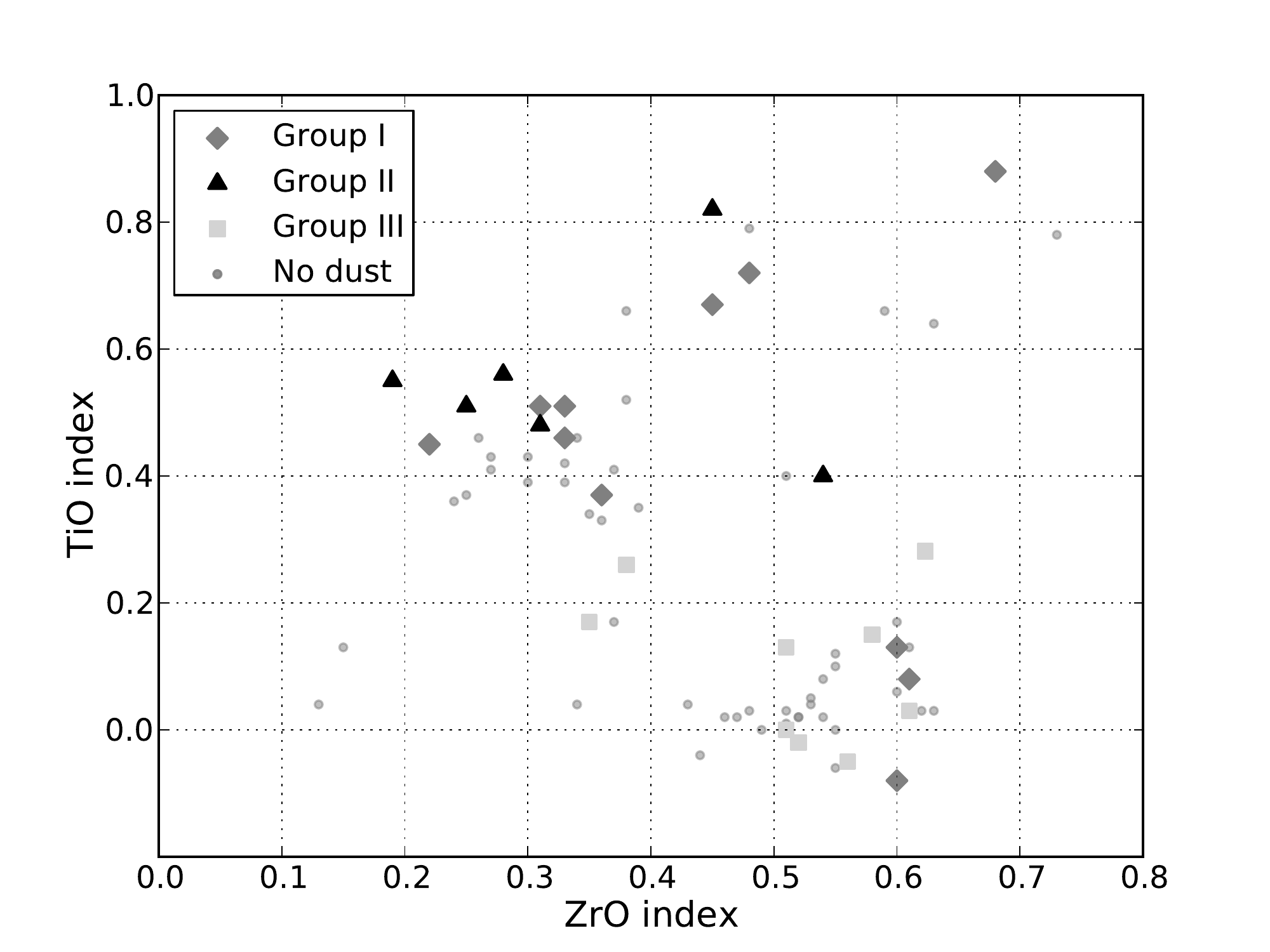}}
  \caption{The ZrO and TiO indices for the stars with different types of dust emission features. Group I, II and III are respectively indicated with dark grey diamonds, black triangles and light grey squares. The stars without dust emission features are shown as dots for comparison.}
  \label{fig:ZrO_TiO_coratio_dustclass}
\end{figure}

In Fig. \ref{fig:ZrO_TiO_coratio_dustclass} we show the ZrO and TiO indices of all stars in our sample with dust emission features, using different symbols for the different dust groups. From this figure it is clear that (i) the stars in Group~I are distributed more or less uniformly throughout this graph, spanning a wide range of ZrO indices, TiO indices and C/O ratios, while (ii) the stars from Group~II are located in the upper left corner of the diagram with low C/O ratios, and (iii) the stars from Group~III clutter in the lower right corner with C/O ratios close to unity. 

Because the number of stars with known C/O ratios in each separate group is small, we used an \emph{unpaired T-test} to check whether there is a significant difference in C/O ratios between these groups. We found that the C/O ratios of stars in Group II are significantly lower than the values for Group III stars ($P < 0.1\%$). This dichotomy shows that there is a clear dependence of the circumstellar dust mineralogy on the chemical composition of the stellar atmosphere in general and more specifically on the C/O ratio.

\subsubsection{Comparison with the Sloan \& Price classification} 
\begin{figure}
  \resizebox{\hsize}{!}{\includegraphics{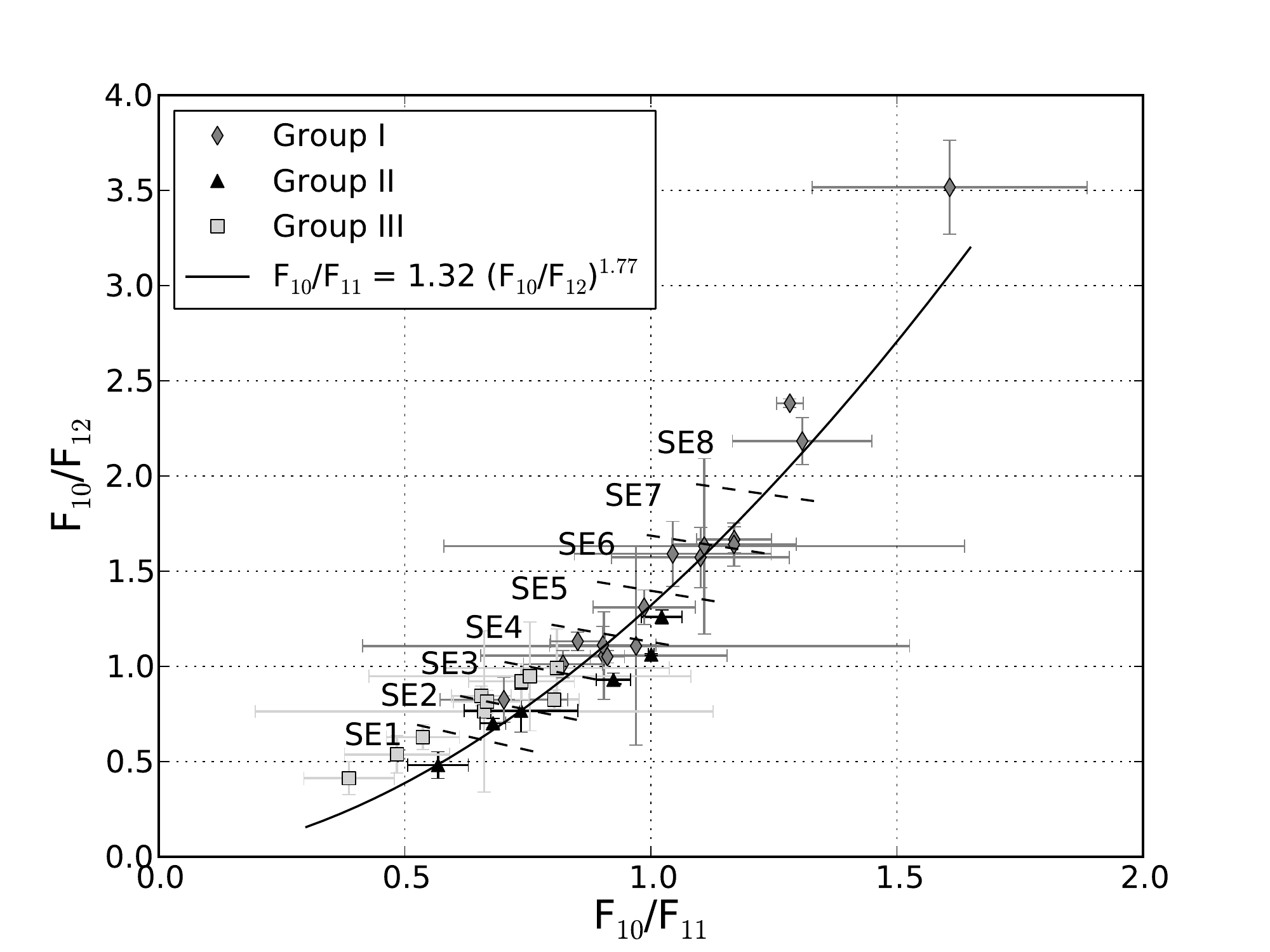}}
  \caption{The classification scheme for IR spectra of AGB stars, presented in \cite{SloanPrice1995, SloanPrice1998}. Group I, II and III are respectively indicated with grey diamonds, black triangles and light grey squares. The solid line shows the powerlaw presented in the original paper. The dashed lines indicate the boundaries between the subsequent classes of the Sloan \& Price classification.}
  \label{fig:sildustseq}
\end{figure}

\cite{SloanPrice1995, SloanPrice1998} also present a classification scheme for IR spectra of AGB stars, using flux ratios at 10, 11, and 12\,\um to quantify the shape of the spectrum in that region. The ratio F$_{10}$/F$_{11}$ is an a proxy for the width of the 10\,\um feature and F$_{10}$/F$_{12}$ for the strength of the red shoulder. \cite{SloanPrice1995} find that these flux ratios are strongly correlated and represent this correlation with a powerlaw. Along this sequence, they subdivide the stars into eight separate classes, where the stars with the lowest F$_{10}$/F$_{11}$ and F$_{10}$/F$_{12}$ ratios fall in the SE1 class and the stars with highest ratios in the SE8 class.

The interpretation of this sequence is that the spectral appearance of the stars in classes SE1--SE3 is dominated by oxides, the stars in classes SE6--SE8 are dominated by amorphous silicates and the intermediate classes show indications of both dust species. This change in spectral appearance with increasing SE-class number has been explained as either an effect of increasing mass-loss rates \citep[for example in ][]{Sloan2003, Pitman2010} or an effect of decreasing C/O ratio \citep{SloanPrice1998}. Both interpretations are based on the fact that alumina grains condense before silicate grains. If the binding of oxygen and alumina uses up the available oxygen, no free atoms will remain to form silicate grains, and alumina grains will dominate the resulting shell \citep{Stencel1990}. The difference between both interpretations is only whether the mass-loss rate or the C/O ratio is the dominating cause of this observed sequence.

In Fig. \ref{fig:sildustseq} we compare this power law with the flux ratios we find for the stars in our sample of S-type stars. Although we subtract the continuum flux in a different way, using the continuum estimates determined by the dust decomposition tool, as a combination of naked stellar photospheres and Planck curves of different temperatures, instead of the adapted Engelke function used by \cite{SloanPrice1998}, we find a similar relation between the flux ratios. 

The majority of the S-type stars in our sample have classification SE1--SE4, and only a third of the stars have classification SE5--SE8, similar to the findings in \cite{SloanPrice1995, SloanPrice1998}. Based on our sample, we find that most stars from Group~I can be classified as SE5--SE8 stars, while stars from Group~II and Group~III make up classes SE1--SE4. Although Group~II and Group~III stars clearly have different IR spectra at longer wavelengths, they have similar 10--12\,\um features and hence there is no difference between the Sloan \& Price classification of both groups. Hence, the stars with C/O ratios close to unity (Group~III) make up the same IR classes as stars with low C/O ratios (Group~II). This is contradictory to the idea that the 10--12\,\um emission is determined by the photospheric C/O ratio for AGB stars. However, there is still a likely difference between the position of these stars in Fig. \ref{fig:sildustseq}. Most stars of Group~III are situated below the powerlaw, while the stars of Group~II are above it. If this effect is real, we can conclude that the C/O ratio does not determine the Sloan \& Price classification, but might explain the spread on this correlation, while the Sloan \& Price class is determined by the mass-loss rate.

\section{Molecular emission features}\label{sct:molemis}
\begin{figure}
  \resizebox{\hsize}{!}{\includegraphics{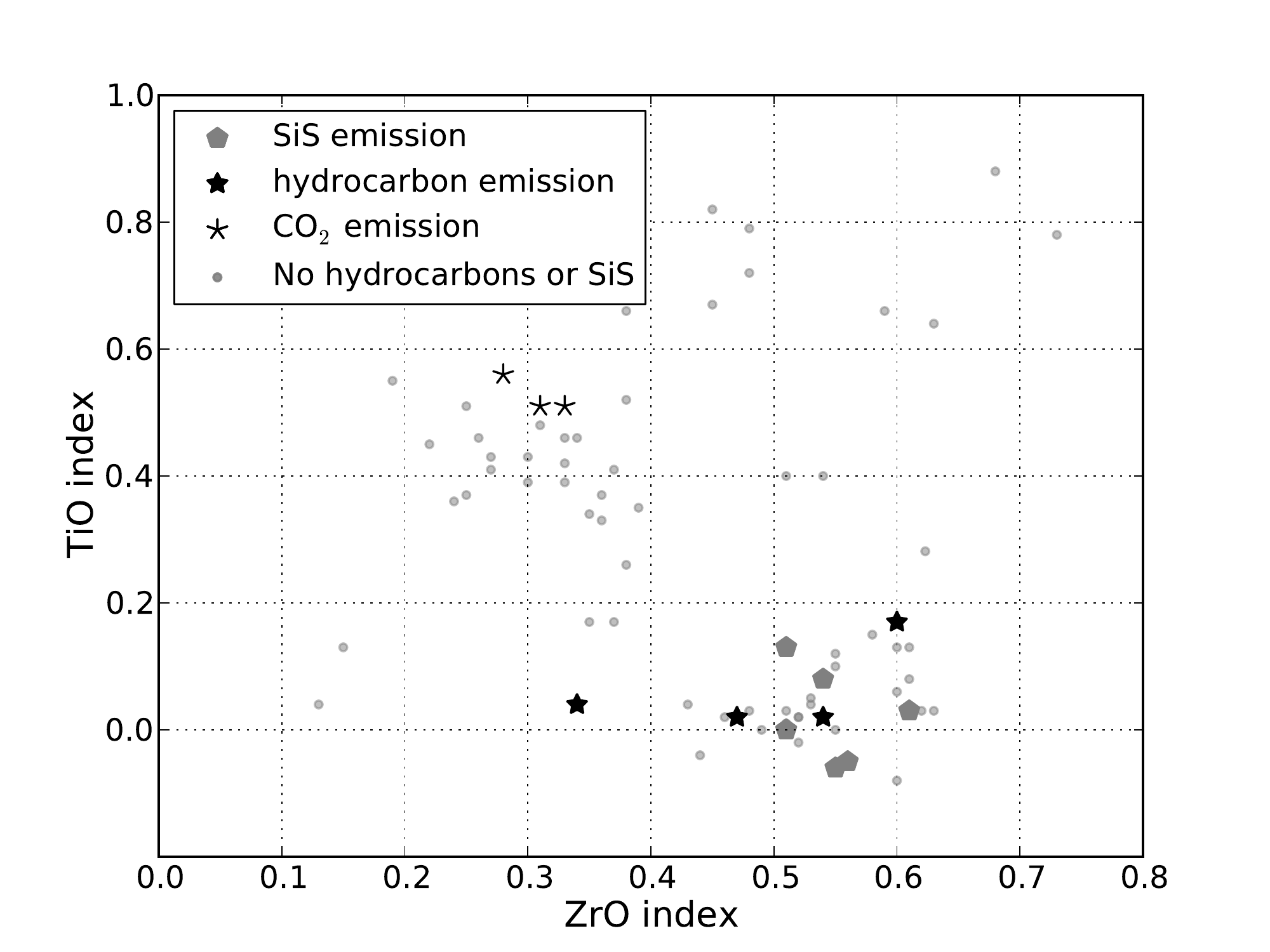}}
  \caption{The ZrO and TiO indices for the stars with emission features due to hydrocarbons, SiS or CO$_2$. The other stars in the sample are shown as dots for comparison.}
  \label{fig:ZrO_TiO_sis_and_pahs}
\end{figure}

\subsection{Hydrocarbon emission}
Hydrocarbon molecules exhibit very characteristic emission features in the 5--14\,\um region, the most prominent at 6.2, 7.9, 8.6, and 11.2\,\um. These features arise from the bending and stretching of the carbon and hydrogen bonds in the large molecules. \cite{Peeters2002} showed that the 7.9~\um PAH feature can be very different from source to source. They classified the PAH spectra in three classes, based on the peak wavelength of this feature. For type C sources, the peak of this feature is shifted to 8.2~\um, marking a clear difference with the 7.9~\um feature of class B PAHs and the 7.6~\um feature of class A PAHs.

\cite{Smolders2010} present the detection and identification of polycyclic aromatic hydrocarbon (PAH) emission in four S-type AGB stars from the present sample (\object{BZ CMa}, \object{KR Cam}, \object{CSS 173} and \object{CSS 757}). Based on the peak-wavelength, \object{BZ CMa} is the only one that shows the B class PAHs resembling PNe, Herbig Ae/Be stars, and red supergiants \citep{Tielens2008}. The other three stars can be classified as rare class C sources resembling some of the PAHs found in post-AGB stars \citep{Clio2009}, the carbon-rich red giant HD~100764 \citep{Sloan2007} and in a disk around the oxygen-rich K giant HD~233517 \citep{Jura2006}.

Furthermore, these relatively red emission features are consistent with the strong correlation found between the centroid wavelength of the 7.9~\um feature and the temperature of the central star \citep{Sloan2007, Keller2008, Acke2010}. The detection of PAHs in four S-type stars extends this correlation towards lower temperatures and more redshifted features. This is consistent with the hypothesis that Class C PAH spectra arise from a mixture of aliphatic and aromatic hydrocarbons, perhaps as isolated PAH molecules embedded in a matrix of aliphatic hydrocarbon bonds, found around stars with weak UV radiation fields \citep[see][Fig.~17]{PendletonAllamandola2002}. The hydrocarbons around \object{CSS~757}, \object{KR~Cam}, and \object{CSS~173} thus represent the composition as condensed in the AGB wind, before entering the interstellar medium where harsh UV radiation alters their chemical structure.

In Fig. \ref{fig:ZrO_TiO_sis_and_pahs} we again show the ZrO and TiO indices of all stars in our sample. From this figure, it is clear that the stars with hydrocarbon emission have C/O ratios that are higher than average. This is consistent with the idea that in stars with a nearly equal amount of carbon and oxygen atoms, non-equilibrium, shock-driven chemistry can form C$_2$H$_2$ molecules in the outer regions of the stellar outflow. These molecules are, further out in the wind, the necessary building blocks for the formation of hydrocarbons \citep{Allamandola1989}.

\begin{figure}
  \resizebox{\hsize}{!}{\includegraphics{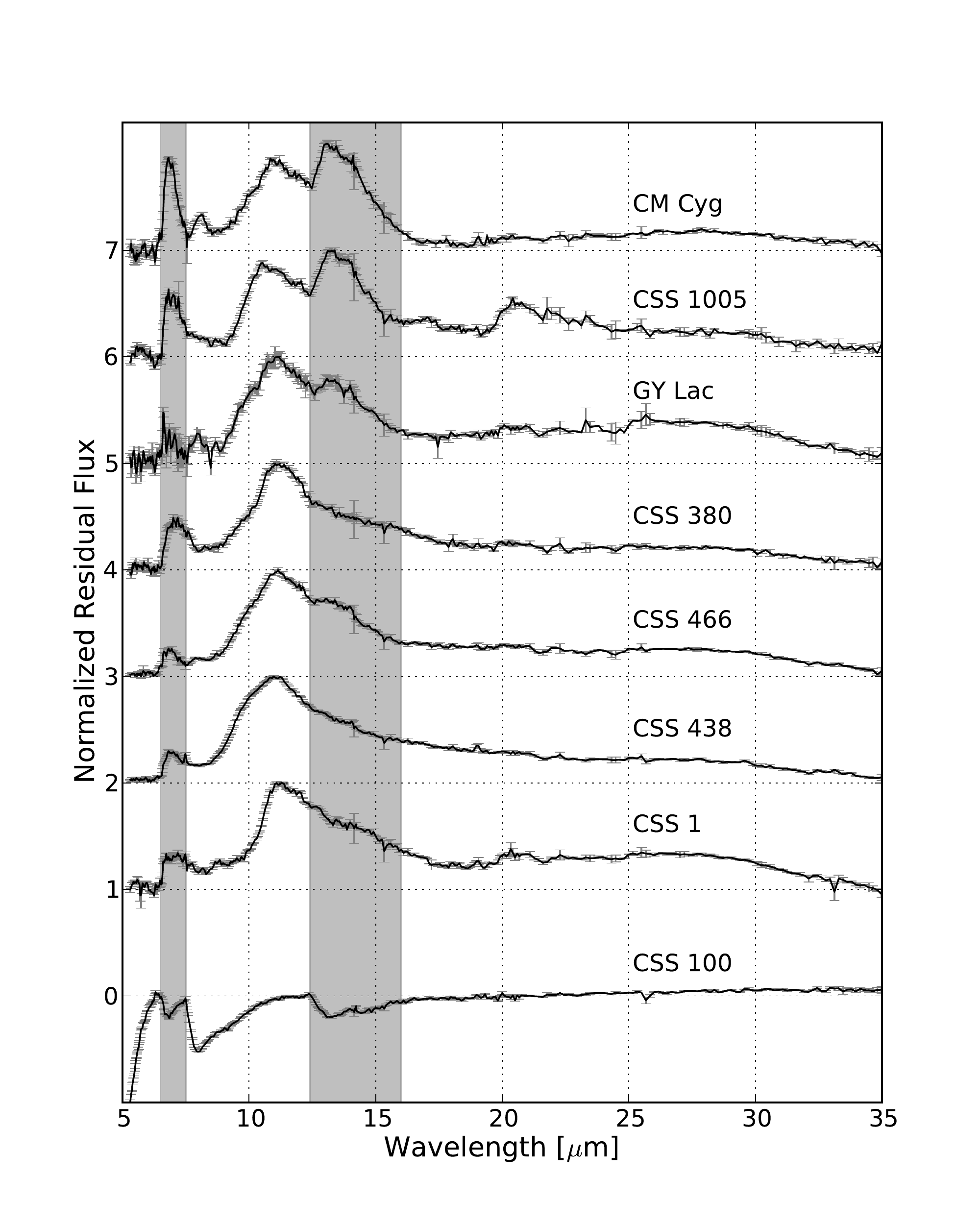}}
  \caption{The spectra of all S-type stars in our sample with suspected SiS emission. The grey areas indicate the position of the 6.6 and 13.5 SiS bands. At the bottom, CSS~100, one of the stars with SiS absorption bands, is shown for comparison.}
  \label{fig:sisemissionstars}
\end{figure}

\subsection{SiS emission}\label{sct:sisemission}
\cite{Sloan2011} identify the emission features at 6.6 and/or at 13.5\,\um as molecular emission of SiS. These features appear at precisely the wavelengths where they are
detected in absorption \citep{Cami2009}. The lack of any wavelength shift strongly suggests that the emission is from SiS in the gas phase. \citet{Sloan2011} only discuss the two stars in our sample with the strongest emission features (\object{CM Cyg} and \object{CSS 1005}), an additional five stars show a significant emission band at either 6.6 or 13.5\,\um, \emph{all} of which are classified as Group~III stars (see Figs.~\ref{fig:sisemissionstars}). The presence of these features in emission show that the SiS molecules are not only abundant in the stellar atmosphere, but also in the outer layers of the AGB wind.

In Sect. \ref{sct:nakedstelatm} and \ref{sct:codichotomy} we have shown that both magnesium sulfide grains and SiS molecules are only observed around S-type stars with C/O ratios close to unity (see Fig.~\ref{fig:sio_sis_co} and Fig.~\ref{fig:ZrO_TiO_coratio_dustclass}). Therefore, this strong relation between the presence of SiS emission and the dust classification could be explained as a shared dependence on the C/O ratio. However, it is possible that the magnesium sulfide grains in these stars are formed by a reaction of SiS molecules with Mg, as discussed in Sect. \ref{sct:MgSdiscussion}.

\subsection{CO$_2$ emission}
The narrow features at 13.9, 15 and/or 16.2\,\um, visible in 9 objects have been identified as CO$_2$ emission lines (more specifically $^{12}$C$^{16}$O$_2$). These lines are observed in approximately 30\% of oxygen-rich AGB stars, most of which also show a 13\,\um feature \citep{Justtanont1998, Sloan2003}. 

\begin{figure}
  \resizebox{\hsize}{!}{\includegraphics{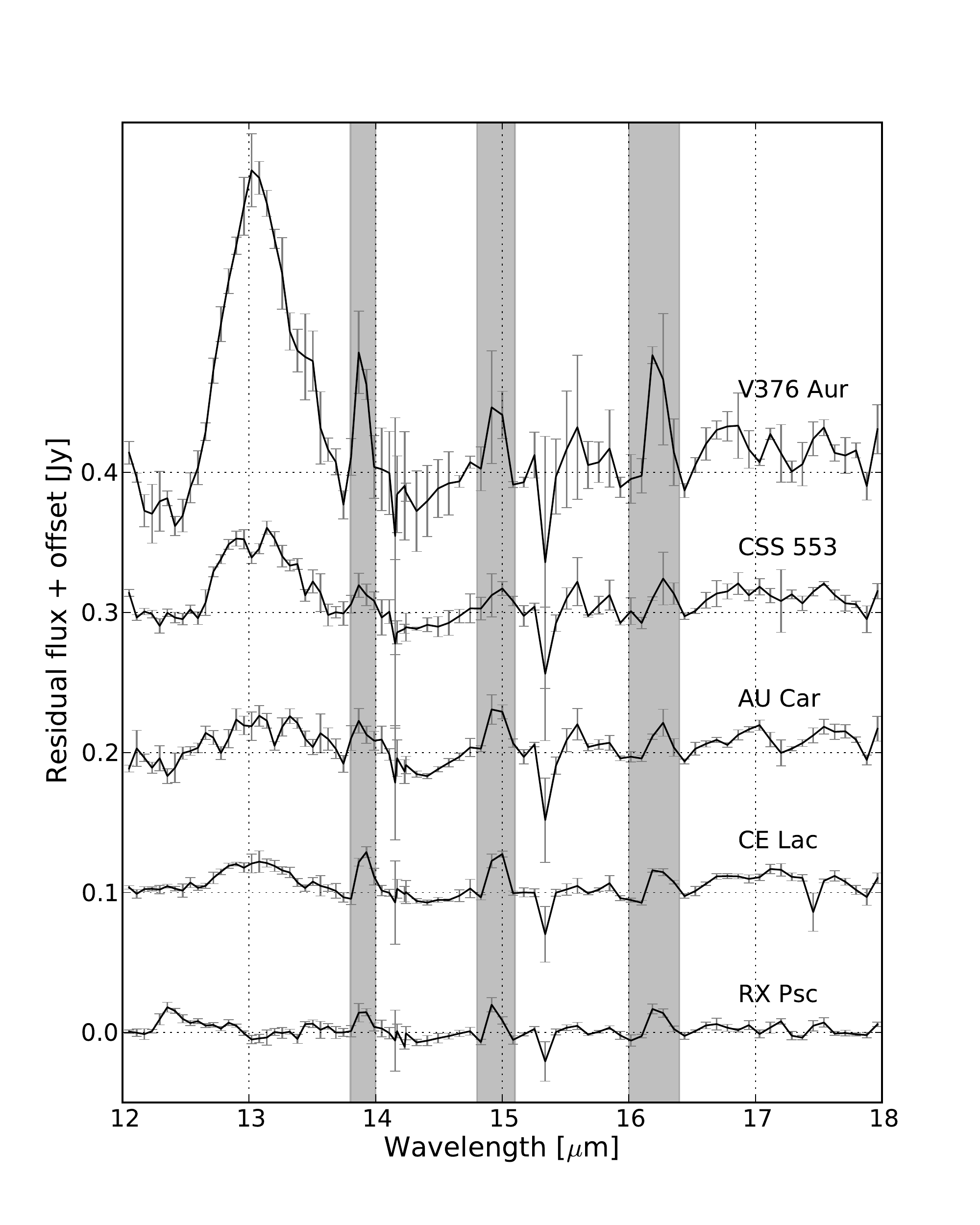}}
  \caption{The residual flux, after subtracting a line, of all stars with a positive detection of the CO$_2$ emission features at 13.9, 15 and/or 16.2\,\um, indicated here with the grey bands.}
  \label{fig:co2}
\end{figure}

To extract these emission features, we fit a line under the features (at 13.7 and 14.1\,\um for the 13.9\,\um feature, 14.7 and 15.2\,\um for the 15\,\um feature, and 15.9 and 16.5\,\um for the 16.2\,\um feature). As a conservative criterion, we consider a detection of CO$_2$ emission if all three features are significant at a 99\% confidence level. In our sample of S-type stars, we detect significant CO$_2$ emission features in the 5 objects shown in Fig. \ref{fig:co2}. The grey bands indicate the location of the emission features.

It is clear from Fig. \ref{fig:ZrO_TiO_sis_and_pahs}, that all stars with a positive detection of CO$_2$ emission lines have low C/O ratios. Because CO$_2$ requires an oxygen-rich environment to form, this dependence on the C/O ratio is not unexpected. Furthermore, although the signal-to-noise ratio and the resolution of the spectroscopic data do not allow for a quantitative analysis of the relationship between the strength of the CO$_2$ features and the 13\,\um feature, we can confirm that the 13\,\um feature is present in most of the stars with CO$_2$ emission (4 out of 5 objects). This relation between the presence of the 13\,\um emission and the detection of CO$_2$ could be explained as a shared dependence on the C/O ratio.

\section{Discussion}
\subsection{Carbon- versus oxygen-chemistry}
We find that almost all S-type AGB stars in our sample show either molecular absorption of molecules such as \h2o or \sio, and/or emission due to dust species such as amorphous alumina or silicates. This mainly oxygen-dominated chemistry in the S-type stars is also confirmed by the stellar parameters, derived from the {\sc marcs} model atmospheres, where we find that {\sc marcs} models with C/O ratios between 0.5 and 0.99 can explain the optical spectra of all stars in our sample. This lack of carbon-rich S-type stars is a direct consequence of the spectral classification: for an AGB star to be classified as an S-type star, the ZrO bands need to be present in the optical spectrum. \citet{Ferrarotti2002} show that the presence of ZrO in the stellar atmosphere critically depends on the C/O ratio of the star. When the C/O ratio becomes larger than unity, the abundance of ZrO in the stellar atmosphere drops several orders of magnitude and hence the strength of the absorption bands drops, until the ZrO bands are not detectable anymore and the star is not classified as an S-type star \citep{Piccirillo1980}.

However, some stars additionally show absorption bands of SiS, hydrocarbon emission features or the broad emission feature of magnesium sulfide dust, which are typical for carbon-rich sources. However, there are no stars in our sample with an IR spectrum that only shows carbon-rich characteristics. Because all stars in our sample have oxygen-dominated stellar atmospheres, it is remarkable that we can observe emission features that are generally only found in carbon-rich environments. The presence of these emission features indicates the (possibly strong) importance of non-equilibrium chemistry in the circumstellar environment of these AGB stars. The best example of this is the presence of hydrocarbons around \object{KR Cam}, \object{CSS 173}, \object{CSS 757}, and \object{BZ CMa}. The thermal equilibrium calculations presented in \citet{Ferrarotti2002} do not show the production of hydrocarbon molecules, because the majority of the carbon atoms would be locked in the stable and abundant CO molecule. Two main scenarios have been proposed to explain the presence of the hydrocarbon emission in these S-type stars. The first scenario assumes the formation of hydrocarbons from the basic atoms, in the AGB wind itself. \citet{Cherchneff2006} shows that, in AGB stars with a C/O ratio close to unity, non-equilibrium chemistry predicts the formation of C$_2$H$_2$ molecules in the stellar outflow \citep{Cherchneff1992, Cau2002}, which are the building blocks for hydrocarbons \citep{Allamandola1989}. Although this mechanism can explain the formation of hydrocarbons around S-type stars, C$_2$H$_2$ has a very distinct and sharp absorption feature at 13.7\,\um, which is not observed in our S-type stars. The second scenario starts from larger, carbonaceous grains that are broken down into smaller structures, such as hydrocarbons. However, this scenario still requires the presence of carbonaceous dust grains in an oxygen-rich stellar environment.

\subsection{Variability characteristics versus dust properties}
The stellar winds in AGB stars are commonly attributed to the coupled effect between large amplitude pulsations and radiation pressure on dust grains \citep{Hofner1998}. This relationship between the pulsations of the stellar atmosphere and the circumstellar environment can be translated into in a relationship between the properties of the infrared spectrum and the variability characteristics of the star. For oxygen-rich stars, a survey of the globular cluster 47 Tuc has shown that not only the mass-loss rate, but also the dust mineralogy changes as the stars evolve towards higher luminosities and longer pulsation periods \citep{Lebzelter2006, vanLoon2006, McDonald2011}.

We were able to derive the period and amplitude of the pulsations for 55 out of the 87 S-type stars in our sample. This gives us a small subset of stars with and without a dusty circumstellar environment with known variability characteristics. An analysis of this subset shows no significant difference between the periods and amplitudes of the stars with and without dust emission. Apparently for these S-type AGB stars, the period and amplitude of the pulsations are not enough to explain why approximately 65\% of the stars have no circumstellar dust, while the other 35\% do. Furthermore, it is remarkable that we have detected 5 Mira-like pulsators without significant dust emission features. Looking at the slope of the IR spectra, we can argue that three of these stars have a continuum excess, which might arise from amorphous carbon or metallic iron (\object{AU~Car}, \object{RX~Car} and \object{CSS~724}). However, CSS~794 and XY~CMi show no dust excess at all.

Furthermore, a multivariate analysis of the variability characteristics, the stellar parameters and the results from the dust decomposition tool found no significant correlation with either amplitude or period. From this, we can conclude that during this short evolutionary phase where a star is visible as an S-type star, the period and amplitude are not indicative for the mineralogy or the dust mass-loss rate.

\subsection{Dust formation around S-type AGB stars}\label{sct:correlations}
\begin{figure}
  \resizebox{\hsize}{!}{\includegraphics{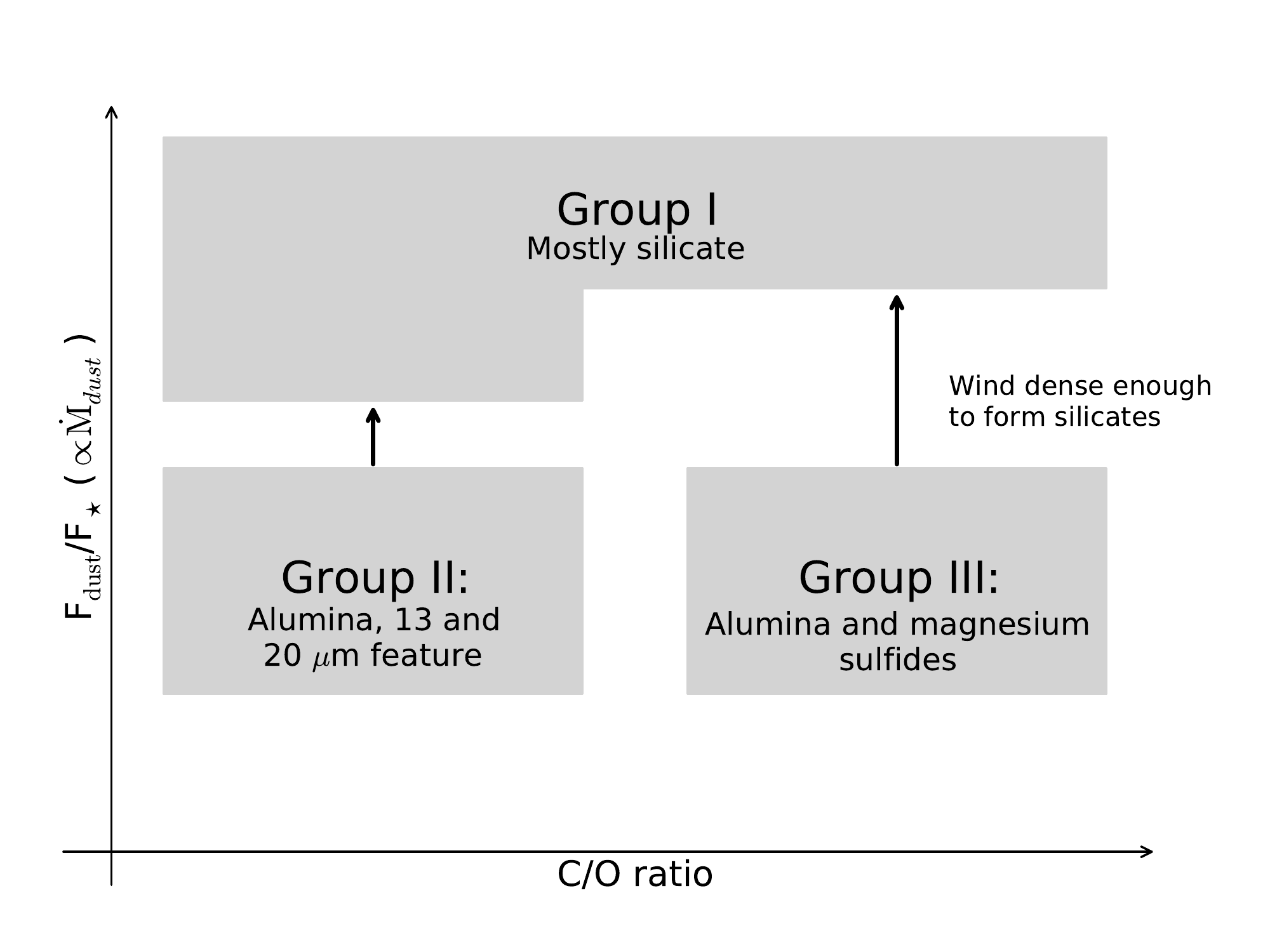}}
  \caption{Schematical overview of the proposed idea for dust formation in S-type AGB stars.}
  \label{fig:dfhypothesis}
\end{figure}

\begin{figure*}[!]
\center
\subfigure[The ratio of the integrated dust emission over the stellar emission versus the relative contribution of silicates to the total dust emission in percentages.]{
\includegraphics[width=8.7cm]{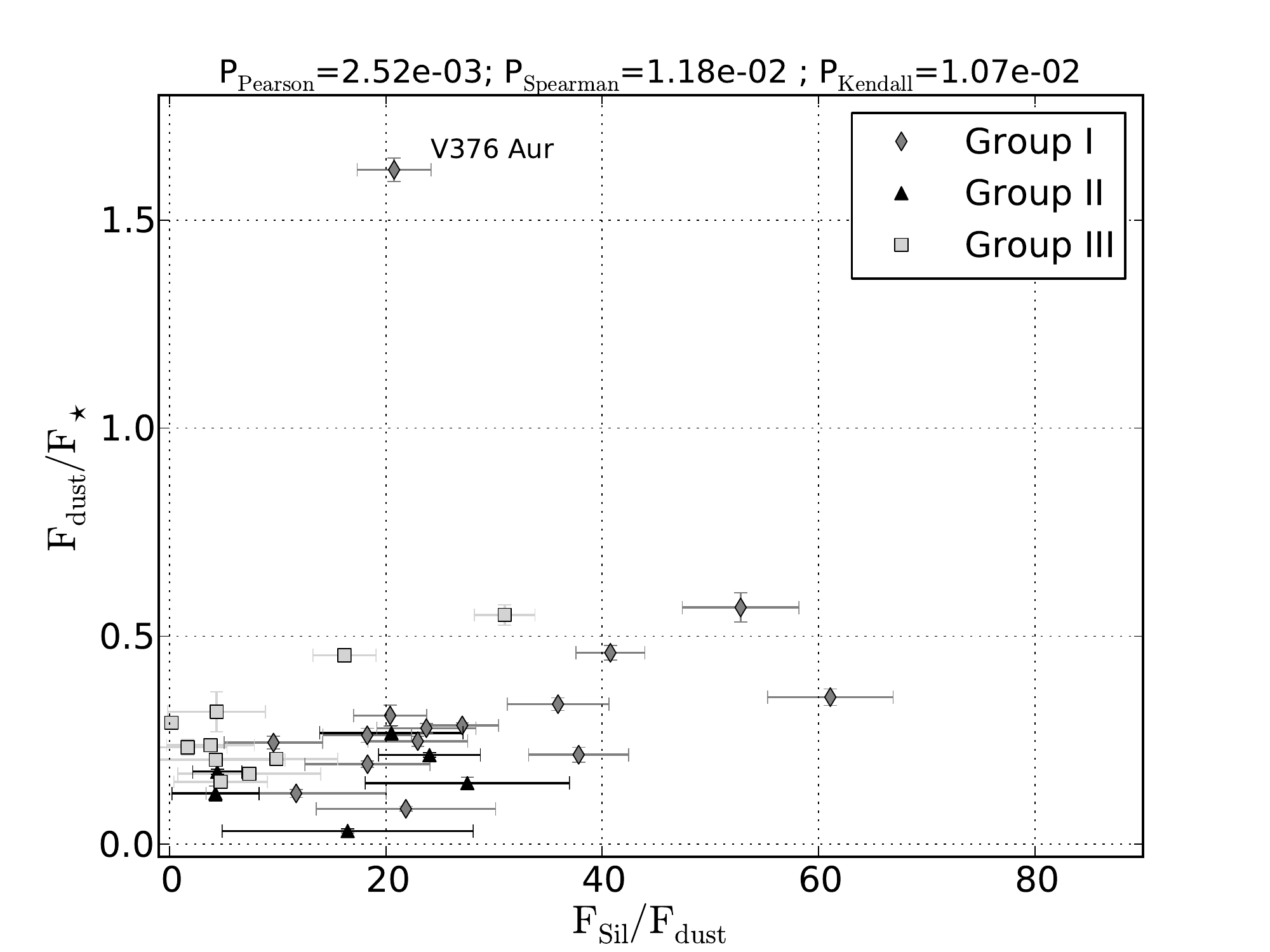}
\label{fig:subfig1}
}
\subfigure[Relative contribution of silicates and alumina to the total dust emission in percentages.]{
\includegraphics[width=8.7cm]{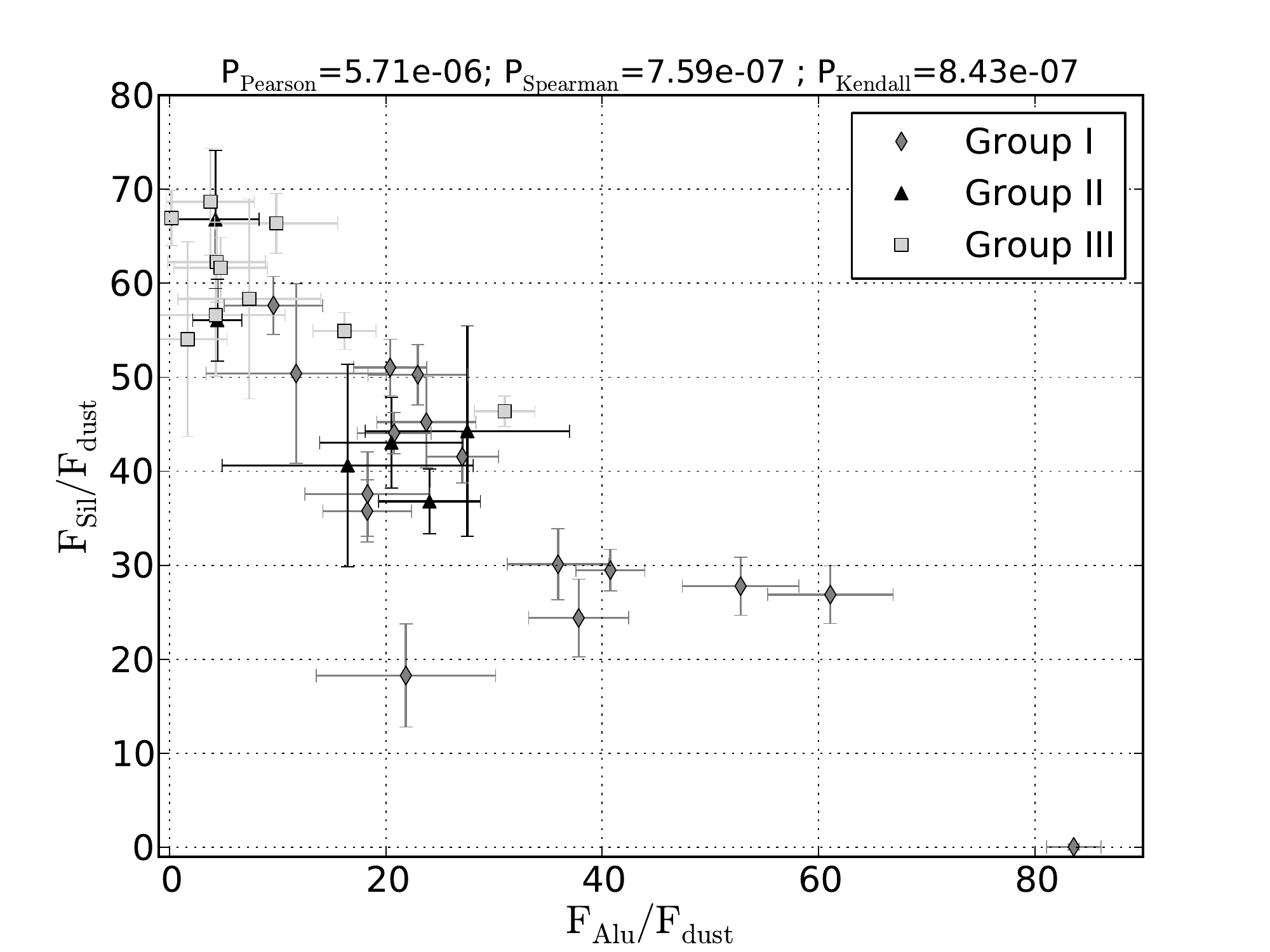}
\label{fig:subfig2}
}
\subfigure[Relative contribution of magnesium sulfides and alumina to the total dust emission in percentages.]{
\includegraphics[width=8.7cm]{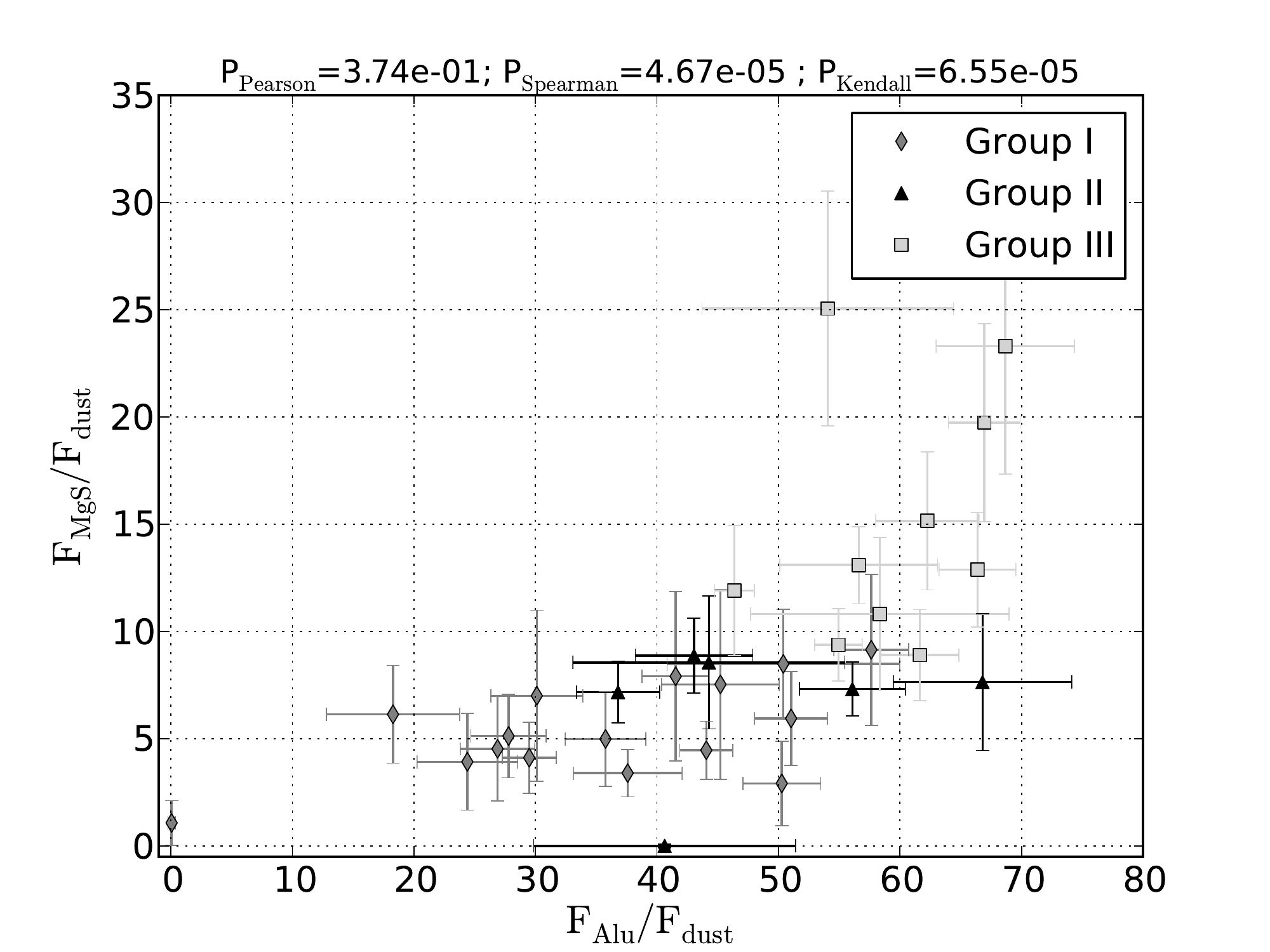}
\label{fig:subfig4}
}
\subfigure[Relative contribution of magnesium sulfides and silicates to the total dust emission in percentages.]{
\includegraphics[width=8.7cm]{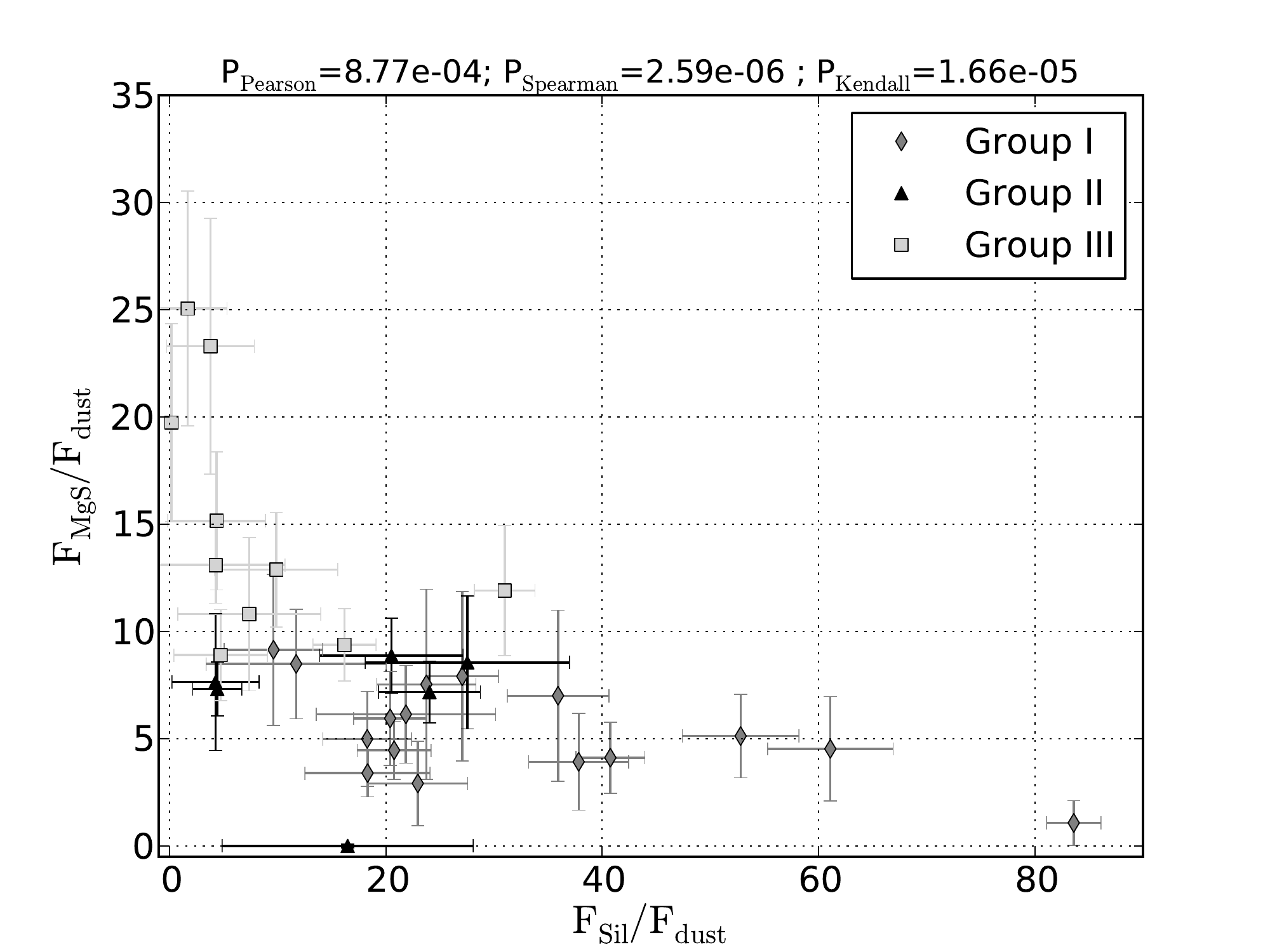}
\label{fig:subfig5}
}
\caption[]{\label{fig:subfigureExample}
All significant correlations presented in Section \ref{sct:correlations}. Above each figure, you can find the Spearman, Kendall and Pearson probabilities of each of the correlations. Please keep in mind that Pearson can only be used for linear relationships, while Spearman and Kendall are both rank-test and thus can be used for any relation between two parameters. The correlations in figures \subref{fig:subfig2}, \subref{fig:subfig4} and \subref{fig:subfig5} are clearly significant, while the correlations in figure \subref{fig:subfig1} is only tentative.}
\end{figure*}

In Sect. \ref{sct:dustemis} the dust decomposition tool was used to quantify the contribution of dust emission features to the infrared flux between 5 and 35\,\um. A multivariate  analysis of these parameters, together with the stellar parameters ($T_\mathrm{eff}$, C/O and [s/Fe]) yields four significant correlations (using a 99\% certainty interval) shown in Fig. \ref{fig:subfigureExample}. As an indication for significance of the correlation, we give the probability of obtaining a Pearson, Spearman or Kendall correlation coefficient at least as extreme as the one that was actually observed, assuming that there is no correlation between the two given variables. In order to interpret these p-values correctly, one has to keep in mind that Pearson assumes a linear correlation, while Spearman and Kendall are ranking coefficients. Because we look at ten parameters and thus effectively investigate 45 possible correlations, we have to acknowledge the possibility for a false-positive correlation. All important correlations are shown in Fig. \ref{fig:subfigureExample}.

\begin{description}
\item[\textbf{Silicates and total dust emission}] If we remove the outlier \object{V376 Aur}, we find a significant correlation between the total dust emission and the flux percentage of silicate features, shown in Fig. \ref{fig:subfig1}. This correlation indicates that stars with a larger dust excess, and thus a higher dust mass-loss rate, also tend to form silicates in their circumstellar environment. This is in close agreement with the study of Galactic bulge AGB stars \citep{Blommaert2006} and the study of AGB stars in the Large Magellanic Cloud presented in \citet{Dijkstra2005}. Here, the authors find that the dust features displayed by oxygen-rich stars are mainly ascribed to alumina for low dust mass-loss rates, while stars with higher dust mass-loss rates show an increasing amount of silicates. Furthermore, if we look at the total dust excess in the three groups, we find that, on average, Group~III stars have a higher dust excess than Group~I and Group~II stars. Although the difference is only tentative, this is still a noteworthy dichotomy between the three groups, because this is what could be expected based on the results of \citet{Blommaert2006} and \citet{Dijkstra2005}.

\item[\textbf{Silicates and alumina}] In Fig. \ref{fig:subfig2} we show the significant and strong inverse correlation between the relative flux contribution of alumina and silicates. This correlation can be explained by the standard dust condensation sequence, if we take into account freeze-out. Alumina seeds are formed early on in the dust-condensation sequence at high temperatures, but if the density in the wind is high enough, eventually silicates will form and hence the relative flux contribution of alumina will decrease as the flux contribution of silicates increases.


\item[\textbf{Magnesium sulfides and alumina}] In Fig. \ref{fig:subfig4} a strong correlation between the relative flux contribution of magnesium sulfides and alumina is shown. This correlation is unexpected because the presence of alumina grains points to an oxygen-rich circumstellar chemistry while magnesium sulfide emission is usually found in carbon-rich stars. Because this is the first time that this correlation is observed, it can be considered as a new constraint on any dust formation hypothesis for S-type AGB stars.

\item[\textbf{Magnesium sulfides and silicates}] In Fig. \ref{fig:subfig5} we show the inverse correlation between the relative flux contribution of silicates and magnesium sulfides. This correlation does not add extra constraints, because it is a consequence of the correlations between the relative flux contribution of (i) silicates and alumina and (ii) magnesium sulfides and alumina.
\end{description}

We propose that the dust formation in S-type stars is critically dependent on the C/O ratio and the total mass-loss rate. Because we expect that silicates only form further out in the wind where the temperature is low enough, the dust condensation sequence can be cut short before silicates start to condense. This could happen because (i) the density in the wind is too low to initiate the necessary chemical reactions (freeze-out) or (ii) there is not enough free oxygen to form the next dust species in the dust condensation sequence (oxygen depletion). At this point, stars with C/O ratios close to unity can only form alumina grains and possibly magnesium sulfide. Stars with lower C/O ratios still have enough free oxygen atoms to form alumina and the carriers of the 13\,\um and 19.8\,\um features. In stars with higher mass-loss rates, the density throughout the wind is high enough to start forming silicates. Because the formation of silicates requires a large number of oxygen atoms, the density at which silicates can condense might depend on the C/O ratio. 

This hypothesis can explain perfectly why we can find Group~I stars with any C/O ratio, as seen in Fig. \ref{fig:ZrO_TiO_coratio_dustclass}, while Group~II stars are clearly oxygen-rich and Group~III stars have C/O ratios close to unity. Furthermore, this hypothesis explains the difference in total dust excess between the groups and it takes into account the tentative correlation between the strength of the silicate emission and the total dust excess.

\subsection{The formation of magnesium sulfides in S-type AGB stars}\label{sct:MgSdiscussion}
A number of observational and theoretical studies on the 26--30\,\um emission feature in AGB and post-AGB stars, and in the environment of planetary nebulae show that magnesium sulfide is a wide-spread dust component in carbon-rich evolved stars \citep{Hony2002, Volk2002, Zijlstra2006, Lagadec2007}. As magnesium sulfide is usually observed in circumstellar environments dominated by carbon chemistry, the chemical models have focused on magnesium sulfide formation in a carbon-rich environment with and without silicon carbide condensation \citep{Zhukovska2008}. There are two main pathways to form magnesium sulfide grains. If the sulfur is locked in SiS molecules, the necessary reaction is:
\begin{displaymath}
 \mathrm{Reaction\ I: \ } \mathrm{SiS} + \mathrm{Mg} \rightarrow \mathrm{MgS\ (solid)} + \mathrm{Si}
\end{displaymath}

If the sulfur is free from the SiS molecules because, for example, the silicon is consumed in the formation of silicon carbide, the free sulfur forms H$_2$S molecules and the main reaction is:
\begin{displaymath}
\mathrm{Reaction\ II:\ } \mathrm{H_2S} + \mathrm{Mg} \rightarrow \mathrm{MgS\ (solid)} + \mathrm{H_2}
\end{displaymath}

\citet{Zhukovska2008} find for carbon-rich AGB stars that only the growth of magnesium sulfide as mantles on SiC cores, following reaction II, can produce enough magnesium sulfide dust to explain the observations. Hence, following this formation mechanism, the production of magnesium sulfide depends on silicon carbide grains (i) to free the sulfur from SiS molecules (so that H$_2$S can be formed) and (ii) as core grains on which the magnesium sulfide can form a mantle. Since we do not see strong silicon carbide emission features in any of our stars (although we cannot completely rule out the presence of a weak emission feature in the 10--12\,\um range) we conclude that alternative seeds are necessary.

If reaction II is the dominant formation path to form magnesium sulfides, we need to free the sulfur from SiS before this reaction can take place. In a mixed chemistry region, there is a much wider variety of silicon-rich dust species as compared to a purely carbon-rich chemistry (see Fig. \ref{fig:dustcondensation}). Already at relatively high temperatures the corundum reacts with circumstellar gas to form gehlenite and further interactions lead to other silicon-bearing dust grains such as akermanite, diopside, anorthite, and eventually silicate grains. All these dust species can consume silicon, hence allowing the free sulfur to form the H$_2$S molecules, necessary for reaction II. 

Although these silicon-bearing grains have emission peaks in the 10--20\,\um range, the infrared spectra of most of the stars with clear magnesium sulfides emission also show amorphous alumina, without clear observational evidence for any silicon-bearing dust species. Hence, the observations indicate that the presence of silicates or other silicon-bearing dust species is not a strict requirement for the formation of magnesium sulfides.

Because many stars with magnesium sulfide emission show additional emission peaks at 6.6 and 13.5\,\um, due to molecular emission of SiS, the observations seem to indicate that reaction I is the main formation mechanism in these S-type AGB stars. Because sulfur is less abundant than magnesium, we can expect that more influx of SiS molecules in the wind would lead to more magnesium sulfides being formed. Furthermore, because SiS molecules are only formed in S-type stars with C/O ratios larger than 0.96, it automatically follows that we will only detect magnesium sulfide dust in stars with similar high C/O ratios, as is shown in Fig. \ref{fig:ZrO_TiO_coratio_dustclass}.

According to \citet{Zhukovska2008}, the formation path of magnesium sulfide is more efficient as heterogeneous growth on a different type of dust particles that already have moderate radii. The dust species that may serve as carrier grains need to form dust grains with a size not much smaller than about 0.1\,\um and need to be formed early on in the wind. The correlation shown in Fig. \ref{fig:subfig4} indicates an (indirect) relation between alumina grains and the magnesium sulfide formation. Because alumina can form very early on in the wind (at temperatures above 1700\,K) and because of the correlation between the relative flux contribution of alumina and magnesium sulfides (see Sect. \ref{sct:correlations}), we propose the hypothesis that magnesium sulfides in S-type AGB stars are formed on amorphous alumina grains. In addition, this formation mechanism can explain the presence of SiS emission bands at 6.6 and/or 13.5\,\um in many Group~III stars, because it is a key ingredient for the formation of magnesium sulfides.

\section{Results and conclusions}
We present a new sample of Spitzer-IRS spectra for 87 galactic S-type AGB stars which were selected based on the presence of a ZrO absorption band at optical wavelengths, pulsation characteristics and K-band magnitude. Although the sample is slightly biased towards lower mass-loss rates, we show that we exclude less than 5\% of the stars with high mass-loss rates. In addition to the IR spectra, we obtained optical spectra (low and/or high resolution) and infrared photometry for the stars in our sample.

We derived the chemical parameters (T$_\mathrm{eff}$, C/O, [s/Fe] and [Fe/H]) for a large subset of the sample, by comparing the available optical spectra to {\sc marcs} model atmospheres.

For a large subset of the sample, we provide pulsation characteristics, based on the ASAS, NSVS and AAVSO light curves. The sample contains 6 irregular pulsators, 38 semi-regular variables and 21 Miras. It is remarkable that 5 of the 17 Miras do not show any dust emission features. This does not imply that these stars have no dusty wind, because we cannot exclude the presence of dust species, such as metallic iron and amorphous carbon.

We find that the presence of the SiS absorption band in the S-type stars without dust emission features is critically dependent on the C/O ratio. Stars with C/O ratios lower than 0.96 do not show significant absorption at 6.6 or 13.5\,\um while all \emph{naked} stellar atmospheres with C/O ratios above 0.96 show significant absorption bands. This opens up the possibility to use the SiS absorption band as an indicator for the C/O ratio.

We present a positive detection and identification of PAH emission in 4 S-type AGB stars, a detailed description and discussion is given by \citet{Smolders2010}.

We use a simple tool to decompose the IR spectra of the dust-rich stars. This tool is used to identify the dust species that are visible in the IR spectra and to quantify the relative flux contribution of each dust type to the total flux. Based on this dust decomposition tool, we are able to classify the IR spectra into three distinct groups: (i) stars with strong silicate/gehlenite features, (ii) stars with strong 13\,\um and 20\,\um emission features and (iii) stars with magnesium sulfides and amorphous alumina. The fact that we find three separate classes is already a remarkable result, for example, no star has both a 13\,\um feature and magnesium sulfide emission.

We also find a strong relation between the C/O ratio and mineralogy. We propose that, for stars with low mass-loss rates, the dust-condensation sequence is cut short and the remaining amount of free oxygen, together with the presence of SiS molecules in the wind determines whether sulfides will be formed (Group III) or whether we will observe a 13\,\um and 20\,\um feature (Group II). Stars with higher mass-loss rates can \emph{complete} the dust formation sequence and form silicate grains (Group I). This scenario is consistent with the correlations shown in Fig.~\ref{fig:subfigureExample}.

Finally, we propose the hypothesis that magnesium sulfides in S-type stars are formed by a reaction of SiS molecules with Mg. This is based (i) on the observation that magnesium sulfides are only observed in stars with C/O ratios close to one, where SiS molecules are formed and (ii) on the detection of SiS emission in half of the stars with magnesium sulfide emission, indicating that the SiS molecules are present in the circumstellar environment. Furthermore, based on the formation mechanism of magnesium sulfides in carbon-rich AGB stars, we propose that magnesium sulfide is formed as mantles around dust grains, for example amorphous alumina.

\begin{acknowledgements}
K. Smolders, J. Blommaert, L. Decin, H. Van Winckel and S. Uttenthaler acknowledge support from the Fund for Scientific Research of Flanders under the grant G.0470.07. S. Uttenthaler acknowledges support from the Austrian Science Fund (FWF) under project P~22911-N16. S. Van Eck is an F.N.R.S Research Associate. MWF and JWM thank the National Research Foundation (NRF) of South Africa for financial support. Based on observations made with the Mercator Telescope, operated on the island of La Palma by the Flemish Community, at the Spanish Observatorio del Roque de los Muchachos of the Instituto de Astrof\'isica de Canarias. Based on observations obtained with the HERMES spectrograph, which is supported by the Fund for Scientific Research of Flanders (FWO), Belgium , the Research Council of K.U.Leuven, Belgium, the Fonds de la Recherche Scientifique (FNRS), Belgium, the Royal Observatory of Belgium, the Observatoire de Gen\`eve, Switzerland and the Th\"uringer Landessternwarte Tautenburg, Germany. This work was partly funded by an Action de recherche concert\'{e} (ARC) from the Direction g\'{e}n\'{e}rale de l'Enseignement non obligatoire et de la Recherche scientifique - Direction de la recherche scientifique - Communaut\'{e} francaise de Belgique. The research leading to these results has received funding from the European Research Council under the European Community's Seventh Framework Programme (FP7/2007--2013)/ERC grant agreement n$^\circ$227224 (PROSPERITY), as well as from the Research Council of K.U.Leuven grant agreement GOA/2008/04
\end{acknowledgements}

\addcontentsline{toc}{chapter}{Bibliography}
\bibliographystyle{aa}
\bibliography{references.bib}

%
\addtocounter{table}{1}
%
\begin{landscape}
\longtab{1}{
\begin{longtable}{cccccccccc}
\caption{\label{table:data} The basic data for the stars in our sample. From left to right we find the CSS number, the most frequently used names, the V magnitude (if possible we used an interpolation of the lightcurve at the time of the SAAO JHKL observations), the new JHKL magnitudes, the ZrO and TiO indices and the available optical data.} \\
\hline\hline
CSS & name & Vmag & SAAO J & SAAO H & SAAO K & SAAO L & I$_\mathrm{TiO}$ & I$_\mathrm{ZrO}$ & Optical Spectra \\
\hline
\endfirsthead
\caption{continued.}\\
\hline\hline
CSS & name & Vmag & SAAO J & SAAO H & SAAO K & SAAO L & I$_\mathrm{TiO}$ & I$_\mathrm{ZrO}$ & Optical Spectra \\
\hline
\endhead
\hline
\endfoot
1 & CSS 1 &  &  &  &  &  &  &  &  \\ 
32 & RX Psc & 12.08 & 6.31 & 5.33 & 5.02 & 4.62 &  &  & Hermes \\ 
38 & CSS 38 &  &  &  &  &  &  &  & Hermes \\ 
91 & NX Per &  &  &  &  &  &  &  &  \\ 
100 & CSS 100 &  &  &  &  &  &  &  &  \\ 
127 & CSS 127 & 12.99 & 6.30 & 5.18 & 4.84 & 4.50 & 0.29 & 0.45 & SAAO, WHT, Hermes \\ 
166 & CSS 166 & 13.40 & 6.83 & 5.64 & 5.27 & 4.93 & 0.26 & 0.46 & SAAO, WHT \\ 
173 & CSS 173 & 11.15 & 6.69 & 5.55 & 5.29 & 4.94 & 0.21 & 0.27 & SAAO, WHT, Hermes \\ 
177 & KR Cam &  &  &  &  &  & 0.26 & 0.44 & WHT, Hermes \\ 
179 & CSS 179 &  &  &  &  &  & 0.24 & 0.52 &  \\ 
191 & GH Aur &  & 5.72 & 4.49 & 4.08 & 3.67 & 0.43 & 0.40 & WHT, Hermes \\ 
196 & V376 Aur &  &  &  &  &  & 0.57 & 0.16 & WHT, Hermes \\ 
202 & CD-39 2449 &  &  &  &  &  & 0.41 & 0.23 & SAAO \\ 
218 & NP Aur &  &  &  &  &  & 0.92 & 0.37 & WHT, Hermes \\ 
220 & CF Gem & 13.56 & 6.64 & 5.37 & 4.97 & 4.60 & 0.22 & 0.43 & WHT, Hermes \\ 
229 & HV 8057 & 13.77 & 6.22 & 5.17 & 4.76 & 4.24 & 0.88 & 0.29 & SAAO \\ 
267 & CSS 267 & 14.09 & 6.64 & 5.37 & 4.97 & 4.60 & 0.36 & 0.46 & SAAO, WHT \\ 
281 & BZ CMa & 12.77 & 6.59 & 5.38 & 4.96 & 4.47 & 0.20 & 0.38 & SAAO, Hermes \\ 
282 & CSS 282 & 13.79 & 6.99 & 5.68 & 5.25 & 4.80 & 0.30 & 0.52 & SAAO, WHT \\ 
298 & CSS 298 & 11.53 & 6.87 & 5.77 & 5.48 & 5.23 &  &  & SAAO, WHT, Hermes \\ 
300 & CPD-19 1672 & 9.90 & 6.31 & 5.34 & 5.14 & 4.97 & 0.22 & 0.13 & SAAO, Coralie, Hermes \\ 
315 & CD-43 2976 & 11.31 & 6.41 & 5.41 & 5.13 & 4.94 & 0.46 & 0.16 & SAAO, Coralie \\ 
325 & XY CMi & 14.51 & 6.79 & 5.74 & 5.37 & 4.97 & 0.55 & 0.15 & SAAO, WHT, Hermes \\ 
328 & CD-30 4223 & 10.38 & 6.62 & 5.65 & 5.44 & 5.21 & 0.26 & 0.30 & SAAO, Coralie \\ 
352 & CSS 352 & 12.98 & 6.89 & 5.70 & 5.35 & 4.96 & 0.43 & 0.29 & SAAO, Hermes \\ 
361 & CSS 361 & 12.58 & 6.21 & 5.08 & 4.75 & 4.39 & 0.30 & 0.50 & SAAO \\ 
371 & CSS 371 & 12.73 & 7.01 & 5.75 & 5.36 & 4.95 & 0.28 & 0.44 & SAAO, Hermes \\ 
380 & CSS 380 & 14.92 & 6.96 & 5.70 & 5.16 & 4.51 & 0.23 & 0.46 & SAAO \\ 
393 & CSS 393 & 11.69 & 6.88 & 5.79 & 5.44 & 5.12 & 0.22 & 0.34 & SAAO, Coralie \\ 
417 & CSS 417 & 13.14 & 6.67 & 5.42 & 5.01 & 4.62 & 0.16 & 0.37 & SAAO, Hermes \\ 
427 & CSS 427 &  & 6.85 & 5.49 & 4.98 & 4.63 & 0.30 & 0.53 & SAAO \\ 
431 & AO Gem &  &  &  &  &  & 0.23 & 0.48 & WHT, Hermes \\ 
438 & CSS 438 &  & 7.00 & 6.00 & 5.48 & 4.89 & 0.27 & 0.50 & SAAO \\ 
443 & EX Pup &  & 7.11 & 6.15 & 5.71 & 5.27 & 0.90 & 0.20 & SAAO \\ 
453 & CSS 453 & 12.27 & 6.35 & 5.17 & 4.80 & 4.53 & 0.40 & 0.15 & SAAO \\ 
454 & CSS 454 & 10.62 & 6.25 & 5.28 & 5.02 & 4.78 & 0.39 & 0.26 & SAAO, WHT, Coralie, Hermes \\ 
457 & CSS 457 & 11.07 & 6.56 & 5.54 & 5.26 & 5.10 & 0.46 & 0.14 & SAAO \\ 
466 & CSS 466 &  & 6.84 & 5.27 & 4.60 & 3.88 & 0.33 & 0.35 & SAAO \\ 
472 & CSS 472 & 12.94 & 6.60 & 5.44 & 5.10 & 4.92 & 0.57 & 0.05 & SAAO \\ 
478 & CSS 478 & 11.20 & 6.55 & 5.44 & 5.15 & 4.97 & 0.39 & 0.13 & SAAO, Coralie \\ 
480 & CSS 480 & 11.43 & 5.93 & 4.88 & 4.56 & 4.34 & 0.54 & 0.08 & SAAO, Coralie \\ 
489 & CD-29 5912 & 10.59 & 6.34 & 5.34 & 5.08 & 4.83 & 0.37 & 0.25 & SAAO, Coralie \\ 
528 & CSS 528 & 12.34 & 6.60 & 5.50 & 5.17 & 4.92 & 0.44 & 0.13 & SAAO \\ 
531 & CSS 531 &  & 7.11 & 5.83 & 5.40 & 5.09 & 0.48 & 0.37 & SAAO \\ 
545 & CO Pyx & 15.00 & 6.73 & 5.82 & 5.40 & 4.88 & 0.81 & 0.10 & SAAO, Hermes \\ 
553 & CSS 553 & 12.70 & 6.13 & 4.93 & 4.56 & 4.34 & 0.56 & 0.13 & SAAO \\ 
558 & CSS 558 & 12.65 & 6.62 & 5.54 & 5.21 & 4.94 & 0.45 & 0.23 & SAAO \\ 
584 & CSS 584 & 13.00 & 6.49 & 5.23 & 4.82 & 4.45 & 0.20 & 0.40 &  \\ 
598 & CSS 598 & 13.41 & 6.85 & 5.69 & 5.36 & 5.07 & 0.34 & 0.54 & SAAO \\ 
603 & CD-30 7391 & 10.67 & 5.90 & 4.87 & 4.60 & 4.46 & 0.48 & 0.10 & SAAO, Coralie \\ 
604 & CSS 604 & 14.26 & 7.20 & 5.98 & 5.60 & 5.31 & 0.32 & 0.53 & SAAO \\ 
613 & CSS 613 & 13.19 & 6.38 & 5.07 & 4.67 & 4.29 & 0.23 & 0.44 & SAAO \\ 
622 & CSS 622 & 13.35 & 6.67 & 5.39 & 4.95 & 4.51 & 0.24 & 0.39 & SAAO \\ 
633 & CSS 633 &  & 6.97 & 5.85 & 5.29 & 4.55 & 0.74 & 0.32 & SAAO \\ 
635 & CSS 635 & 12.61 & 6.97 & 5.77 & 5.42 & 5.04 & 0.18 & 0.42 & SAAO \\ 
640 & CSS 640 & 13.68 & 6.96 & 5.70 & 5.34 & 5.05 & 0.33 & 0.52 & SAAO \\ 
650 & CSS 650 & 10.98 & 6.40 & 5.36 & 5.07 & 4.95 & 0.41 & 0.20 & SAAO \\ 
657 & CSS 657 & 12.46 & 6.78 & 5.51 & 4.99 & 4.33 & 0.26 & 0.46 & SAAO \\ 
661 & CSS 661 & 12.76 & 6.49 & 5.15 & 4.63 & 4.05 & 0.25 & 0.19 & SAAO \\ 
665 & CSS 665 & 13.22 & 6.91 & 5.79 & 5.47 & 5.26 & 0.28 & 0.51 & SAAO \\ 
679 & AU Car &  & 5.89 & 4.73 & 4.27 & 3.88 &  &  & SAAO \\ 
682 & CSS 682 & 11.71 & 6.19 & 5.07 & 4.74 & 4.45 & 0.44 & 0.18 & SAAO \\ 
690 & RX Car &  & 6.54 & 5.42 & 4.90 & 4.39 & 0.80 & 0.26 & SAAO \\ 
718 & CSS 718 & 12.18 & 6.36 & 5.25 & 4.93 & 4.66 & 0.43 & 0.15 & SAAO \\ 
719 & CSS 719 & 12.27 & 6.53 & 5.34 & 4.99 & 4.64 & 0.25 & 0.44 & SAAO \\ 
723 & CSS 723 & 13.41 & 7.09 & 5.95 & 5.40 & 4.94 & 0.07 & 0.04 & SAAO \\ 
724 & CSS 724 & 13.07 & 6.20 & 5.19 & 4.83 & 4.46 &  &  & SAAO \\ 
739 & CSS 739 & 13.35 & 6.81 & 5.95 & 5.50 & 4.81 & 0.80 & 0.21 & SAAO \\ 
749 & CSS 749 & 13.49 & 6.52 & 5.21 & 4.80 & 4.36 & 0.33 & 0.46 & SAAO \\ 
757 & CSS 757 & 13.08 & 6.68 & 5.37 & 4.93 & 4.54 & 0.20 & 0.45 & SAAO \\ 
763 & CSS 763 & 12.48 & 6.67 & 5.40 & 5.03 & 4.67 & 0.47 & 0.16 & SAAO \\ 
783 & CSS 783 & 11.16 & 6.66 & 5.65 & 5.36 & 5.15 & 0.39 & 0.12 & SAAO \\ 
794 & CSS 794 &  & 6.88 & 5.70 & 5.26 & 4.86 & 0.79 & 0.73 & SAAO \\ 
806 & CSS 806 & 14.68 & 7.31 & 5.70 & 5.13 & 4.75 & 0.31 & 0.11 & SAAO \\ 
881 & CSS 881 & 12.88 & 6.47 & 5.09 & 4.62 & 4.26 & 0.27 & 0.37 & SAAO \\ 
921 & CSS 921 & 12.01 & 5.99 & 4.88 & 4.54 & 4.32 & 0.46 & 0.10 & SAAO \\ 
987 & CSS 987 &  & 7.68 & 6.19 & 5.57 & 5.00 & 1.13 & 1.13 & SAAO \\ 
1005 & CSS 1005 & 13.30 & 6.26 & 5.05 & 4.62 & 4.14 & 0.22 & 0.47 & SAAO \\ 
1046 & CSS 1046 &  & 6.40 & 5.27 & 4.92 & 4.58 &  &  & WHT \\ 
1077 & CSS 1077 & 11.61 & 6.40 & 5.37 & 5.07 & 4.83 & 0.46 & 0.18 & SAAO, WHT \\ 
1087 & TU Dra &  &  &  &  &  &  &  & WHT \\ 
1137 & V734 Cyg &  &  &  &  &  & 0.61 & 0.32 & WHT \\ 
1138 & CSS 1138 &  &  &  &  &  &  &  & WHT \\ 
1160 & CSS 1160 & 11.69 & 7.03 & 5.99 & 5.49 & 4.79 &  &  & WHT \\ 
1180 & CM Cyg &  &  &  &  &  & 0.29 & 0.49 & WHT \\ 
1181 & V899 Aql &  & 7.00 & 6.07 & 5.56 & 4.90 & 1.29 & 3.88 & WHT \\ 
1305 & CE Lac &  &  &  &  &  & 0.62 & 0.12 & WHT \\ 
1307 & GY Lac &  &  &  &  &  & 0.22 & 0.48 & WHT \\ 
1311 & BD+02 4571 & 9.15 & 6.42 & 5.60 & 5.42 & 5.19 & 0.07 & 0.05 & WHT, Hermes \\ 
1336 & CSS 1336 &  &  &  &  &  & 0.50 & 0.16 & WHT \\ 
\end{longtable} 
\end{landscape}}

\begin{landscape}
\begin{table*}[!ht]
\caption{The results of the period analysis for all stars as described in Section \ref{sct:periodanalysis}. A dash indicates that there was no significant pulsation period found or that the timeseries did not cover a timespan longer than the found period.}
\smallskip
\label{table:periods}  
\centering

\begin{tabular}{rccccrcccc}
\noalign{\smallskip}
name  & Type & Period & Type (GCVS) & Period (GCVS)  & name  & Type & Period  & Type (GCVS) & Period (GCVS)  \\
      &      & days   &             &  days          &       &      &  days   &             &   days         \\
\hline\hline  
 CSS 1 &   &  -  &  -  &  -  & CSS 553 &   &  -  &  -  &  - \\
 RX Psc & Mira & 281 & Mira & 281 & CSS 558 & SR & 142 &  -  &  - \\
 CSS 38 &   &  -  &  -  &  -  & CSS 584 & SR & 301 &  -  &  - \\
 NX Per &   &  -  & Mira &  -  & CSS 598 & Irr &  -  &  -  &  - \\
 CSS 100 &   &  -  &  -  &  -  & CD-30 7391 &   &  -  &  -  &  - \\
 CSS 127 & SR & 239 &  -  &  -  & CSS 604 &   &  -  &  -  &  - \\
 CSS 166 & SR & 262 &  -  &  -  & CSS 613 & SR & 127 &  -  &  - \\
 CSS 173 & SR & 139 &  -  &  -  & CSS 622 & SR & 159 &  -  &  - \\
 KR Cam &   &  -  & SR &  -  & CSS 633 &   &  -  &  -  &  - \\
 CSS 179 &   &  -  &  -  &  -  & CSS 635 & SR & 150 &  -  &  - \\
 GH Aur &   &  -  & Mira & 388 & CSS 640 & SR & 125 &  -  &  - \\
 V376 Aur &   &  -  &   &  -  & CSS 650 & SR & 102 &  -  &  - \\
 CD-39 2449 & SR & 167 &  -  &  -  & CSS 657 & Mira & 409 &  -  &  - \\
 NP Aur &   &  -  & Mira & 334 & CSS 661 & Mira & 409 & Mira & 418\\
 CF Gem & SR & 153 &   &  -  & CSS 665 & SR & 110 &  -  &  - \\
 HV 8057 & Mira & 370 & Mira & 370 & AU Car & Mira & 347 & Mira & 332\\
 CSS 267 & Irr &  -  &  -  &  -  & CSS 682 & Irr &  -  &  -  &  - \\
 BZ CMa & Mira & 333 & Mira & 320 & RX Car & Mira & 336 & Mira & 333\\
 CSS 282 & Irr &  -  &  -  &  -  & CSS 719 & SR & 278 &  -  &  - \\
 CSS 298 & SR & 142 &  -  &  -  & CSS 723 & SR & 223 &  -  &  - \\
 CD-43 2976 & SR & 74 &  -  &  -  & CSS 724 & Mira & 281 & Mira & 280\\
 XY CMi & Mira & 271 &   &  -  & CSS 739 & Mira & 336 &  -  &  - \\
 CD-30 4223 & SR & 198 &  -  &  -  & CSS 749 & SR & 148 &  -  &  - \\
 CSS 352 & SR & 115 &  -  &  -  & CSS 757 &   &  -  &  -  &  - \\
 CSS 361 & SR & 125 &  -  &  -  & CSS 763 & SR & 83 &  -  &  - \\
 CSS 371 & SR & 146 &  -  &  -  & CSS 783 & Irr &  -  &  -  &  - \\
 CSS 380 & Mira & 363 &  -  &  -  & CSS 794 & Mira & 307 & Mira & 305\\
 CSS 393 & SR & 92 &  -  &  -  & CSS 806 &   &  -  &  -  &  - \\
 CSS 417 & SR & 167 &  -  &  -  & CSS 881 & SR & 158 &  -  &  - \\
 CSS 427 &   &  -  &  -  &  -  & CSS 921 & Irr &  -  &  -  &  - \\
 AO Gem &   &  -  & Mira & 313 & CSS 987 & SR & 286 &  -  &  - \\
 CSS 438 & Mira & 405 &  -  &  -  & CSS 1005 &   &  -  &  -  &  - \\
 EX Pup & SR & 290 & Mira & 289 & CSS 1046 &   &  -  &  -  &  - \\
 CSS 453 & SR & 61 &  -  &  -  & CSS 1077 & SR & 70 &  -  &  - \\
 CSS 454 & SR & 79 &  -  &  -  & TU Dra &   &  -  & Mira & 345\\
 CSS 457 & SR & 69 &  -  &  -  & V734 Cyg &   &  -  & Mira & 310\\
 CSS 466 & Mira & 385 &  -  &  -  & CSS 1138 &   &  -  &  -  &  - \\
 CSS 472 & SR & 95 &  -  &  -  & CSS 1160 & Mira & 357 & Mira & 360\\
 CSS 478 & SR & 62 &  -  &  -  & CM Cyg &   &  -  & Mira & 255\\
 CSS 480 & SR & 117 &  -  &  -  & V899 Aql & Mira & 374 & SR & 100\\
 CD-29 5912 & SR & 103 &  -  &  -  & CE Lac &   &  -  & SR & 107\\
 CSS 528 & SR & 107 &  -  &  -  & GY Lac &   &  -  & Mira &  - \\
 CSS 531 &   &  -  &  -  &  -  & CSS 1336 &   &  -  &  -  &  - \\
 CO Pyx & Mira & 329 & Mira &  -  &   &   &   &  -  &  - \\
\hline\hline  
\end{tabular}
\end{table*}
\end{landscape}

\begin{table*}[!ht]
\caption{Effective temperature, C/O ratio and gross s-process overabundance of S stars. Hen 4-NNN lists S stars as in the Simbad database from an unpublished list of Henize \citep[see e.g.][]   {VanEck1999b}.The ``uncertain'' label in the comment column means that the convergence to the quoted set of physical parameters was not as consistent as in all the other cases.}
\smallskip
\label{table:1}  
\centering

\begin{tabular}{rcccc}
\noalign{\smallskip}
CSS  & ${\rm T}_{\rm eff}$ & C/O &  [s/Fe]  &  Comments \\
\hline\hline  
127  & 3300 $\pm$ 100 & 0.95 $\pm$ 0.025 & 1.5 $\pm$ 0.5 & \\
166  & 3300 $\pm$ 100 & 0.98 $\pm$ 0.010 & 1.7 $\pm$ 0.3 & \\
173  & 3600 $\pm$ 100 & 0.97 $\pm$ 0.020 & 1.5 $\pm$ 0.5 & uncertain \\
177  & 3350 $\pm$ 150 & 0.96 $\pm$ 0.010 & 1.0 $\pm$ 0.5 & KR Cam \\
179  & 3250 $\pm$ 50  & 0.98 $\pm$ 0.010 & 1.5 $\pm$ 0.5 & \\
191  & 3200 $\pm$ 100 & 0.84 $\pm$ 0.088 & 1.5 $\pm$ 0.5 & GH Aur ; H$_\alpha$ emission  ;  Mira ; uncertain \\
196  & 3250 $\pm$ 50  & 0.84 $\pm$ 0.088 & 0.5 $\pm$ 0.5 & V376 Aur \\
202  & 3500 $\pm$ 100 & 0.70 $\pm$ 0.200 & 1.0 $\pm$ 0.3 & CD-39 2449 ; Hen 4-8 ; \\
220  & 3300 $\pm$ 100 & 0.98 $\pm$ 0.010 & 1.5 $\pm$ 0.5 & CF Gem \\
267  & 3350 $\pm$ 50  & 0.99 $\pm$ 0.010 & 1.5 $\pm$ 0.5 & \\
281  & 3300 $\pm$ 100 & 0.99 $\pm$ 0.010 & 1.0 $\pm$ 0.5 & BZ CMa \\
282  & 3300 $\pm$ 100 & 0.97 $\pm$ 0.020 & 1.5 $\pm$ 0.5 & \\
315  & 3450 $\pm$ 50  & 0.63 $\pm$ 0.125 & 0.6 $\pm$ 0.4 & CD-43 2976 \\
325  & 3200 $\pm$ 100 & 0.71 $\pm$ 0.210 & 0.5 $\pm$ 0.5 & XY CMi \\
328  & 3450 $\pm$ 150 & 0.93 $\pm$ 0.025 & 1.0 $\pm$ 0.3 & CD-30 4223 ; Hen 4-19 ; \\
352  & 3400 $\pm$ 100 & 0.70 $\pm$ 0.200 & 1.3 $\pm$ 0.3 & \\
361  & 3300 $\pm$ 100 & 0.97 $\pm$ 0.020 & 1.5 $\pm$ 0.5 & Hen 4-23  \\
371  & 3400 $\pm$ 100 & 0.98 $\pm$ 0.010 & 1.5 $\pm$ 0.5 & Hen 4-26 ; uncertain \\
380  & 3150 $\pm$ 250 & 0.98 $\pm$ 0.010 & 1.0 $\pm$ 0.7 & Hen 4-27 ; H$_\alpha$ emission \\
393  & 3450 $\pm$ 50  & 0.97 $\pm$ 0.020 & 1.5 $\pm$ 0.5 & Hen 4-29 \\
417  & 3450 $\pm$ 50  & 0.98 $\pm$ 0.010 & 1.5 $\pm$ 0.5 & H$_\alpha$ emission     \\
427  & 3200 $\pm$ 100 & 0.97 $\pm$ 0.020 & 1.5 $\pm$ 0.5 & \\
431  & 3300 $\pm$ 100 & 0.96 $\pm$ 0.010 & 1.5 $\pm$ 0.5 & AO Gem   ;  H$_\alpha$ emission \\
438  & 3150 $\pm$ 150 & 0.96 $\pm$ 0.010 & 1.5 $\pm$ 0.5 & \\
443  & 2900 $\pm$ 200 & 0.71 $\pm$ 0.210 & 0.5 $\pm$ 0.5 & EX Pup ; H$_\alpha$ emission  \\
453  & 3400 $\pm$ 100 & 0.70 $\pm$ 0.200 & 0.5 $\pm$ 0.5 & \\
454  & 3400 $\pm$ 100 & 0.71 $\pm$ 0.210 & 1.0 $\pm$ 0.2 & \\
457  & 3450 $\pm$ 50  & 0.63 $\pm$ 0.125 & 0.5 $\pm$ 0.5 & Hen 4-38  ; uncertain \\
472  & 3350 $\pm$ 50  & 0.63 $\pm$ 0.125 & 0.1 $\pm$ 0.1 & MS star \\
478  & 3400 $\pm$ 100 & 0.71 $\pm$ 0.210 & 0.5 $\pm$ 0.5 & \\
480  & 3350 $\pm$ 50  & 0.63 $\pm$ 0.125 & 0.4 $\pm$ 0.3 & \\
489  & 3450 $\pm$ 50  & 0.63 $\pm$ 0.125 & 1.0 $\pm$ 0.3 & CD-29 5912 ; Hen 4-44 \\
528  & 3450 $\pm$ 50  & 0.83 $\pm$ 0.075 & 0.4 $\pm$ 0.4 & \\
531  & 3250 $\pm$ 50  & 0.91 $\pm$ 0.012 & 1.2 $\pm$ 0.2 & \\
553  & 3300 $\pm$ 100 & 0.70 $\pm$ 0.200 & 0.4 $\pm$ 0.3 & \\
558  & 3300 $\pm$ 100 & 0.71 $\pm$ 0.213 & 0.9 $\pm$ 0.3 & H$_\alpha$ emission     \\
584  & 3300 $\pm$ 100 & 0.99 $\pm$ 0.010 & 1.5 $\pm$ 0.5 & Hen 4-62  ; uncertain \\
598  & 3250 $\pm$ 150 & 0.97 $\pm$ 0.010 & 1.5 $\pm$ 0.5 & \\
603  & 3400 $\pm$ 100 & 0.70 $\pm$ 0.200 & 0.4 $\pm$ 0.4    & CD-30 7391 \\
604  & 3300 $\pm$ 100 & 0.98 $\pm$ 0.010 & 1.5 $\pm$ 0.5 & \\
613  & 3450 $\pm$ 50  & 0.98 $\pm$ 0.010 & 1.7 $\pm$ 0.3 & \\
622  & 3200 $\pm$ 100 & 0.99 $\pm$ 0.010 & 0.5 $\pm$ 0.4 & \\
633  & 3300 $\pm$ 50  & 0.91 $\pm$ 0.012 & 0.2 $\pm$ 0.1 & Hen 4-73 ; uncertain \\
635  & 3400 $\pm$ 100 & 0.98 $\pm$ 0.010 & 1.5 $\pm$ 0.5 & H$_\alpha$ emission     \\
640  & 3250 $\pm$ 50  & 0.97 $\pm$ 0.010 & 1.0 $\pm$ 0.3 & Hen 4-75  \\
650  & 3450 $\pm$ 50  & 0.70 $\pm$ 0.200 & 0.8 $\pm$ 0.5 & Hen 4-77 \\
657  & 3300 $\pm$ 100 & 0.97 $\pm$ 0.010 & 1.5 $\pm$ 0.5 & Hen 4-81  ;  H$_\alpha$ emission \\
665  & 3300 $\pm$ 100 & 0.97 $\pm$ 0.020 & 1.5 $\pm$ 0.5 & \\
682  & 3400 $\pm$ 100 & 0.71 $\pm$ 0.210 & 0.8 $\pm$ 0.3 & Hen 4-92 \\
690  & 3150 $\pm$ 150 & 0.71 $\pm$ 0.210 & 0.3 $\pm$ 0.3 & RX Car ; H$_\alpha$ emission  ;  Mira ; uncertain \\
718  & 3300 $\pm$ 100 & 0.71 $\pm$ 0.213 & 0.5 $\pm$ 0.5 & Hen 4-106 ; uncertain \\
719  & 3400 $\pm$ 100 & 0.99 $\pm$ 0.010 & 1.5 $\pm$ 0.5 & \\
739  & 3200 $\pm$ 100 & 0.91 $\pm$ 0.012 & 0.1 $\pm$ 0.1 & Mira  \\
757  & 3300 $\pm$ 100 & 0.98 $\pm$ 0.010 & 1.5 $\pm$ 0.5 & \\
763  & 3300 $\pm$ 100 & 0.71 $\pm$ 0.213 & 0.5 $\pm$ 0.5 & \\
783  & 3450 $\pm$ 150 & 0.71 $\pm$ 0.210 & 0.5 $\pm$ 0.5 & \\
794  & 3150 $\pm$ 150 & 0.96 $\pm$ 0.010 & 0.5 $\pm$ 0.5 & Hen 4-122 \\
806  & 3650 $\pm$ 50  & 0.83 $\pm$ 0.075 & 0.5 $\pm$ 0.5 & high extinction; uncertain \\
881  & 3250 $\pm$ 150 & 0.99 $\pm$ 0.010 & 1.0 $\pm$ 0.5 & \\
921  & 3350 $\pm$ 50  & 0.63 $\pm$ 0.125 & 0.3 $\pm$ 0.3 & \\
1005 & 3350 $\pm$ 150 & 0.97 $\pm$ 0.020 & 1.5 $\pm$ 0.5 & \\
1180 & 3300 $\pm$ 100 & 0.99 $\pm$ 0.010 & 1.0 $\pm$ 0.3 & CM Cyg ; strong H$_\alpha$ emission  \\
1305 & 3250 $\pm$ 150 & 0.70 $\pm$ 0.200 & 0.3 $\pm$ 0.3 & CE Lac  \\
1336 & 3300 $\pm$ 100 & 0.63 $\pm$ 0.125 & 0.4 $\pm$ 0.3 & \\
\hline\hline  
\end{tabular}
\end{table*}

\Online

\begin{appendix} 

\section{Results of the dust decomposition tool}
\begin{figure*}
\centering
\subfigure{
\includegraphics[width=8.0cm]{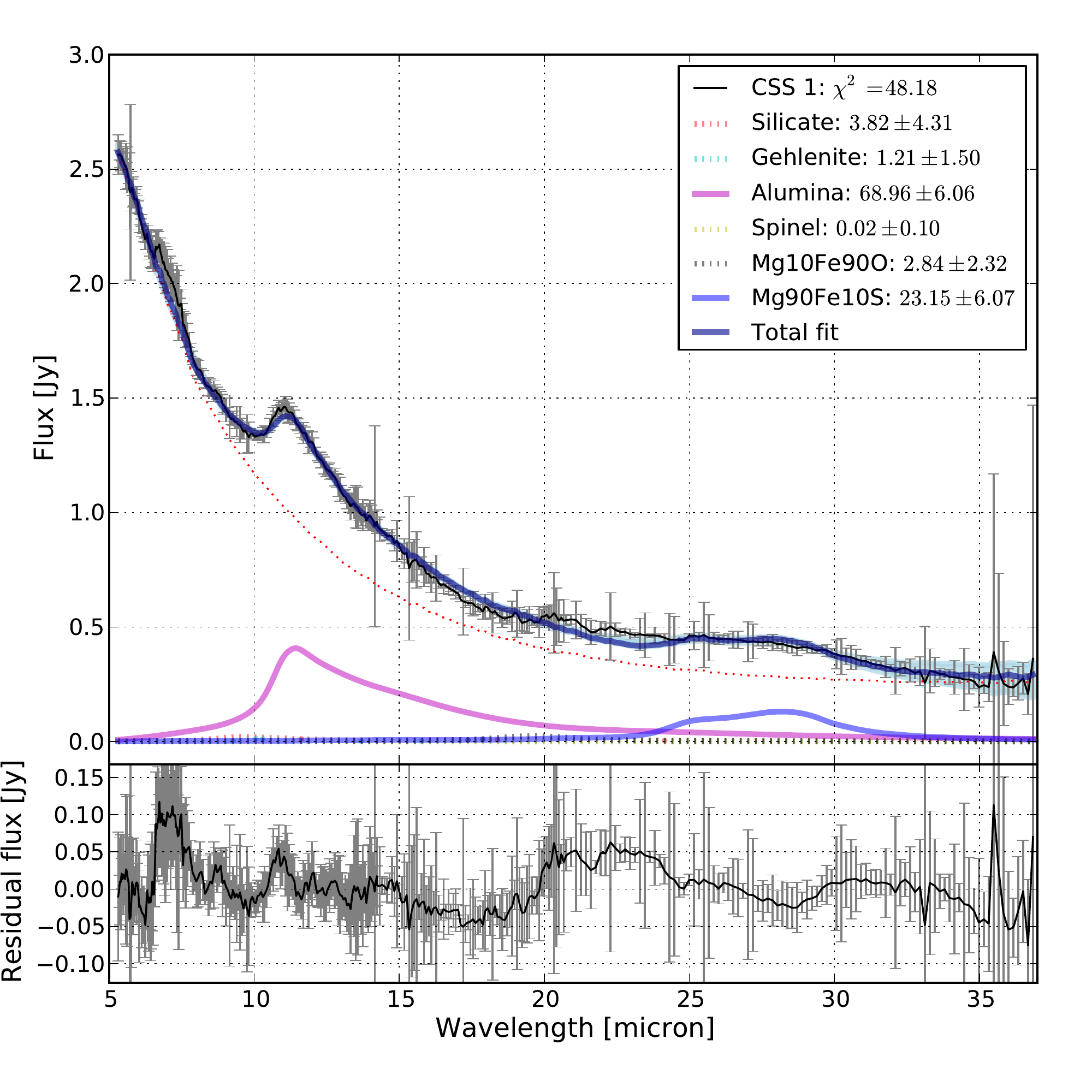}
\label{fig:subfigure_appendix1a}
}
\subfigure{
\includegraphics[width=8.0cm]{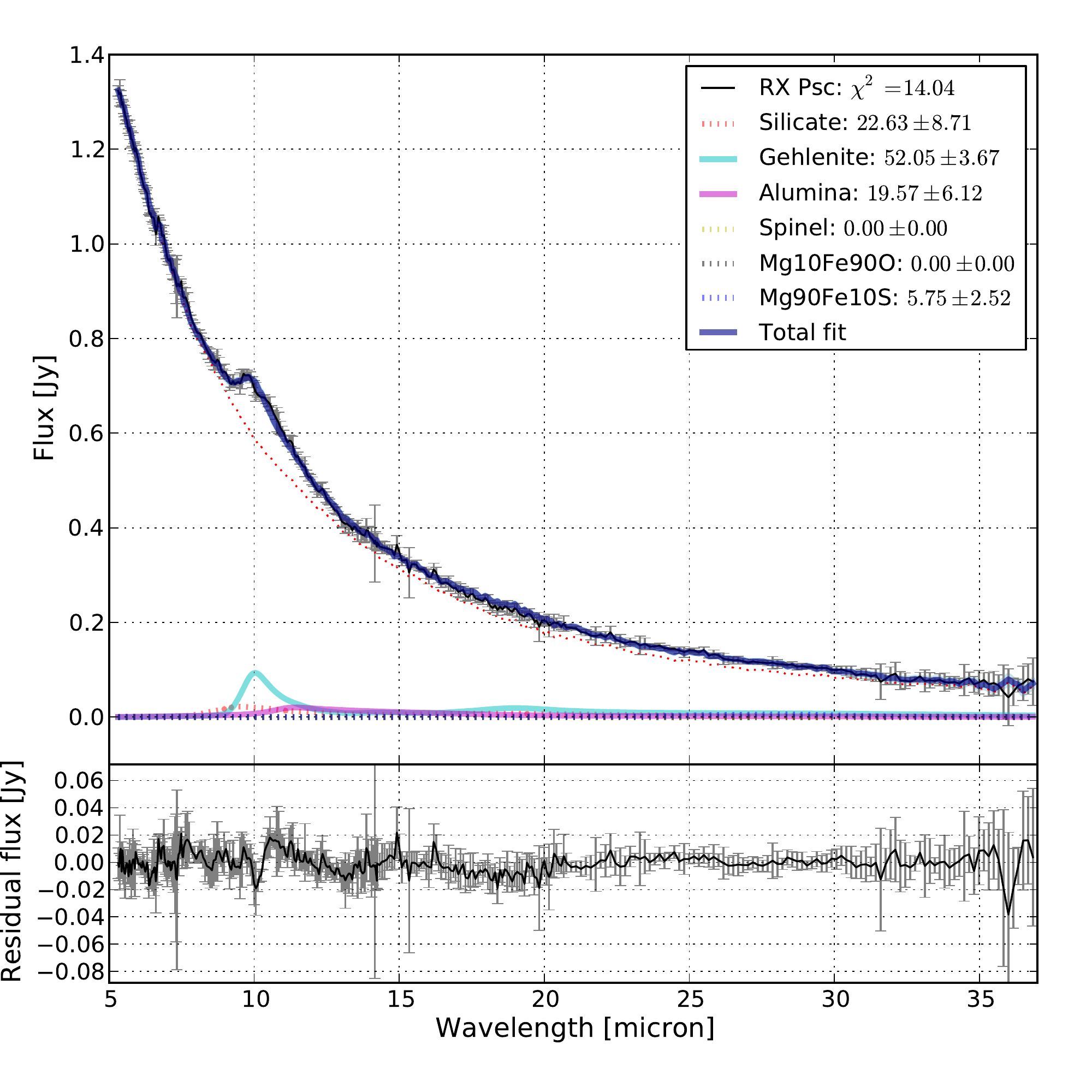}
\label{fig:subfigure_appendix1b}
}
\subfigure{
\includegraphics[width=8.0cm]{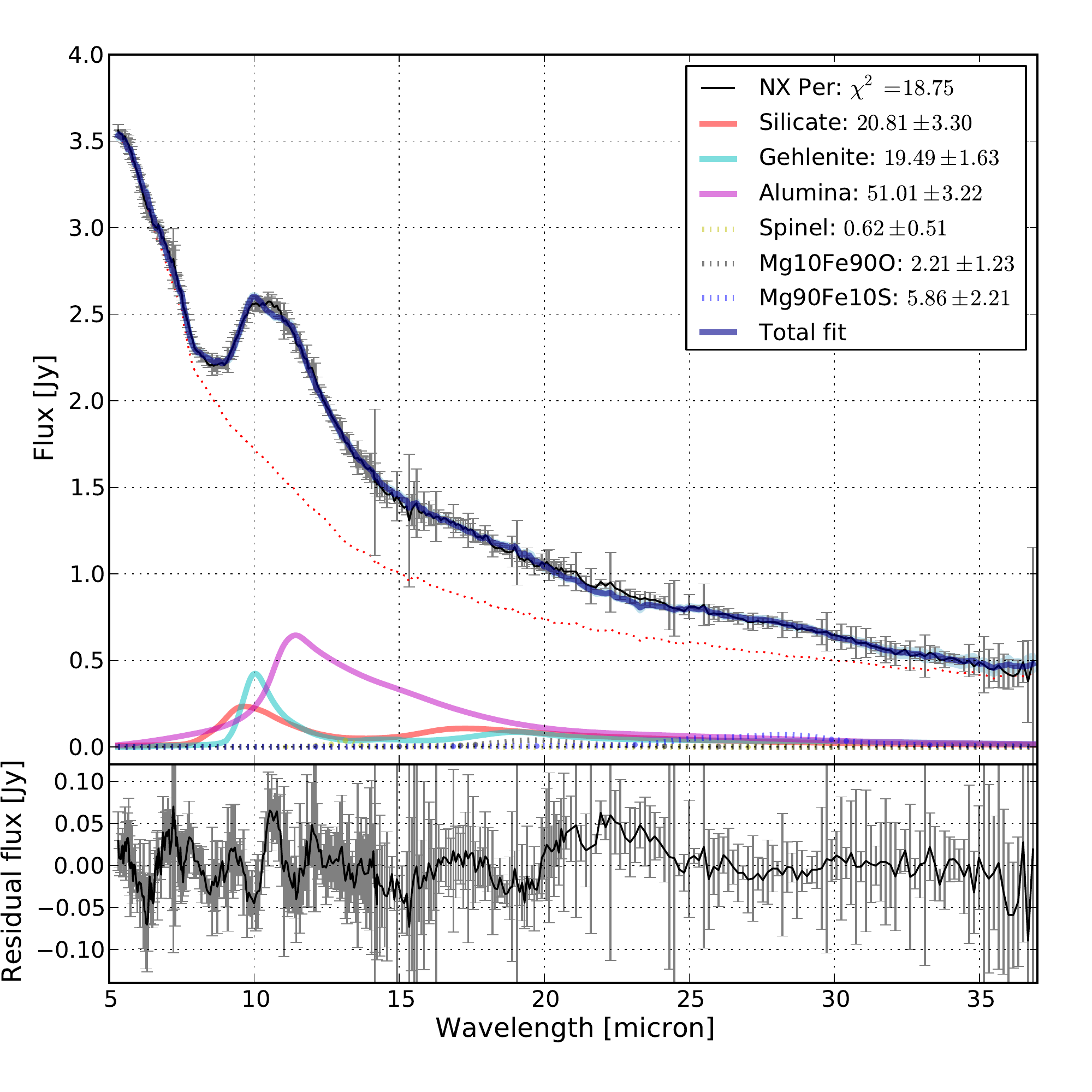}
\label{fig:subfigure_appendix1c}
}
\subfigure{
\includegraphics[width=8.0cm]{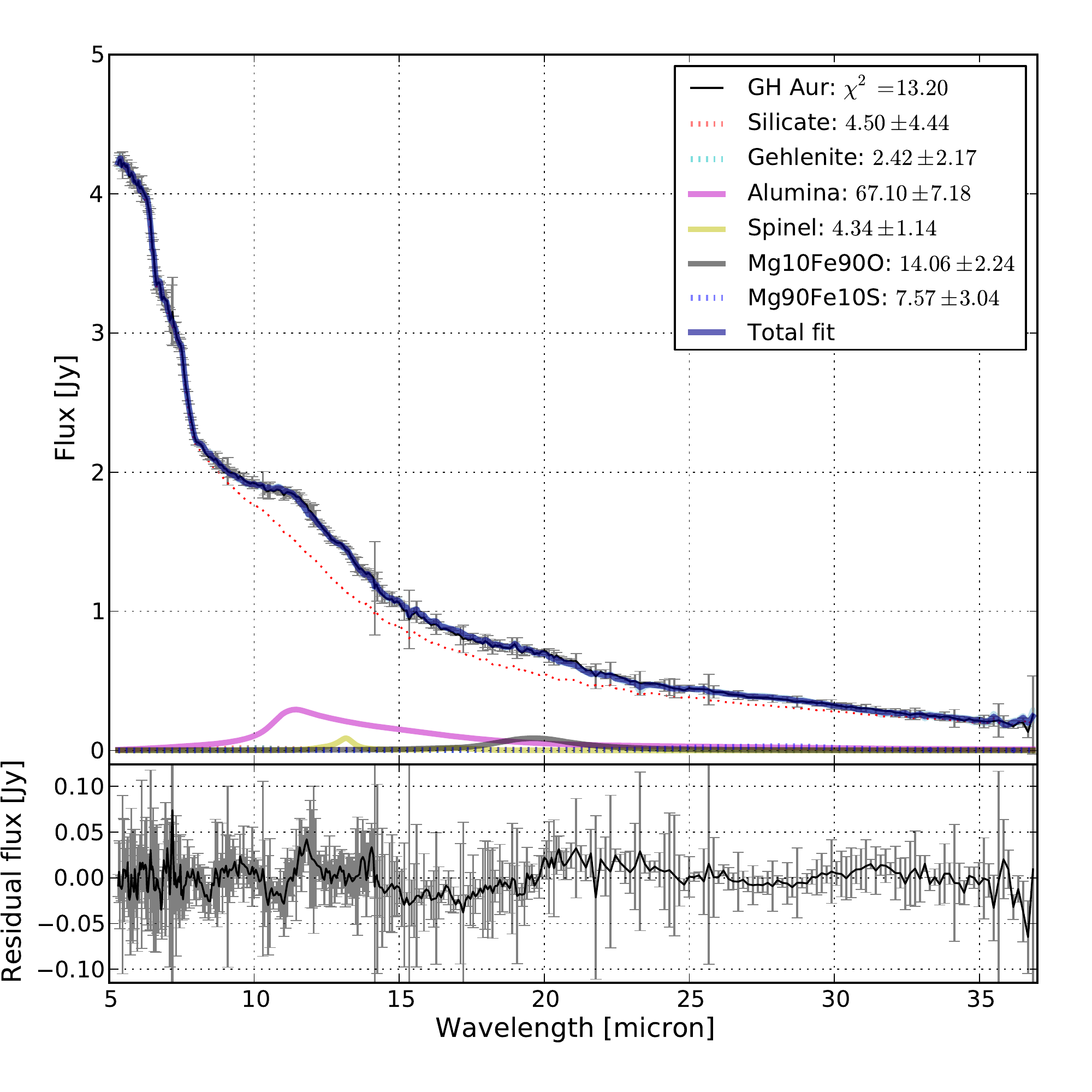}
\label{fig:subfigure_appendix1d}
}
\subfigure{
\includegraphics[width=8.0cm]{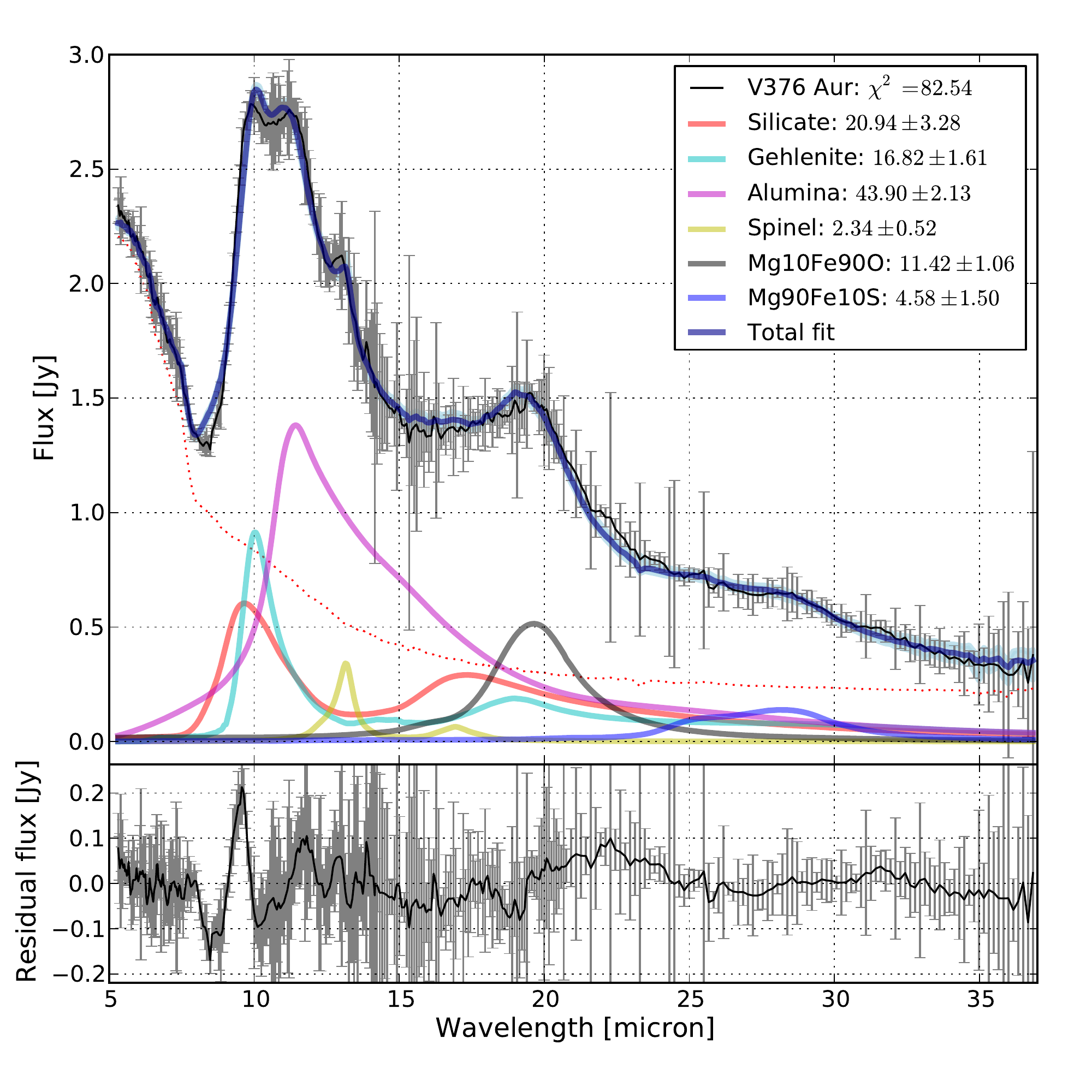}
\label{fig:subfigure_appendix1e}
}
\subfigure{
\includegraphics[width=8.0cm]{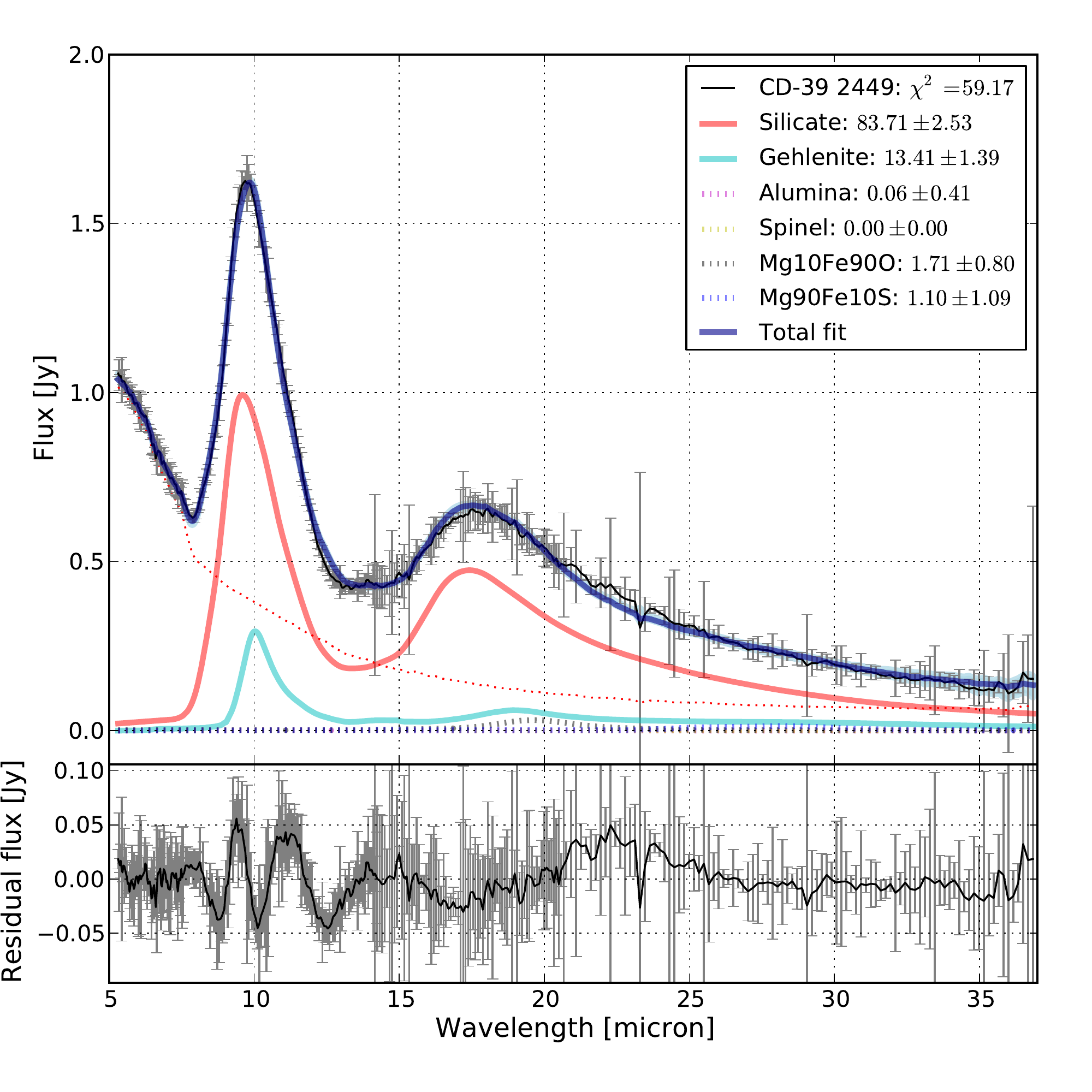}
\label{fig:subfigure_appendix1f}
}
\caption[]{\label{fig:subfigure_appendix1}
The results of the dust decomposition tool for all stars with dust features. Top panel: the black line shows the Spitzer spectrum, the red dotted line the fit the underlying estimate for the stellar continuum, as derived from the dust decomposition tool and the solid, dark blue line indicates the total fit to the spectrum. Solid lines shown in the legend indicate significant contributions to the spectrum, the dotted lines the non-significant contributions. Bottom panel: the residuals after subtracting the total fit.}
\end{figure*}

\begin{figure*}
\centering
\subfigure{
\includegraphics[width=8.0cm]{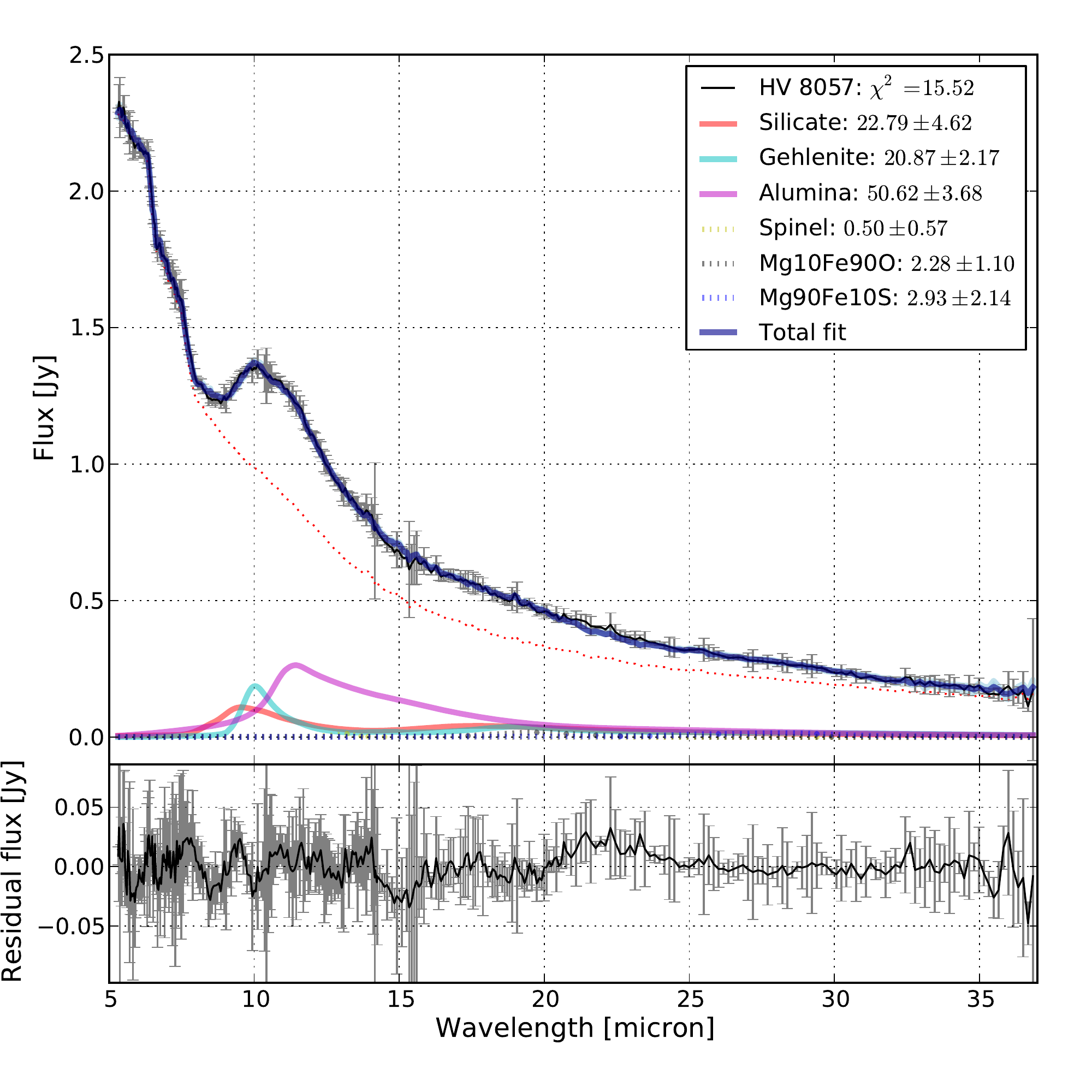}
\label{fig:subfigure_appendix2a}
}
\subfigure{
\includegraphics[width=8.0cm]{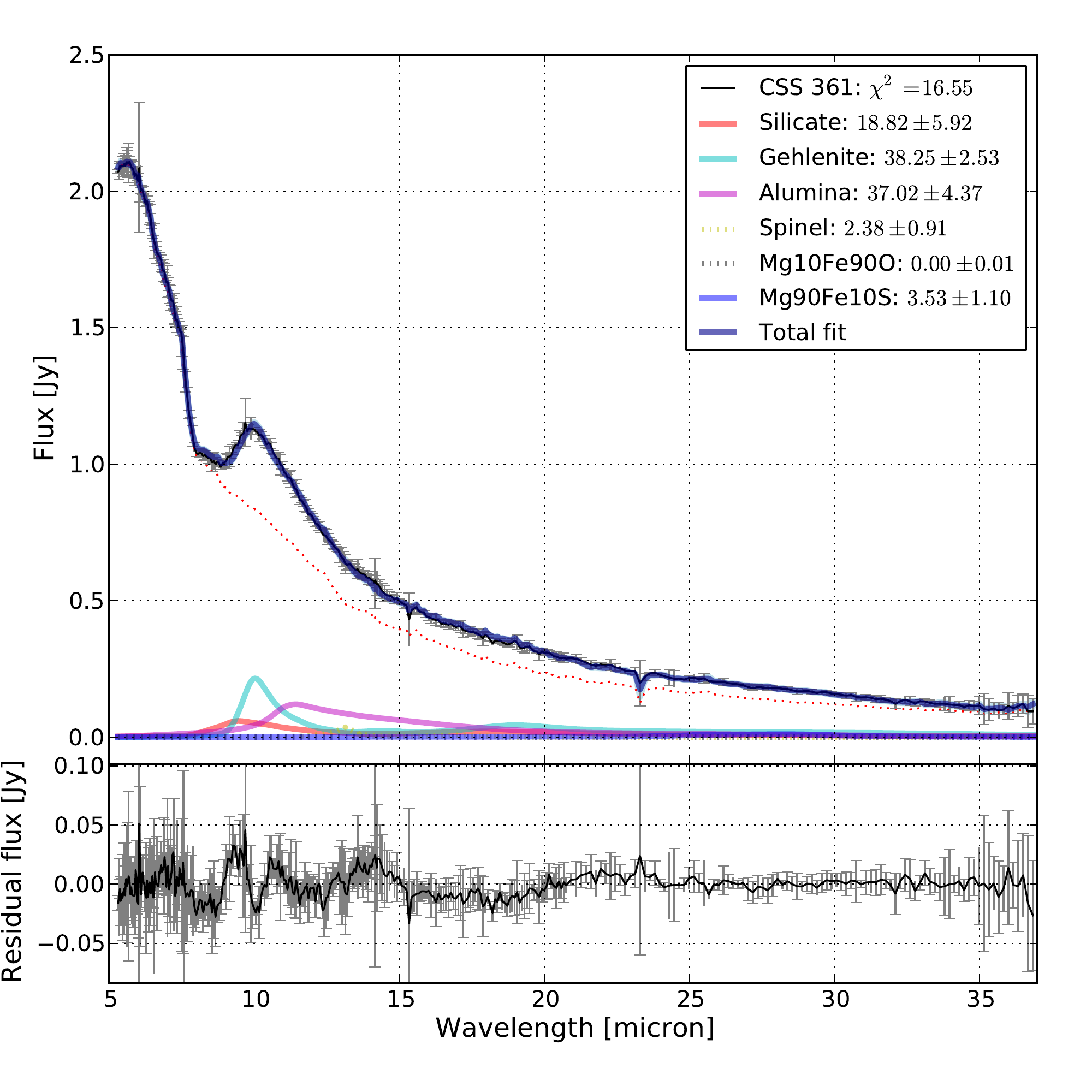}
\label{fig:subfigure_appendix2b}
}
\subfigure{
\includegraphics[width=8.0cm]{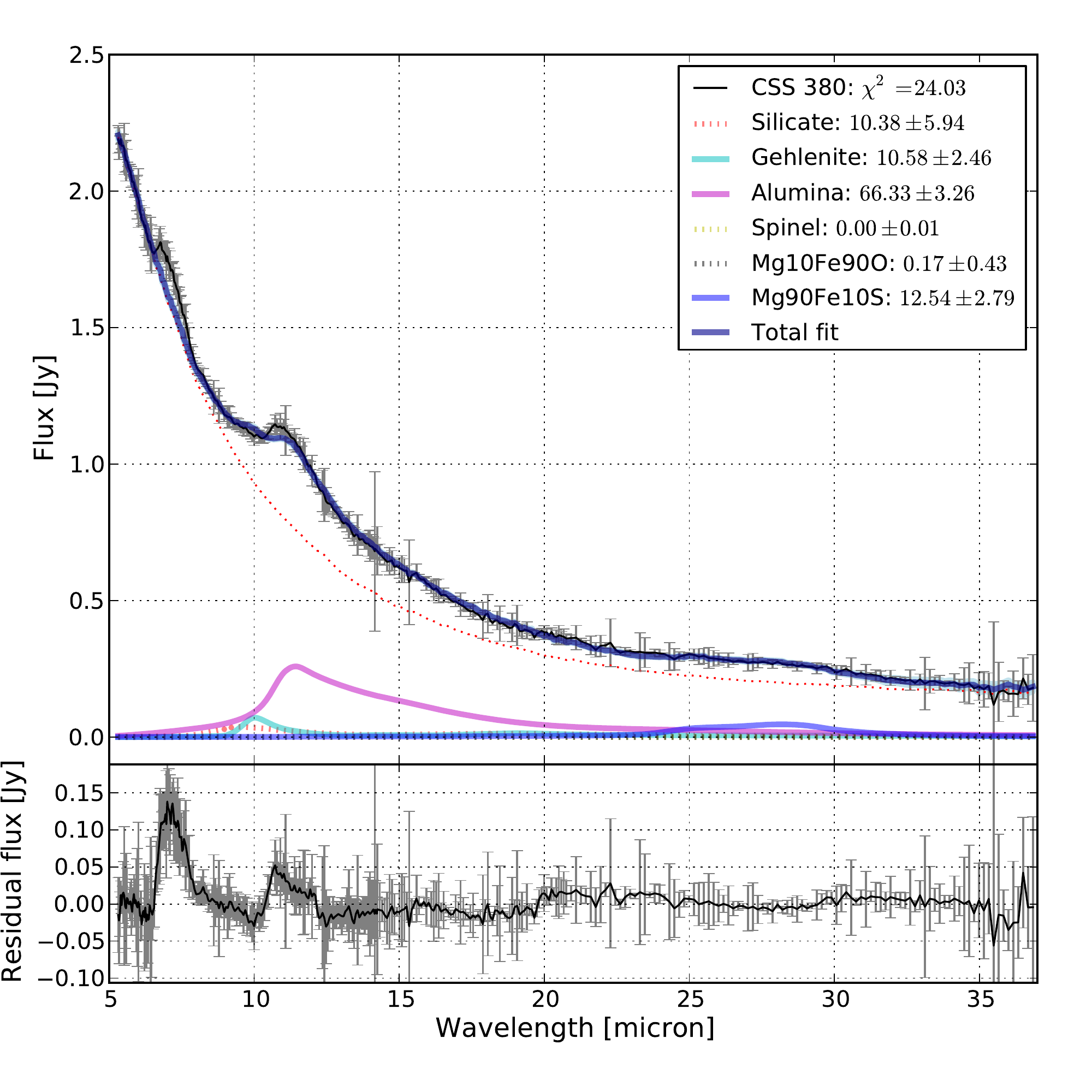}
\label{fig:subfigure_appendix2c}
}
\subfigure{
\includegraphics[width=8.0cm]{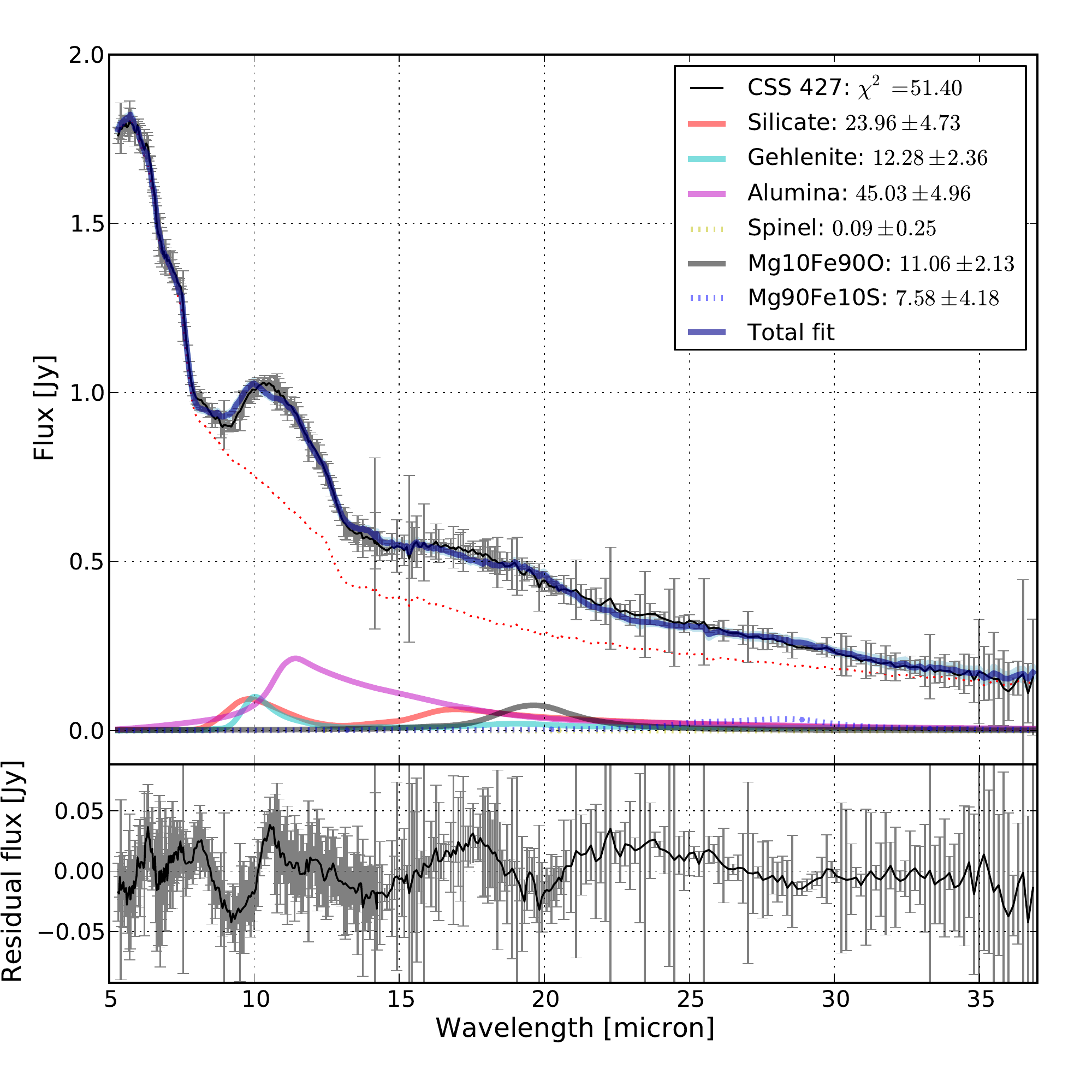}
\label{fig:subfigure_appendix2d}
}
\subfigure{
\includegraphics[width=8.0cm]{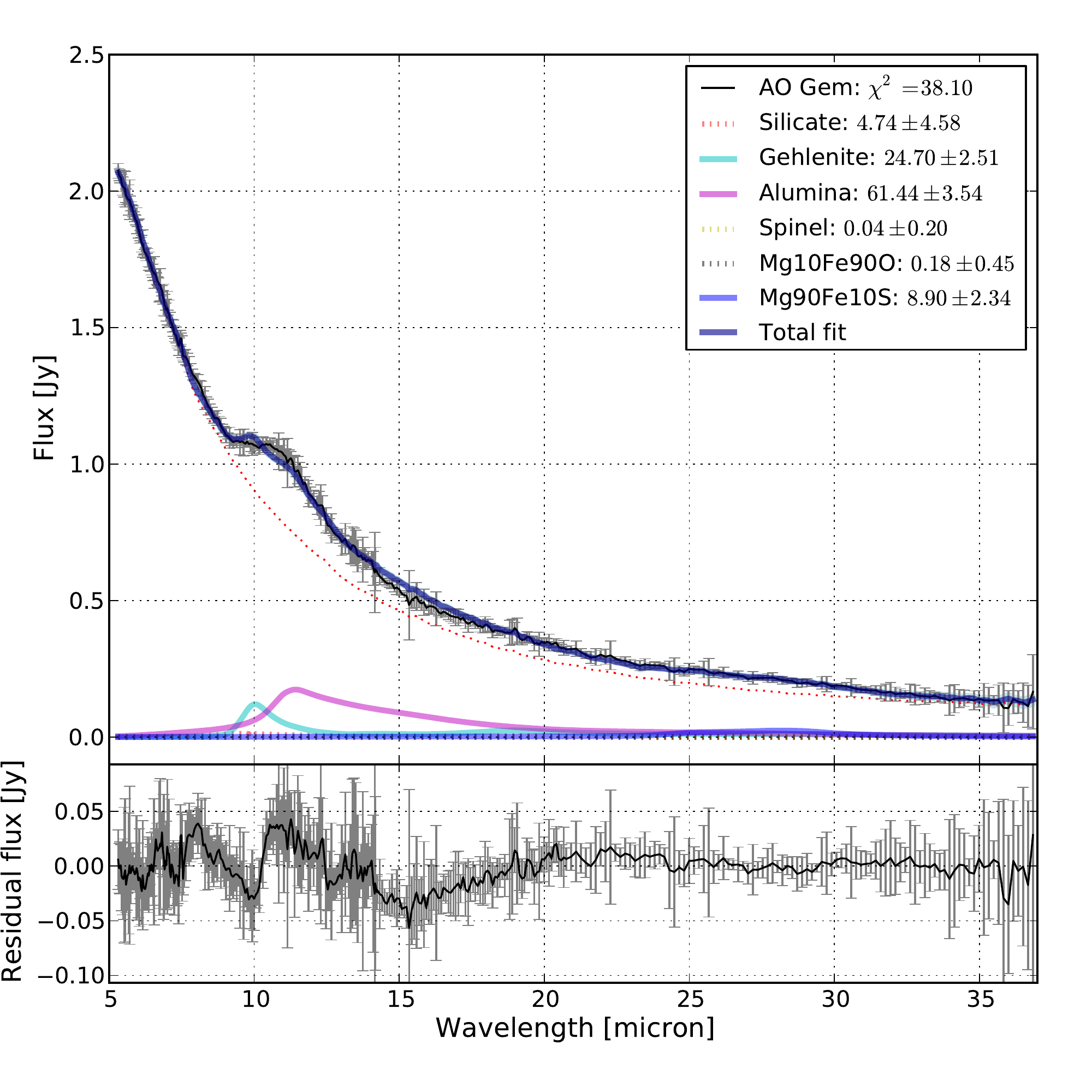}
\label{fig:subfigure_appendix2e}
}
\subfigure{
\includegraphics[width=8.0cm]{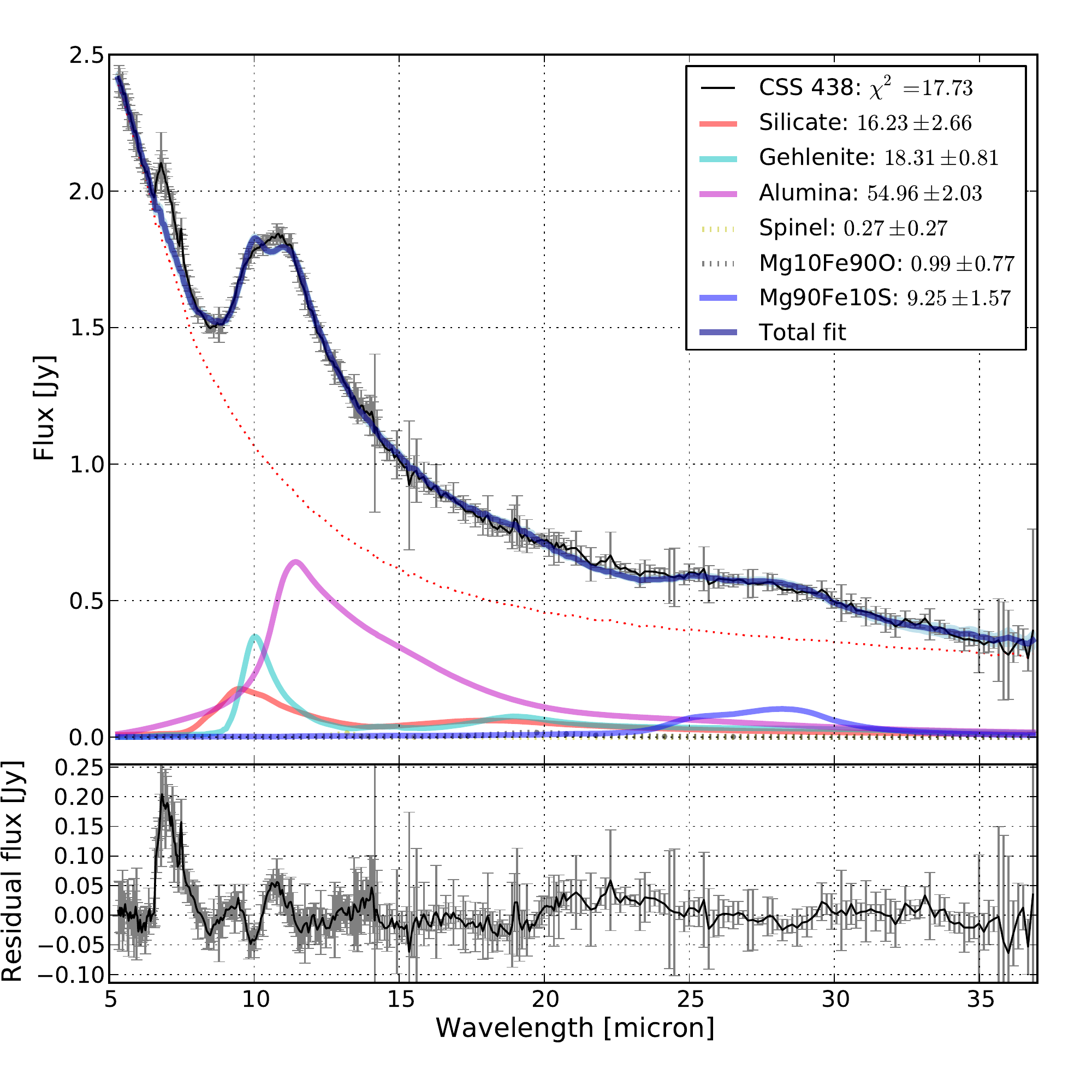}
\label{fig:subfigure_appendix2f}
}
\caption[]{\label{fig:subfigure_appendix2}
Continued}
\end{figure*}

\begin{figure*}
\centering
\subfigure{
\includegraphics[width=8.0cm]{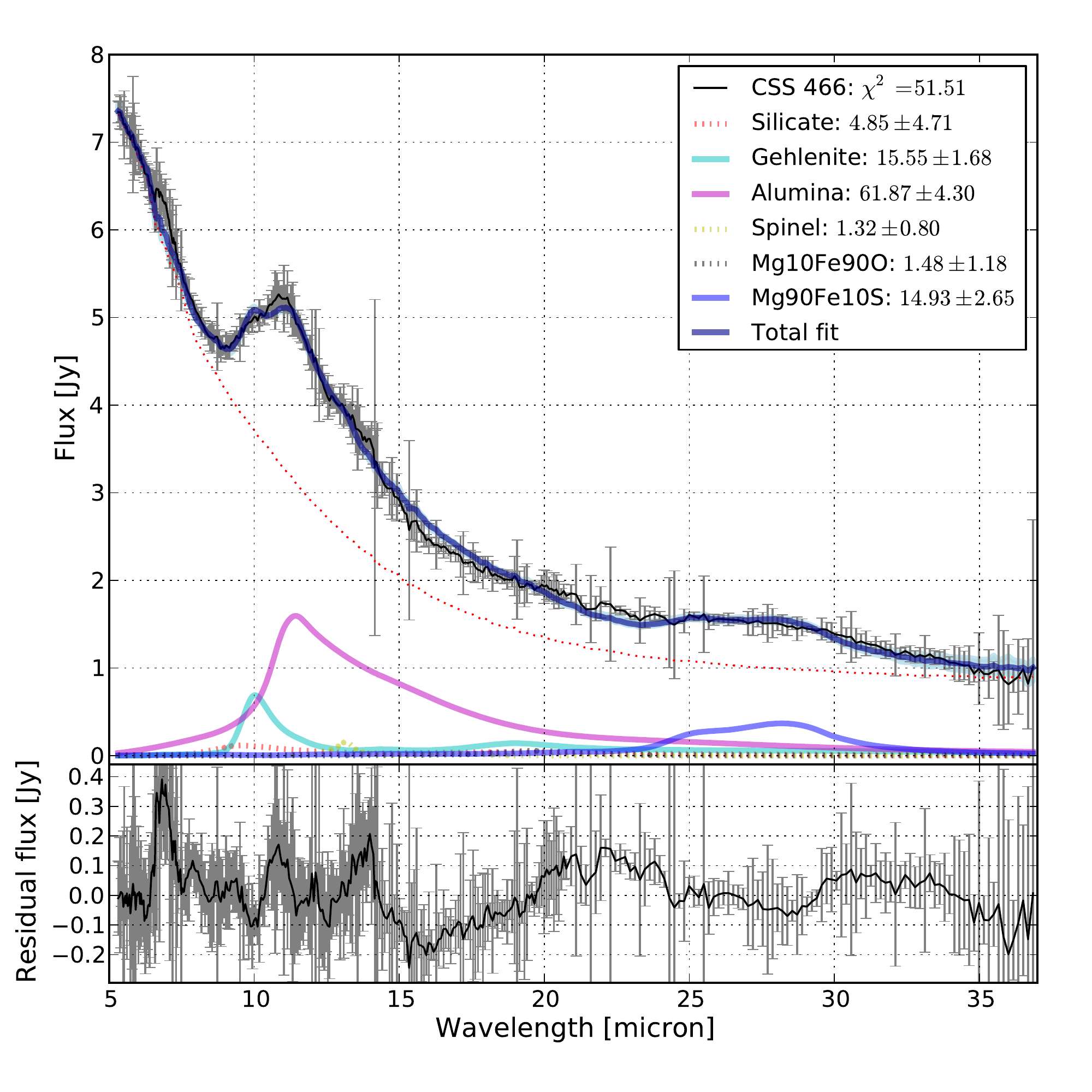}
\label{fig:subfigure_appendix3a}
}
\subfigure{
\includegraphics[width=8.0cm]{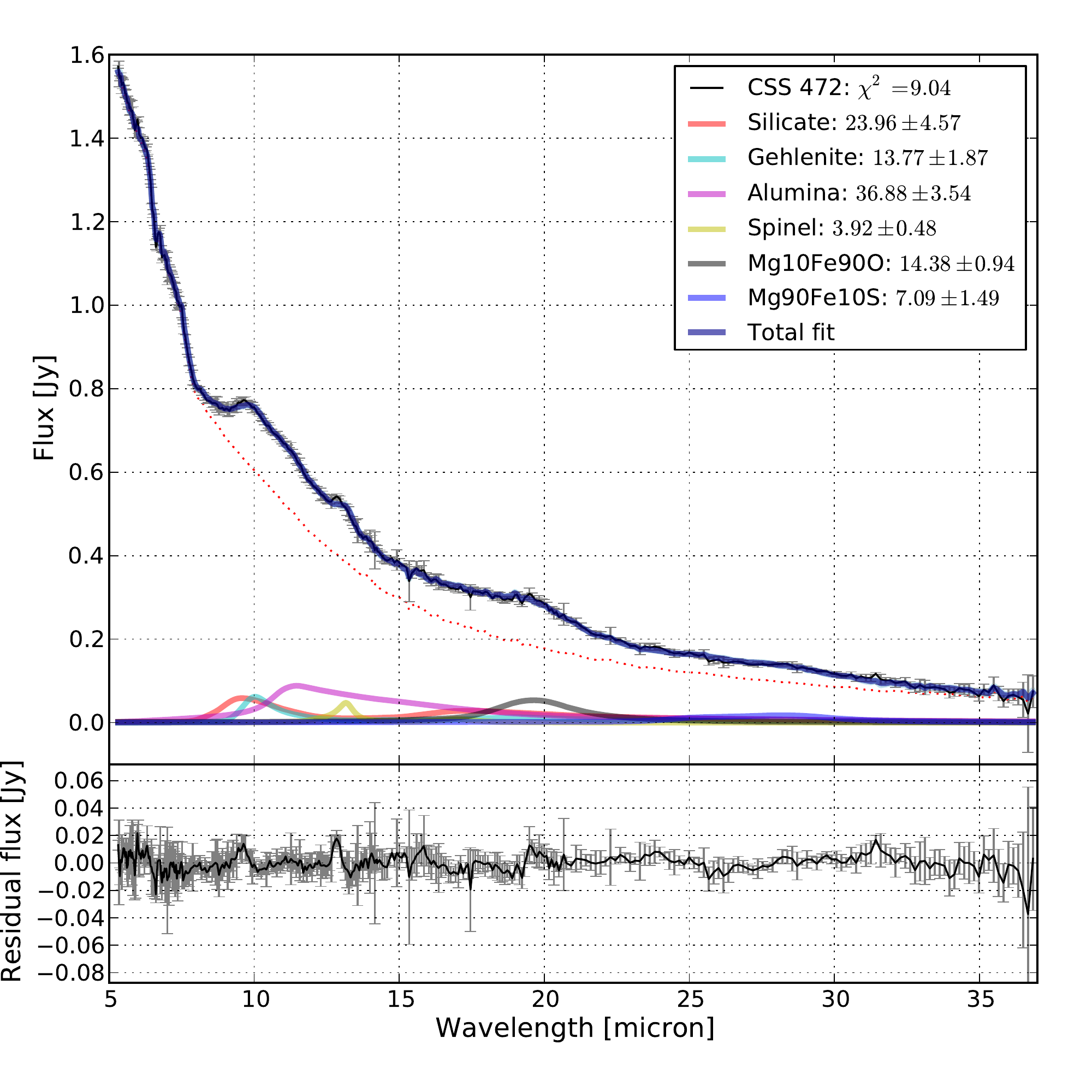}
\label{fig:subfigure_appendix3b}
}
\subfigure{
\includegraphics[width=8.0cm]{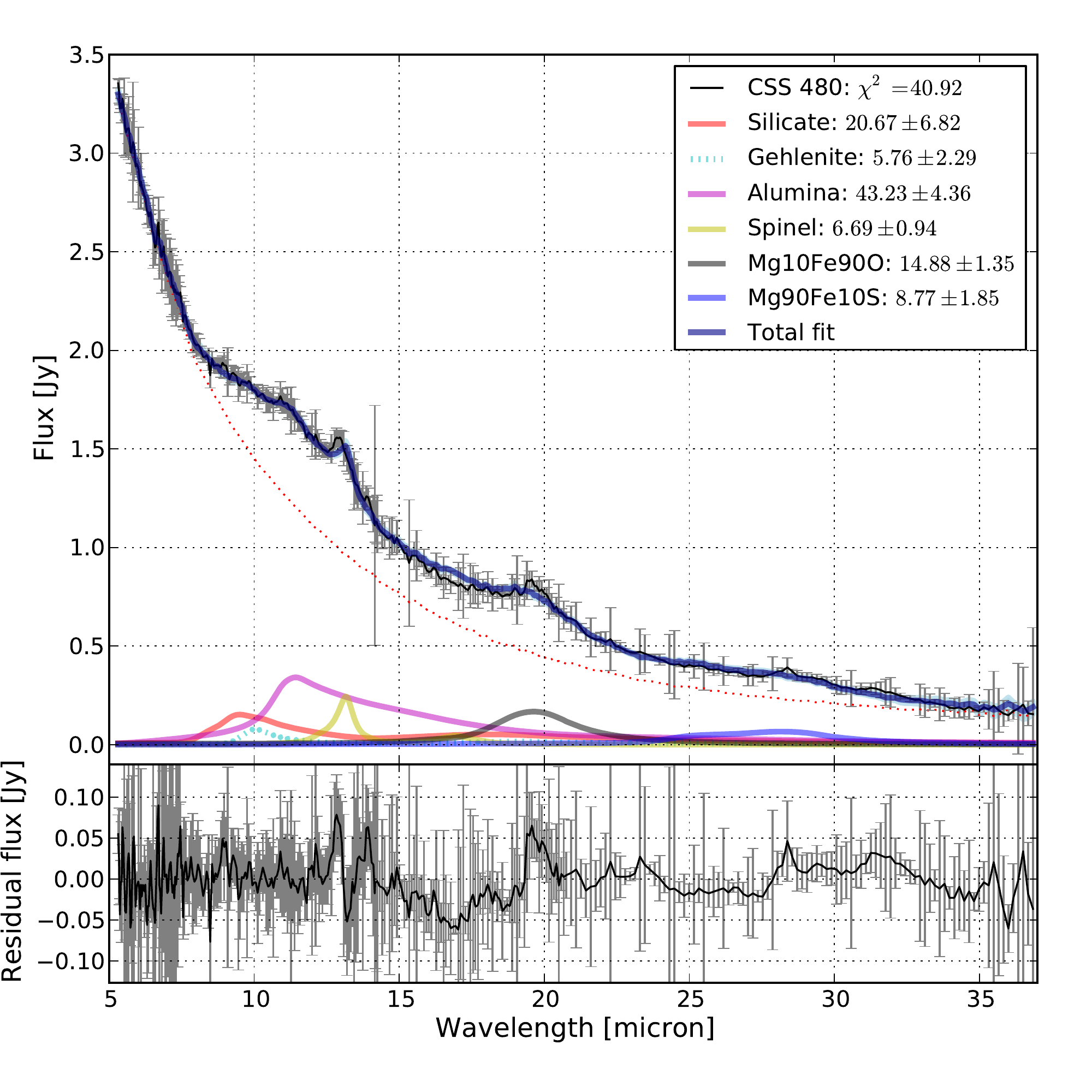}
\label{fig:subfigure_appendix3c}
}
\subfigure{
\includegraphics[width=8.0cm]{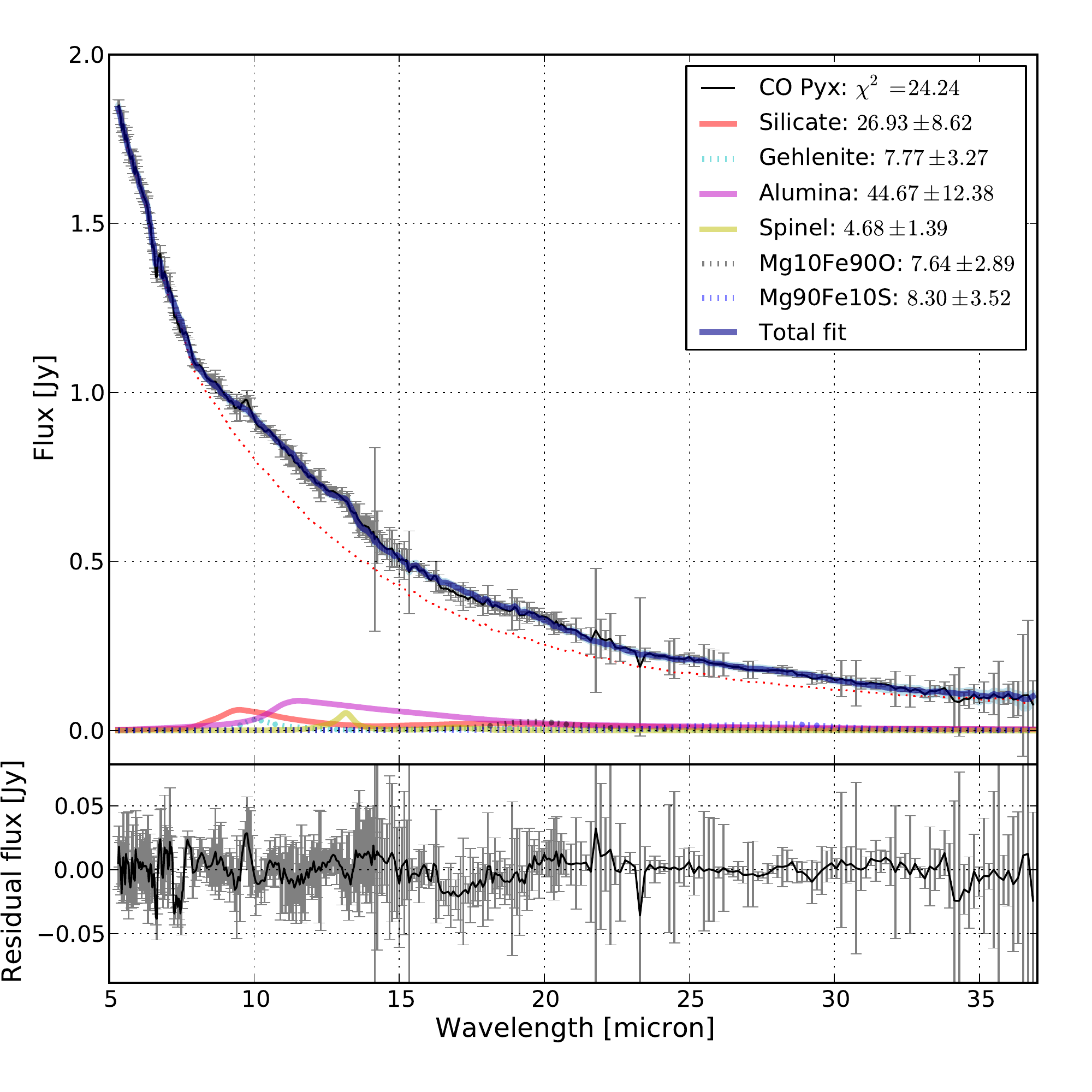}
\label{fig:subfigure_appendix3d}
}
\subfigure{
\includegraphics[width=8.0cm]{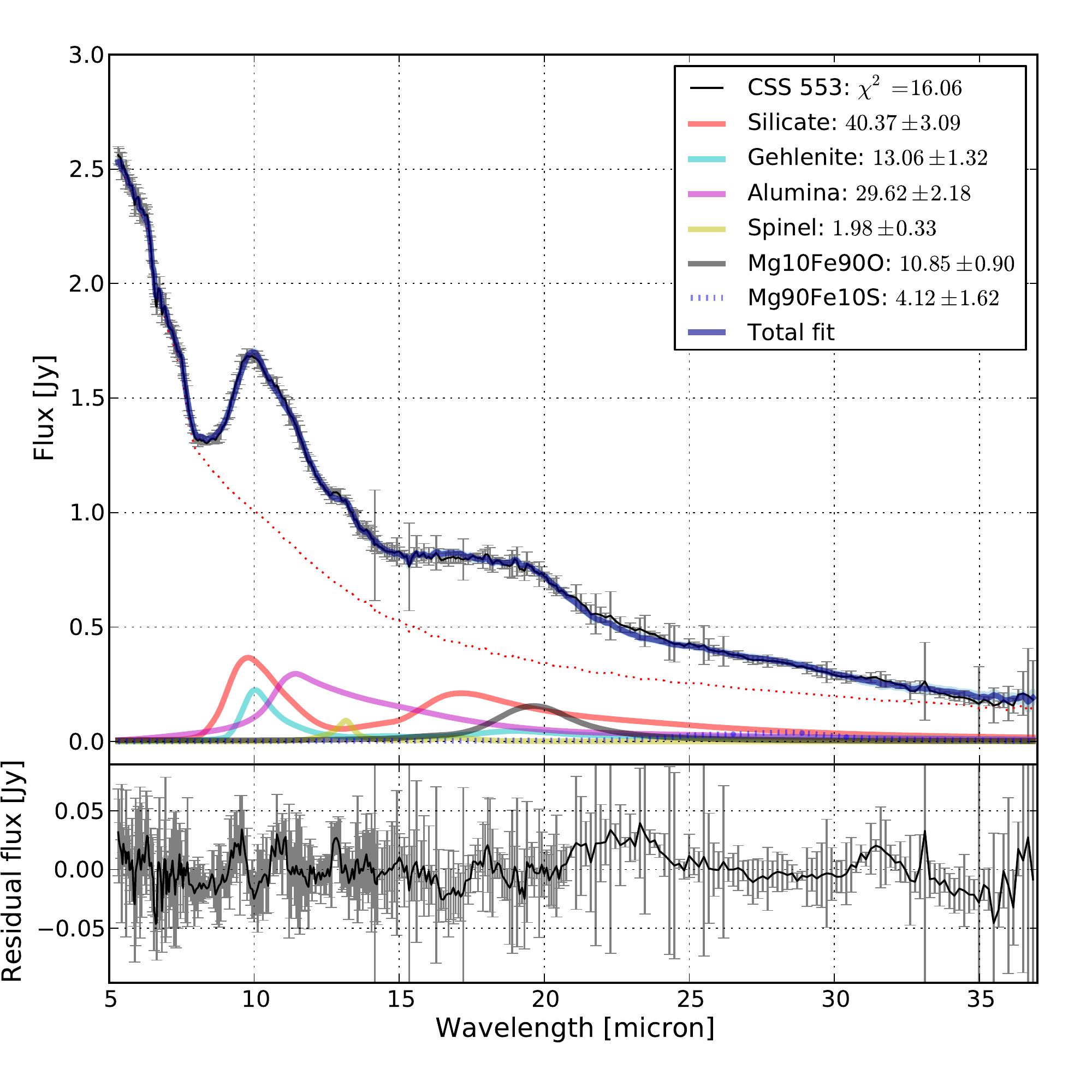}
\label{fig:subfigure_appendix3e}
}
\subfigure{
\includegraphics[width=8.0cm]{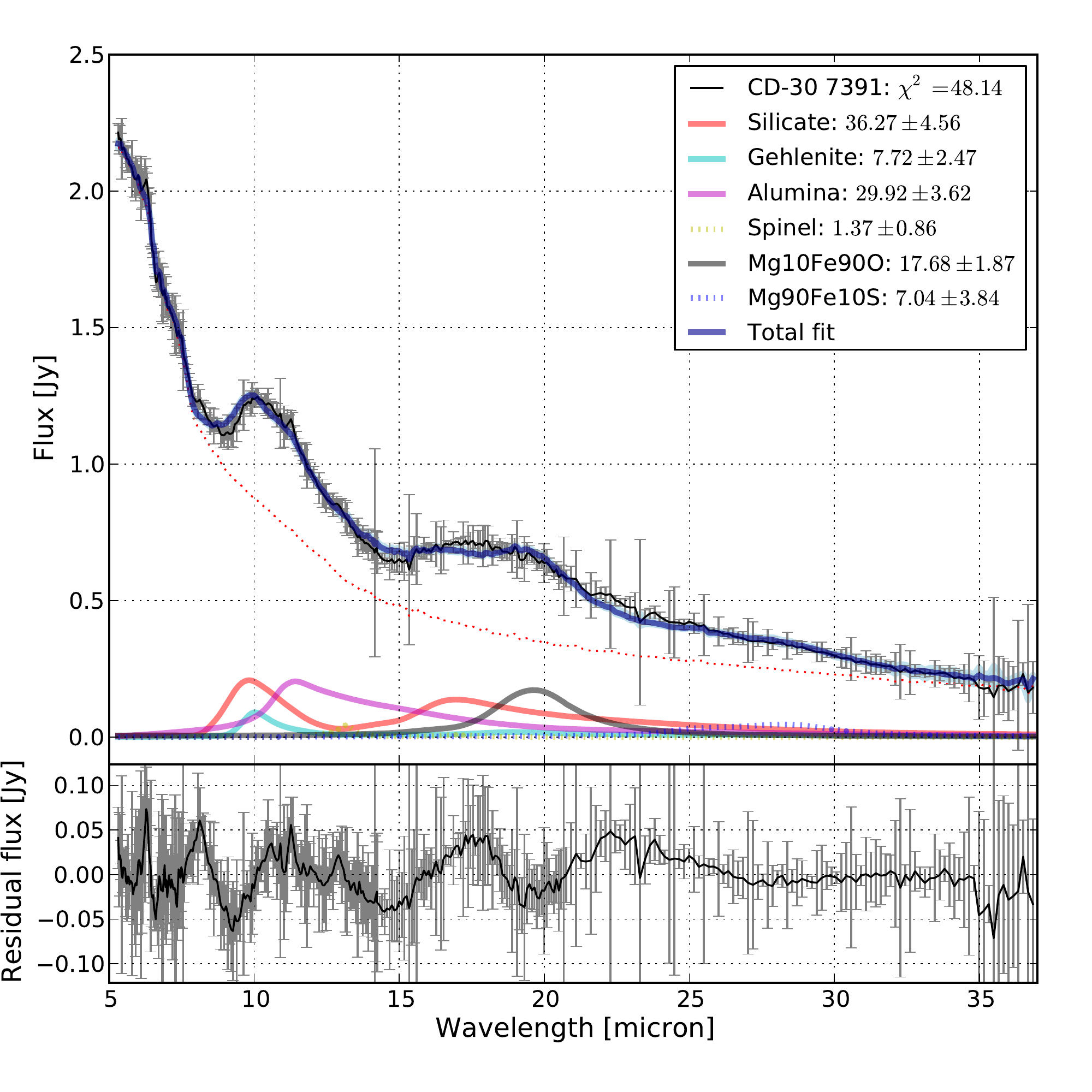}
\label{fig:subfigure_appendix3f}
}
\caption[]{\label{fig:subfigure_appendix3}
Continued}
\end{figure*}

\begin{figure*}
\centering
\subfigure{
\includegraphics[width=8.0cm]{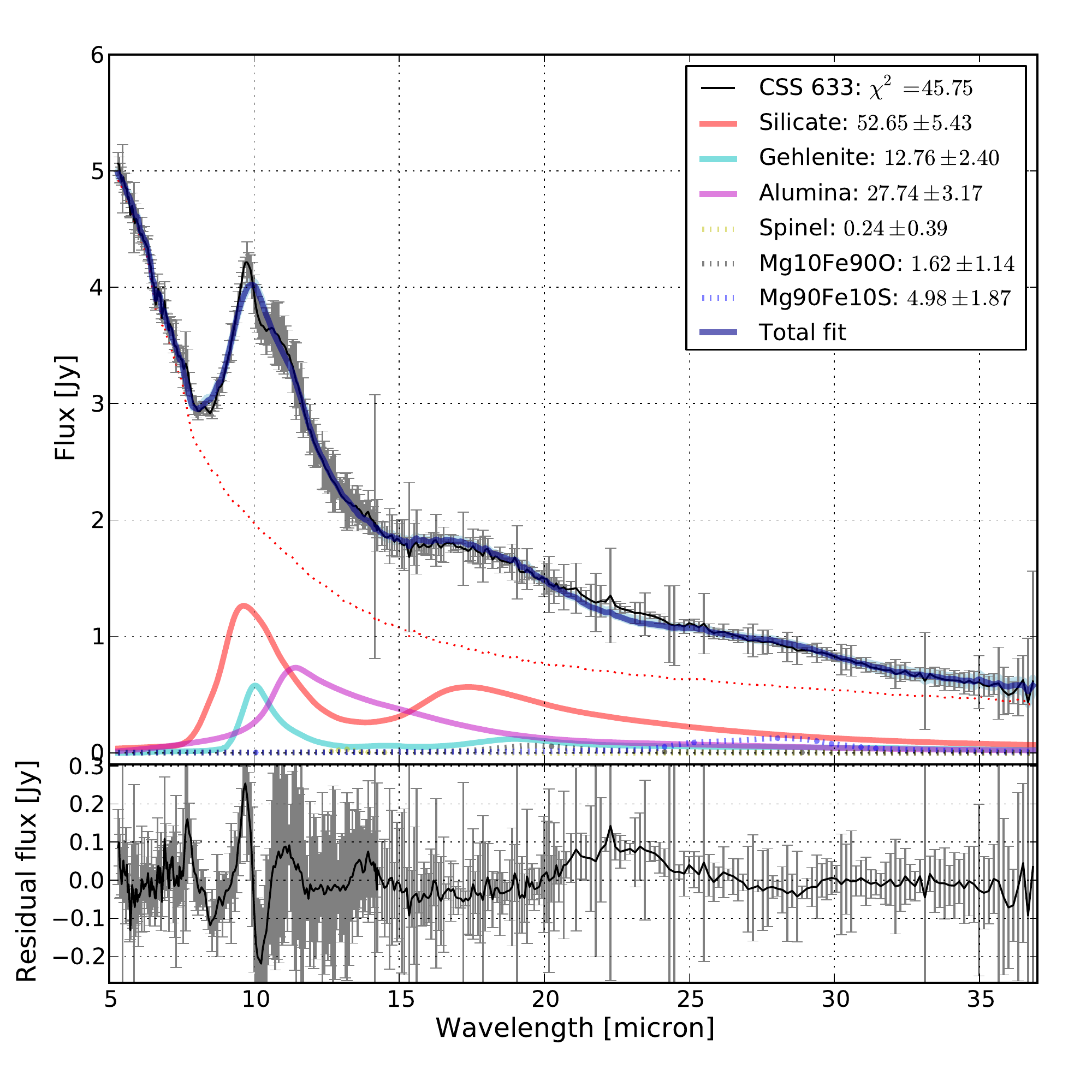}
\label{fig:subfigure_appendix4a}
}
\subfigure{
\includegraphics[width=8.0cm]{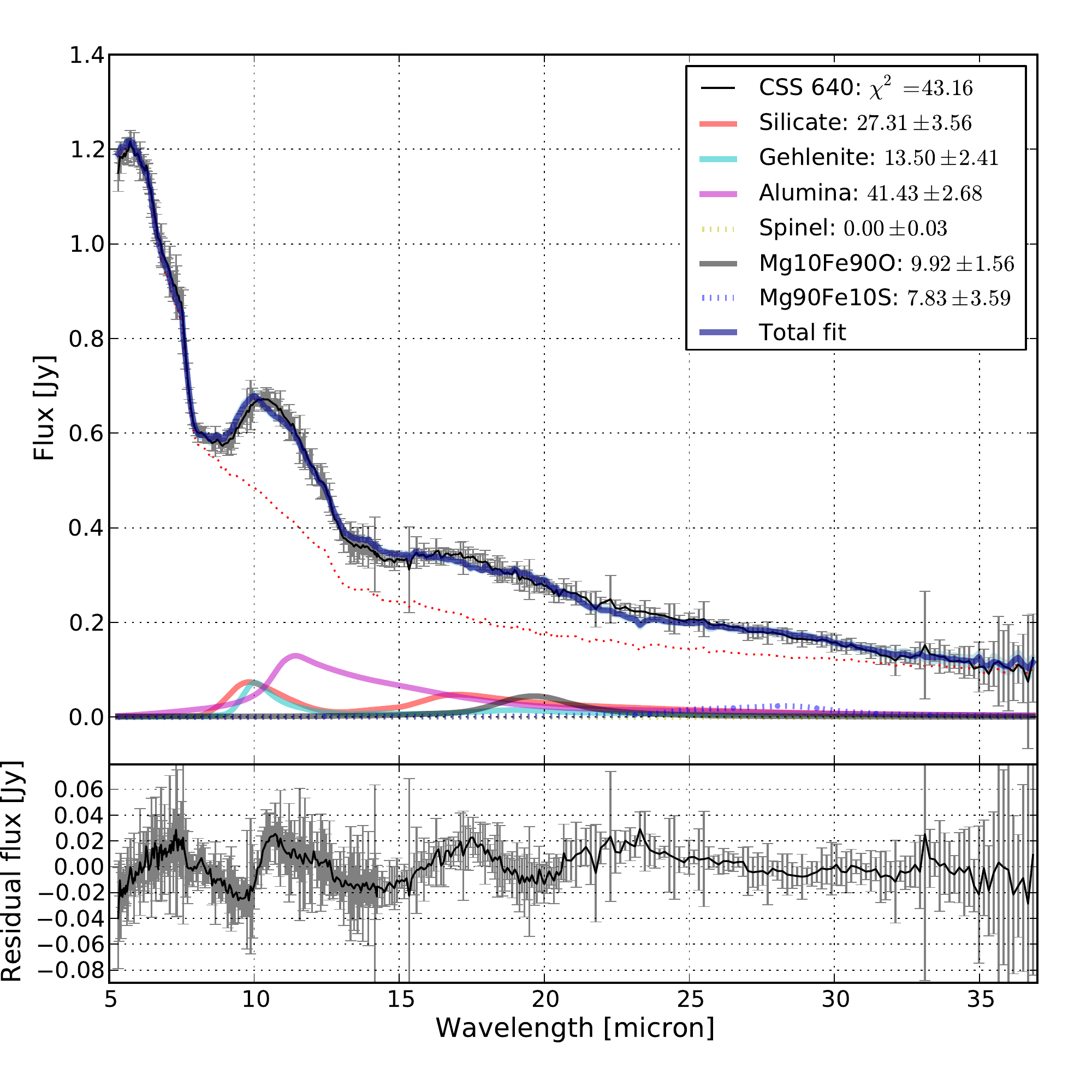}
\label{fig:subfigure_appendix4b}
}
\subfigure{
\includegraphics[width=8.0cm]{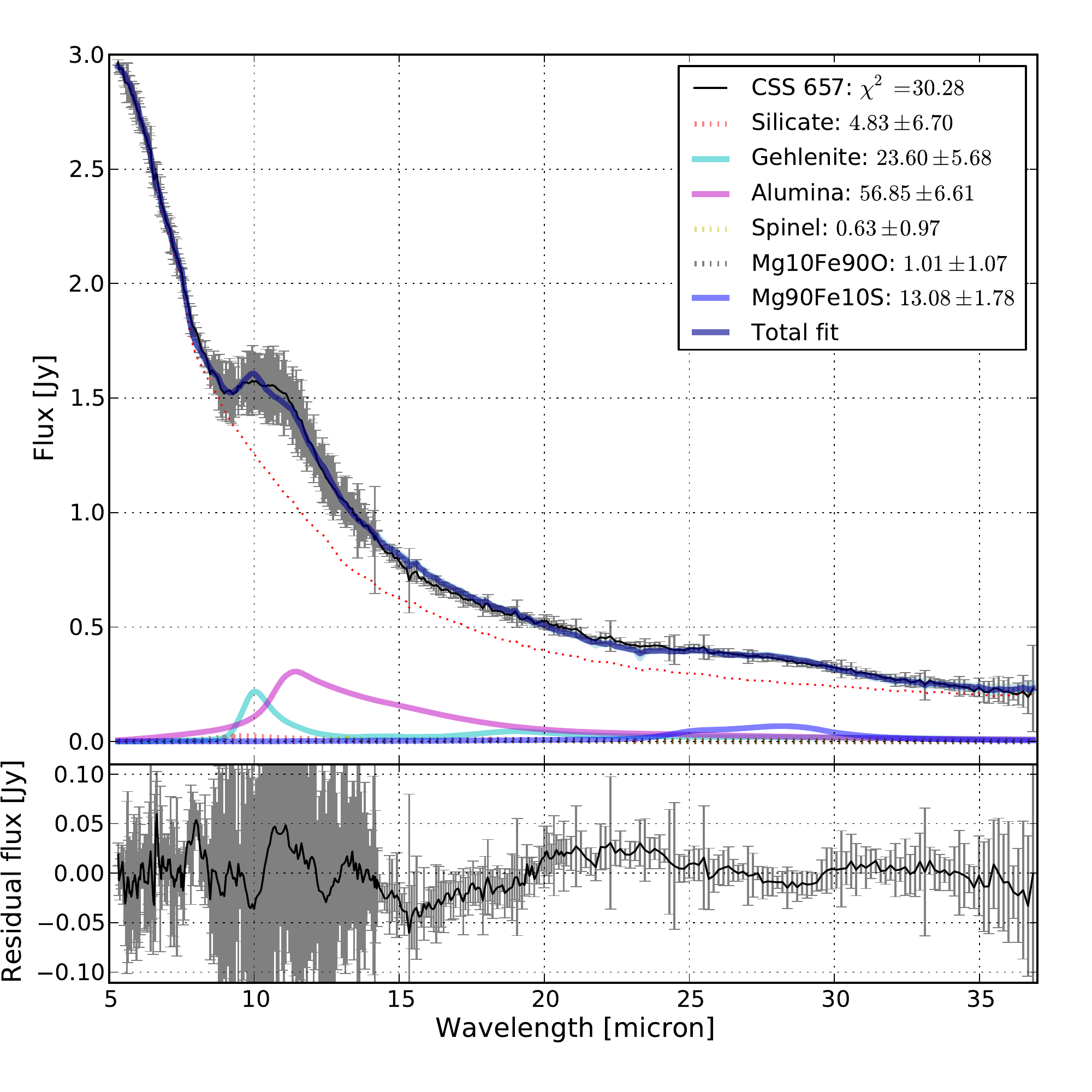}
\label{fig:subfigure_appendix4c}
}
\subfigure{
\includegraphics[width=8.0cm]{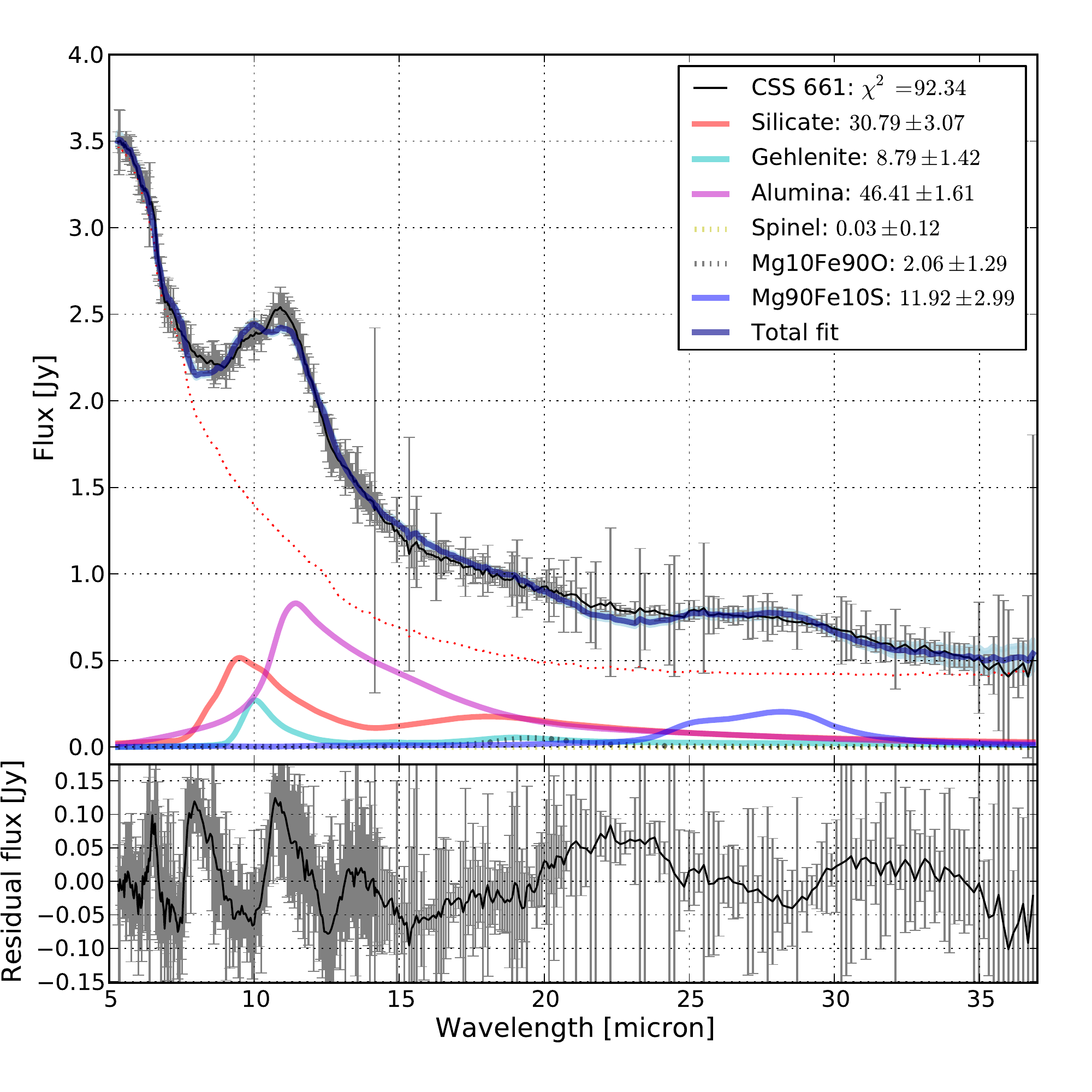}
\label{fig:subfigure_appendix4d}
}
\subfigure{
\includegraphics[width=8.0cm]{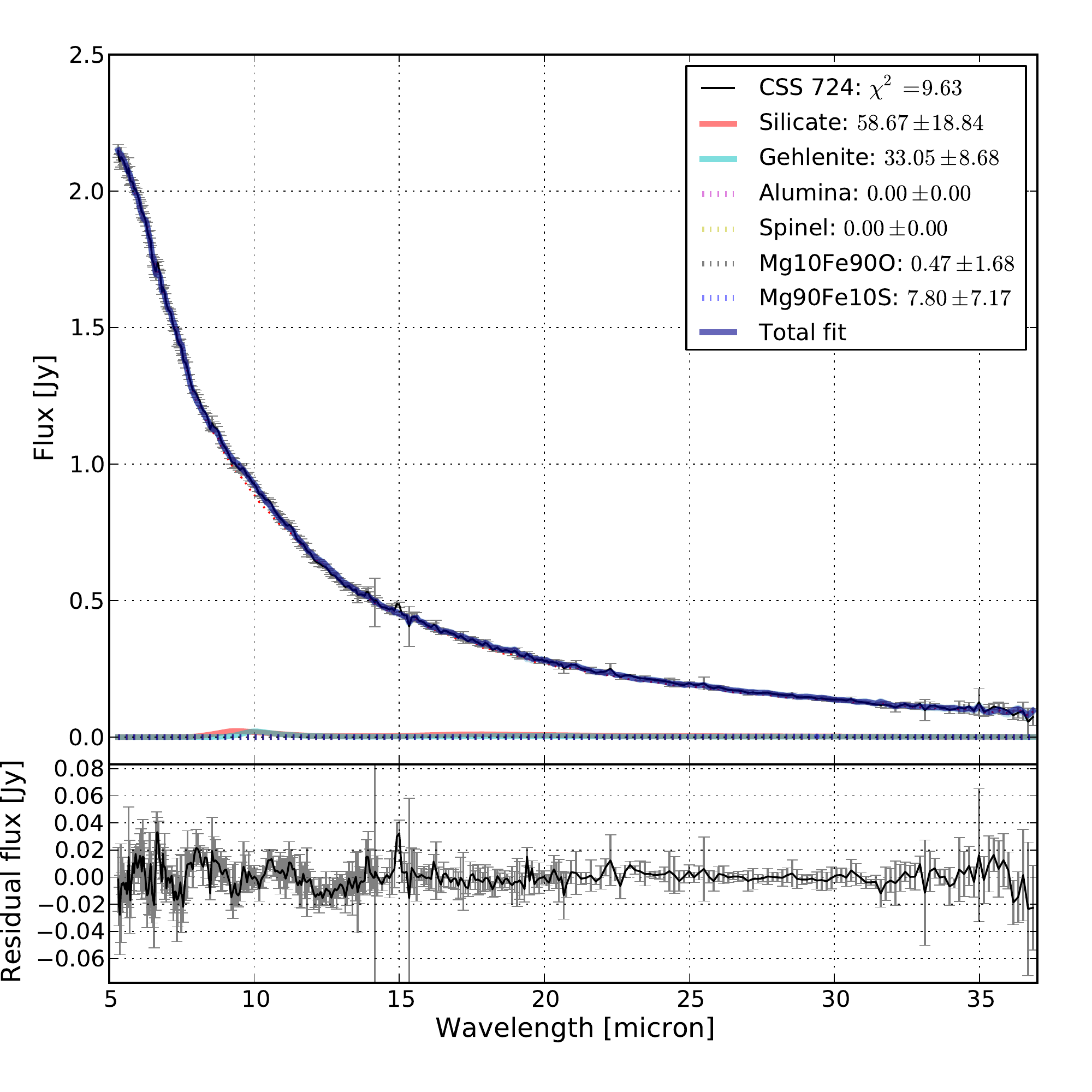}
\label{fig:subfigure_appendix4e}
}
\subfigure{
\includegraphics[width=8.0cm]{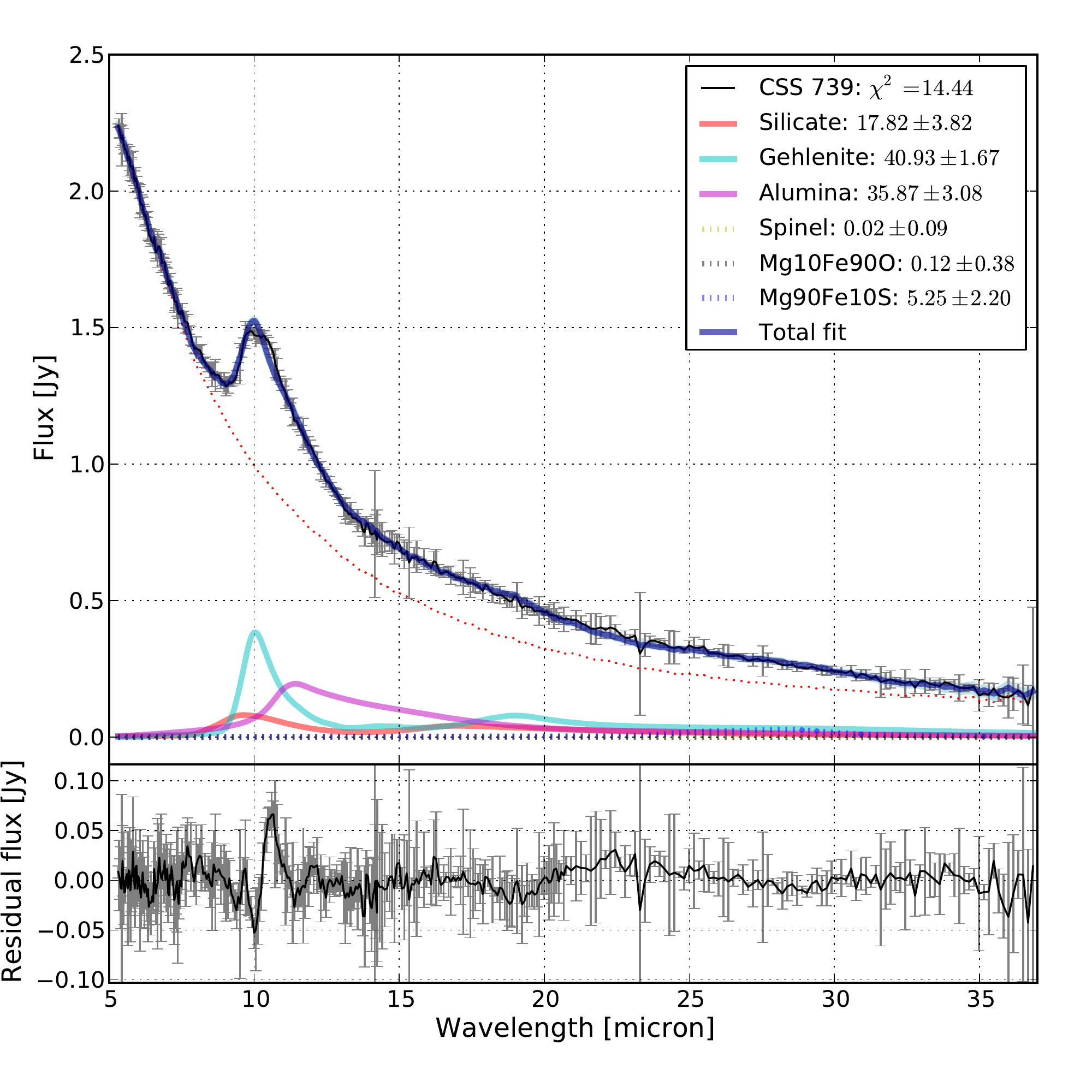}
\label{fig:subfigure_appendix4f}
}
\caption[]{\label{fig:subfigure_appendix4}
Continued}
\end{figure*}

\begin{figure*}
\centering
\subfigure{
\includegraphics[width=8.0cm]{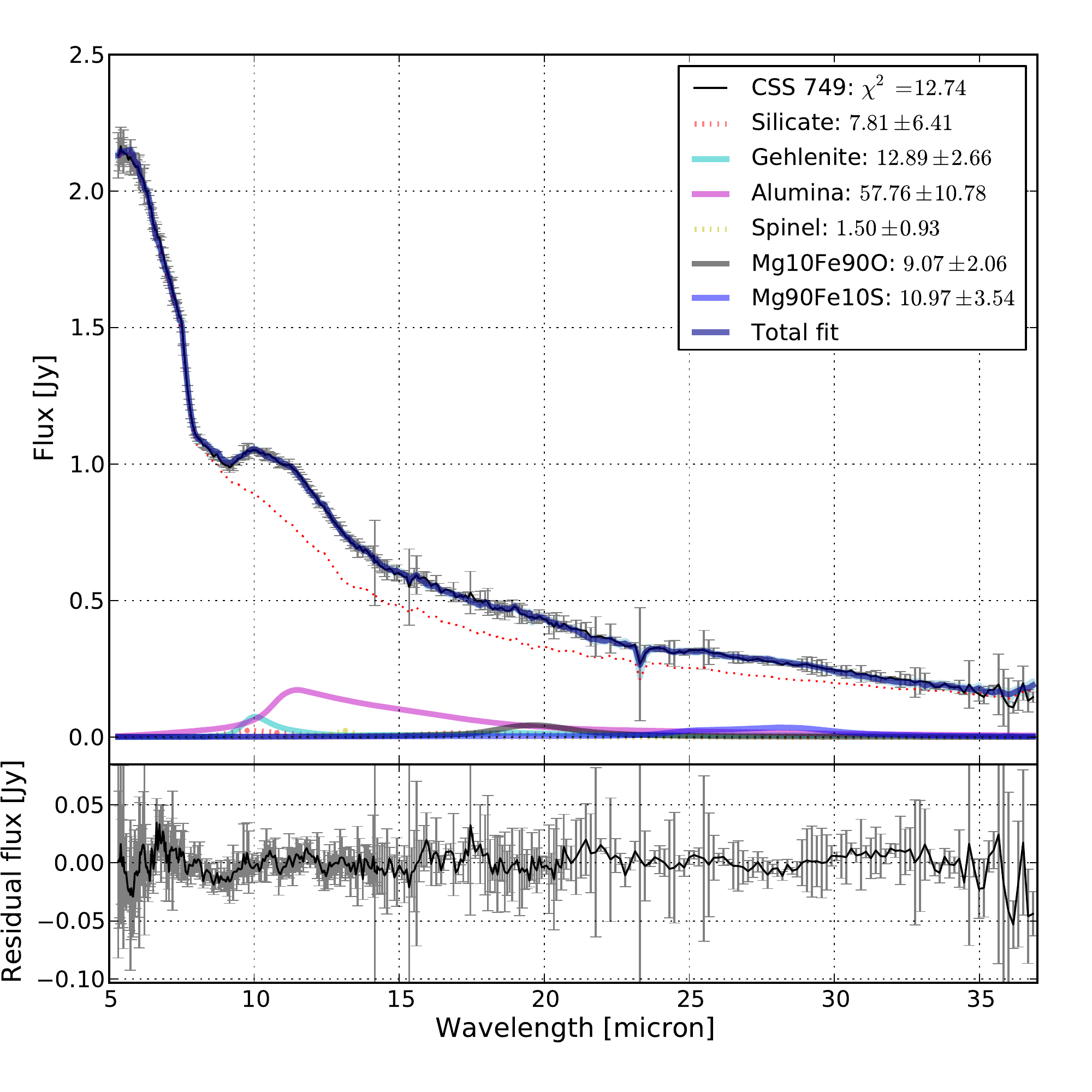}
\label{fig:subfigure_appendix5a}
}
\subfigure{
\includegraphics[width=8.0cm]{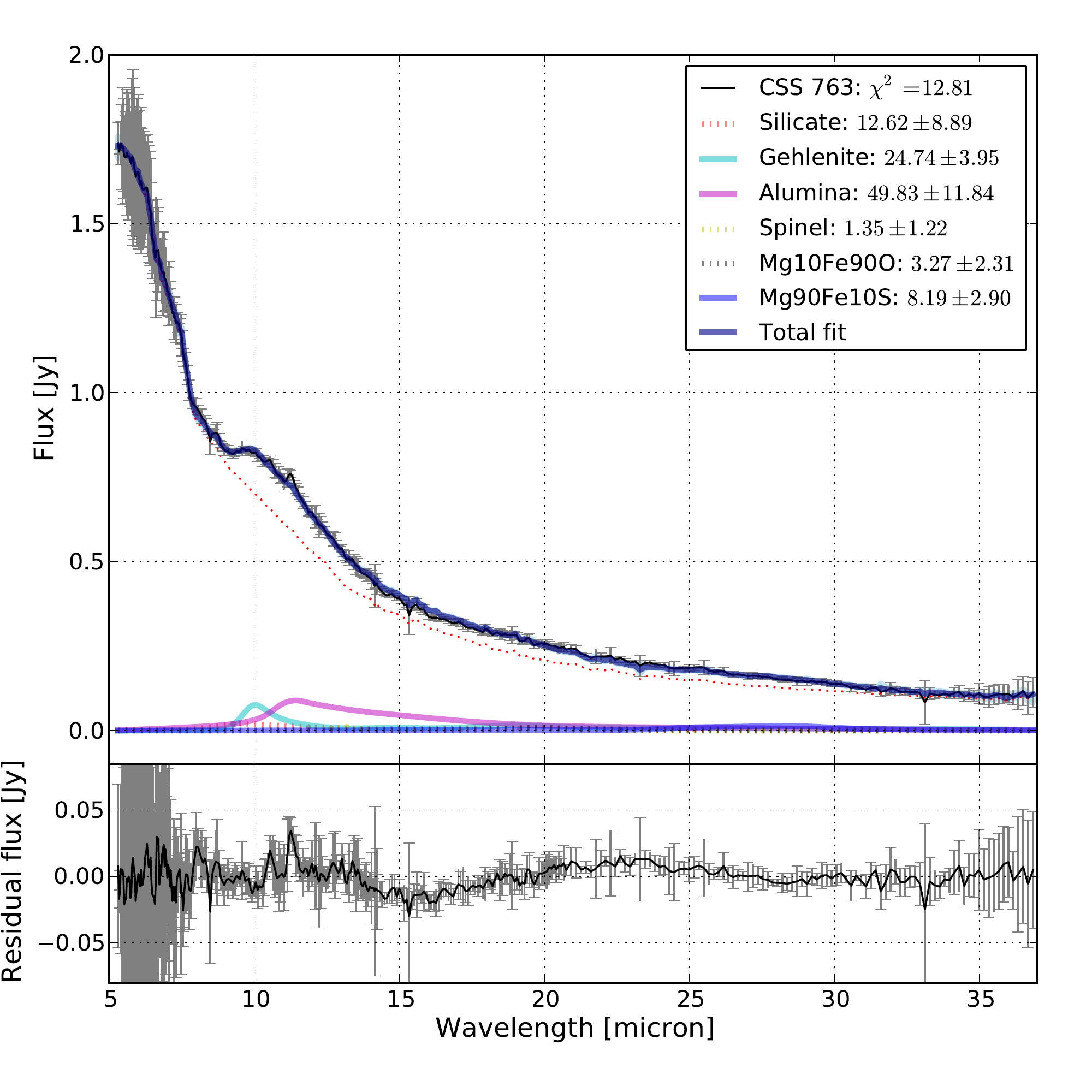}
\label{fig:subfigure_appendix5b}
}
\subfigure{\includegraphics[width=8.0cm]{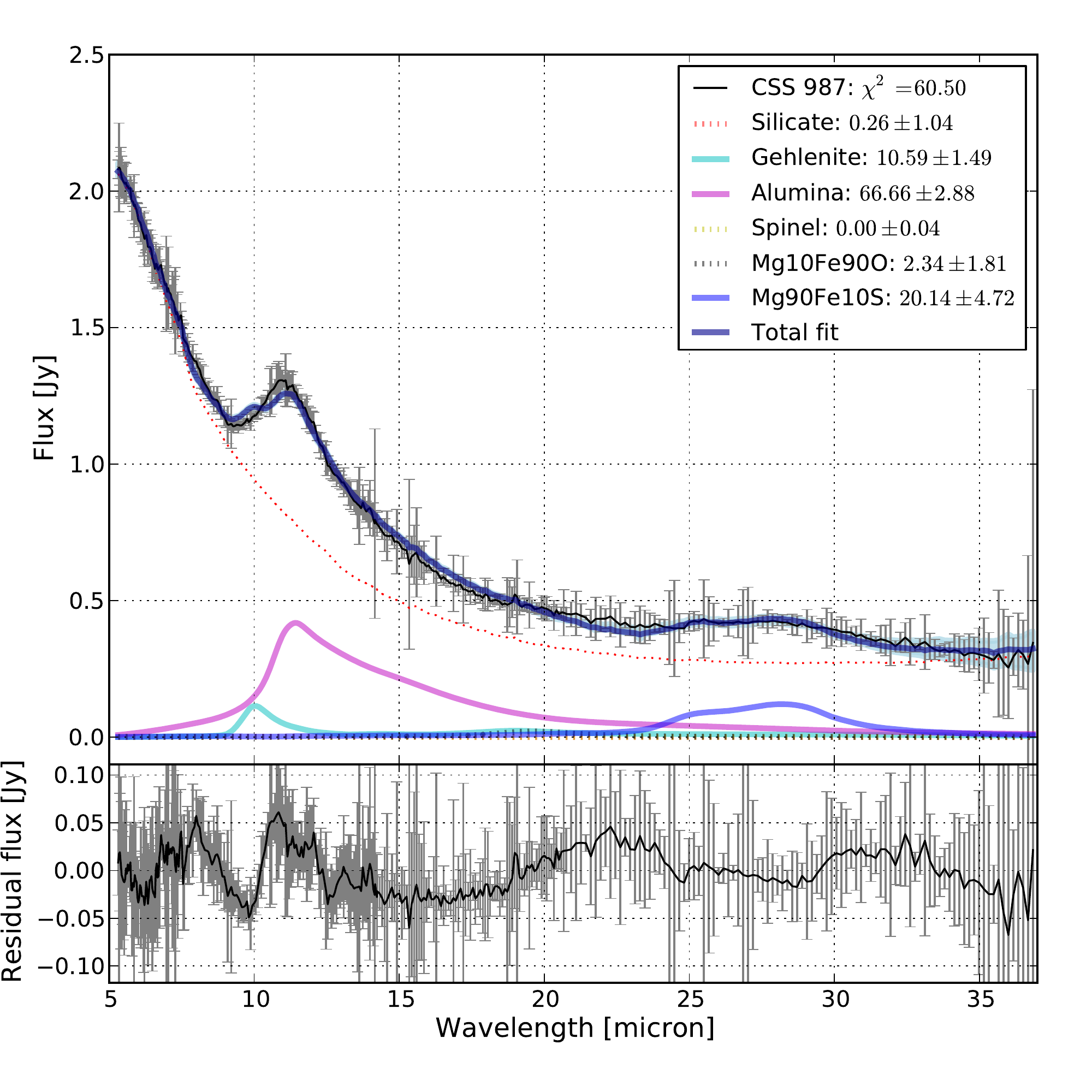}
\label{fig:subfigure_appendix5c}
}
\subfigure{
\includegraphics[width=8.0cm]{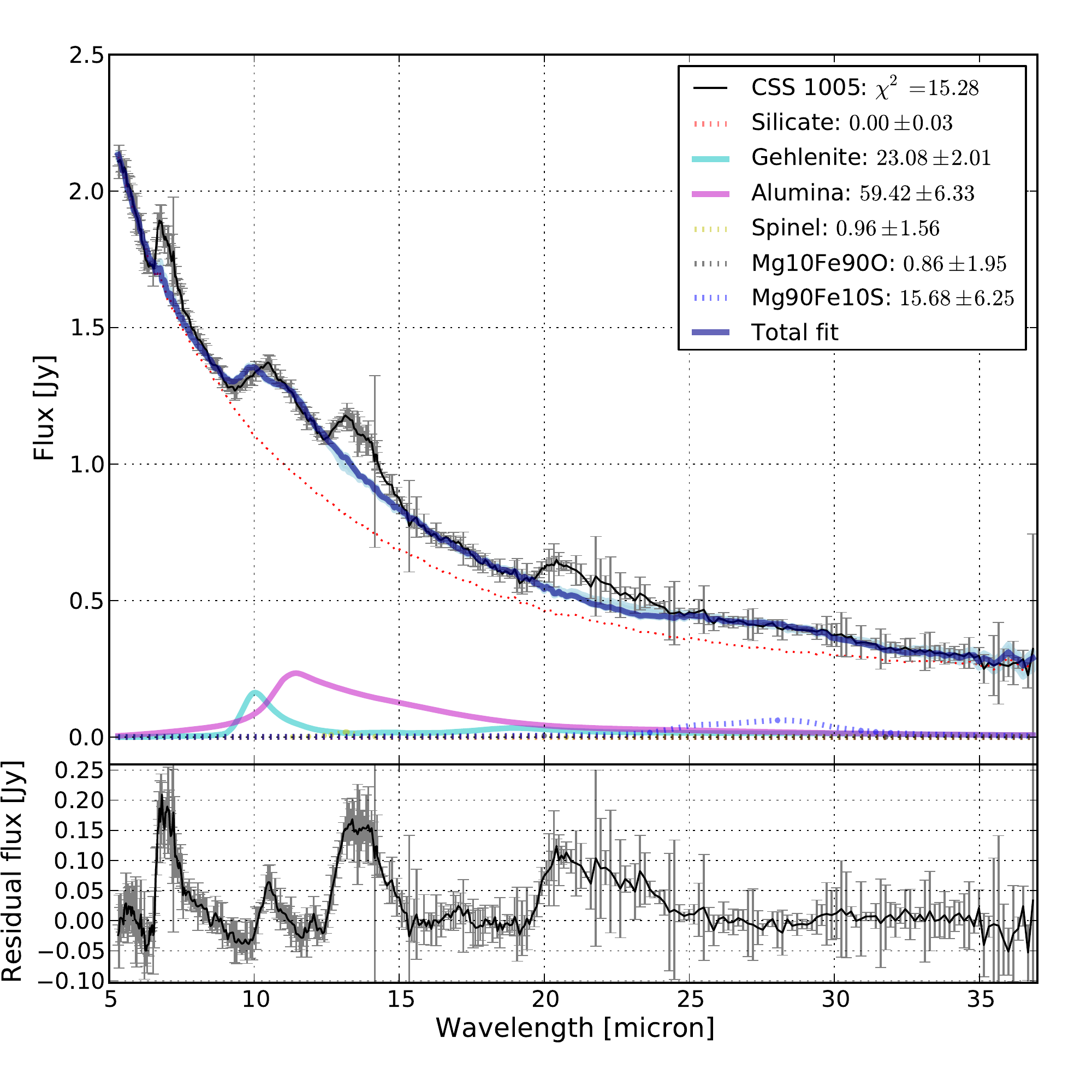}
\label{fig:subfigure_appendix5d}
}
\subfigure{
\includegraphics[width=8.0cm]{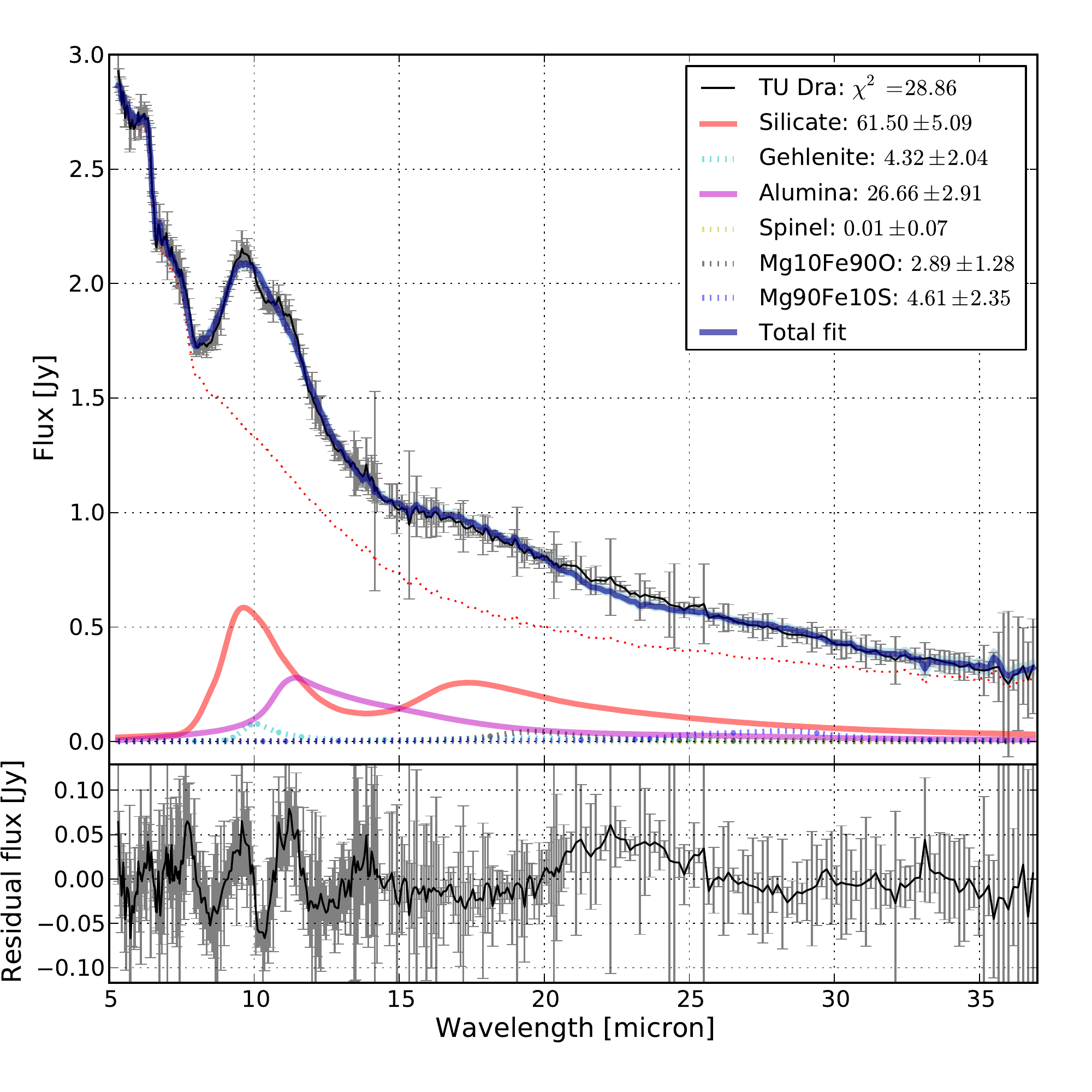}
\label{fig:subfigure_appendix5e}
}
\subfigure{
\includegraphics[width=8.0cm]{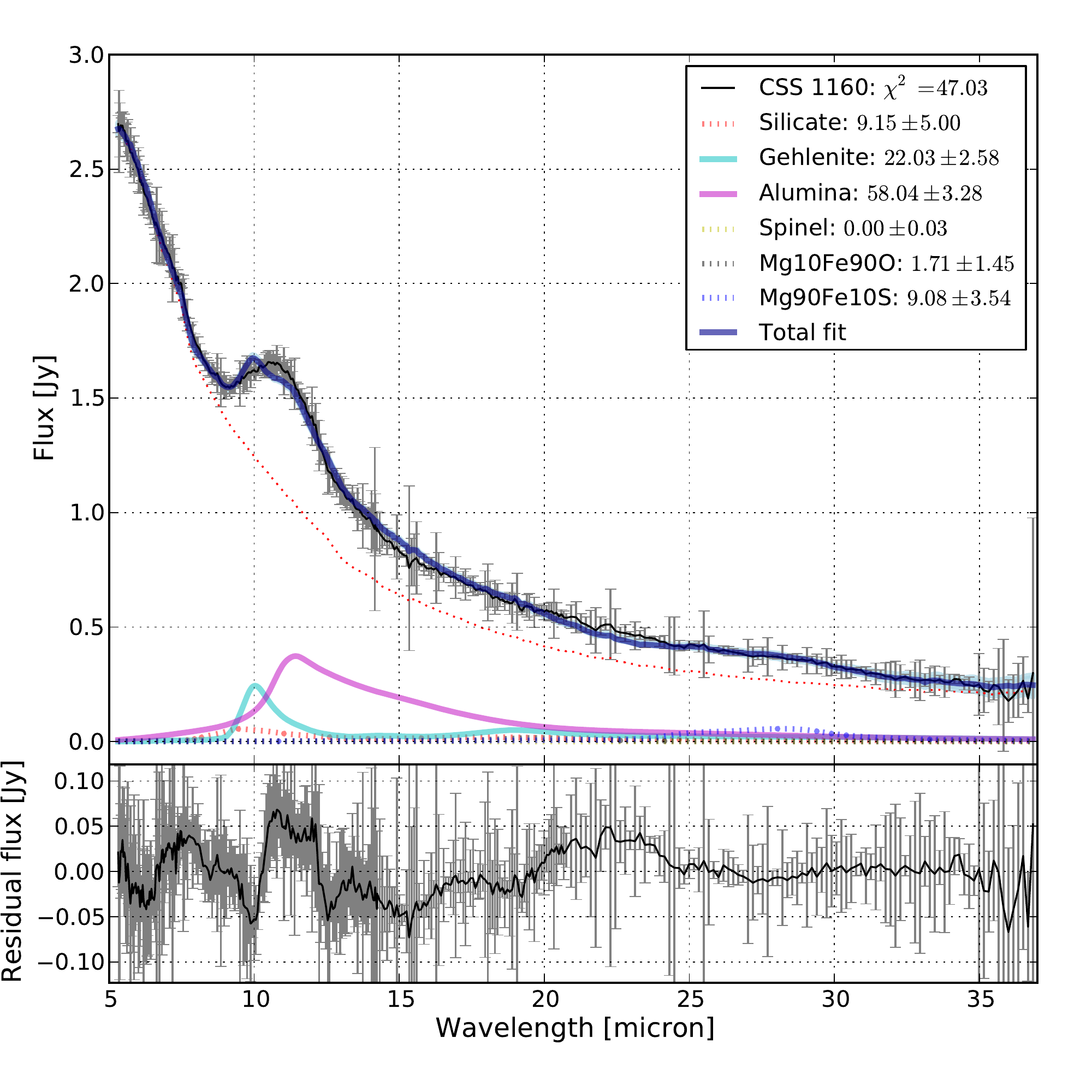}
\label{fig:subfigure_appendix5f}
}
\caption[]{\label{fig:subfigure_appendix5}
Continued}
\end{figure*}

\begin{figure*}
\centering
\subfigure{
\includegraphics[width=8.0cm]{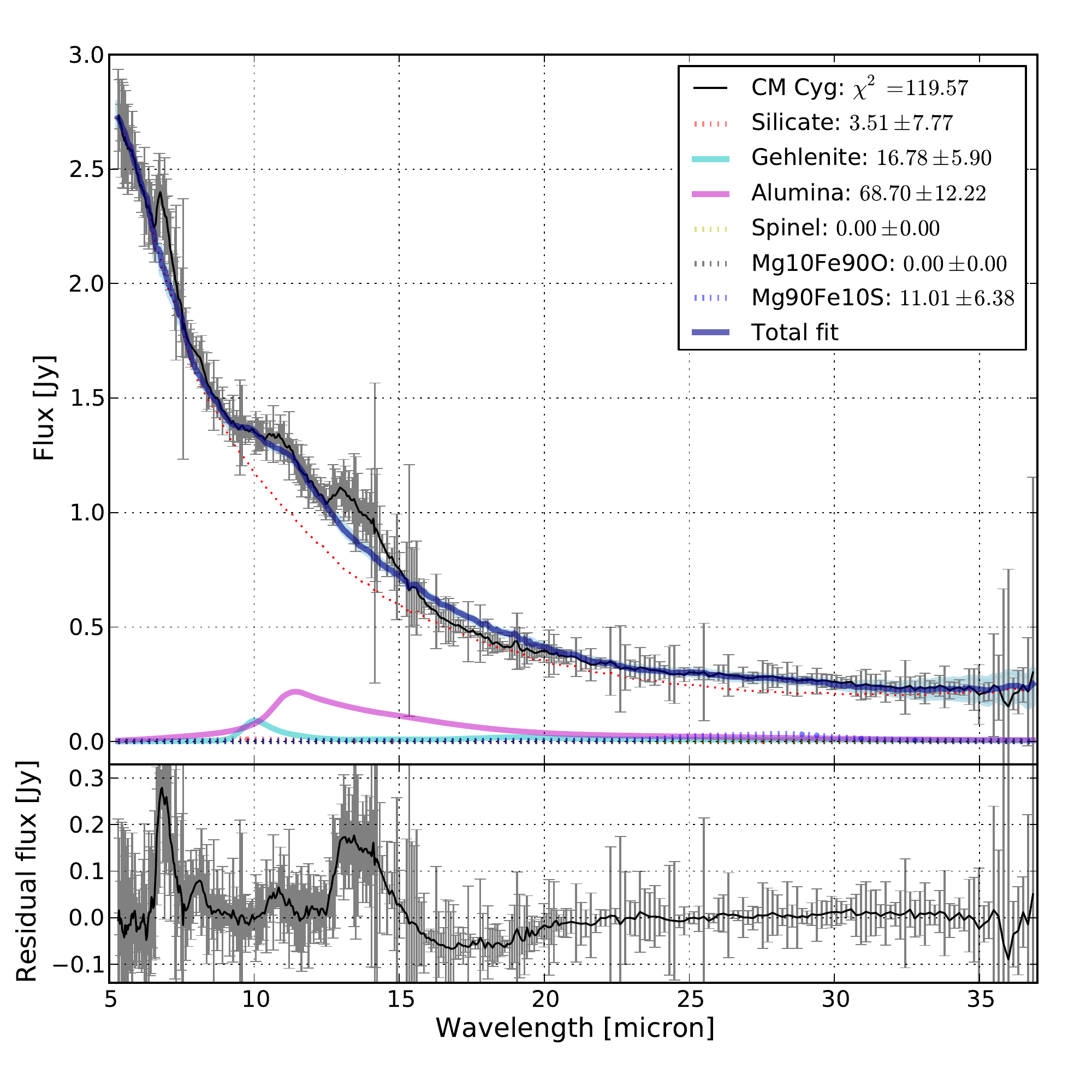}
\label{fig:subfigure_appendix6a}
}
\subfigure{
\includegraphics[width=8.0cm]{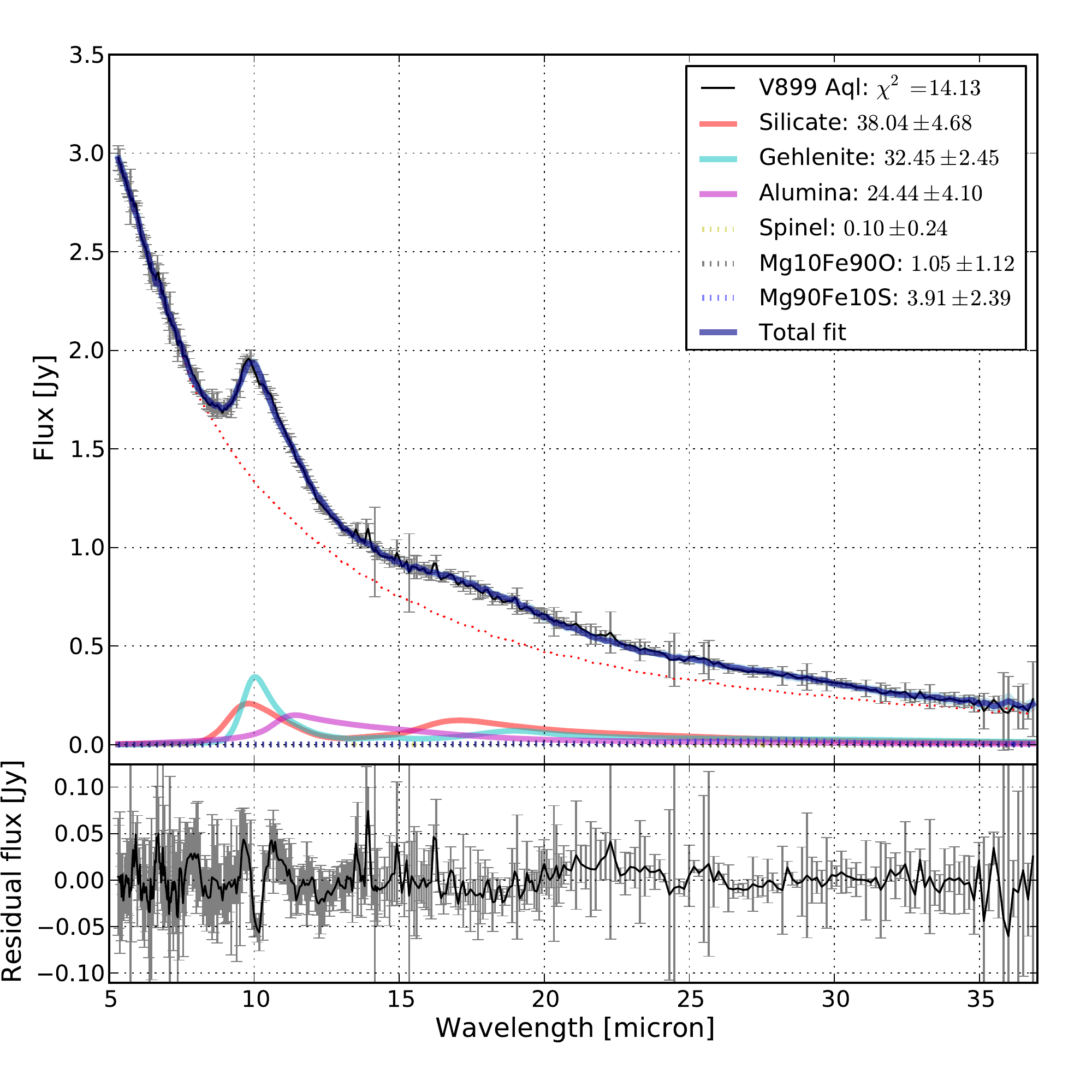}
\label{fig:subfigure_appendix6b}
}
\subfigure{
\includegraphics[width=8.0cm]{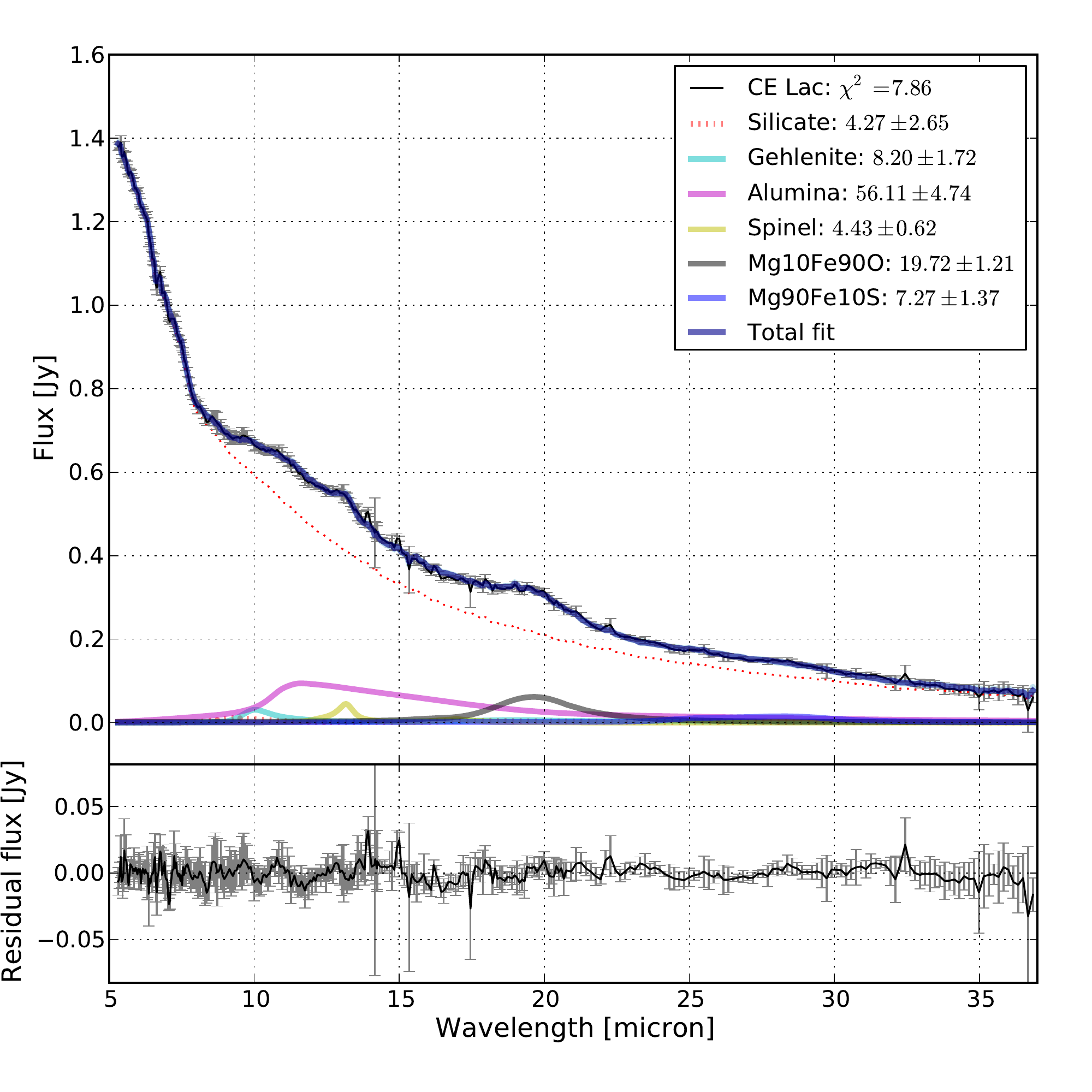}
\label{fig:subfigure_appendix6c}
}
\subfigure{
\includegraphics[width=8.0cm]{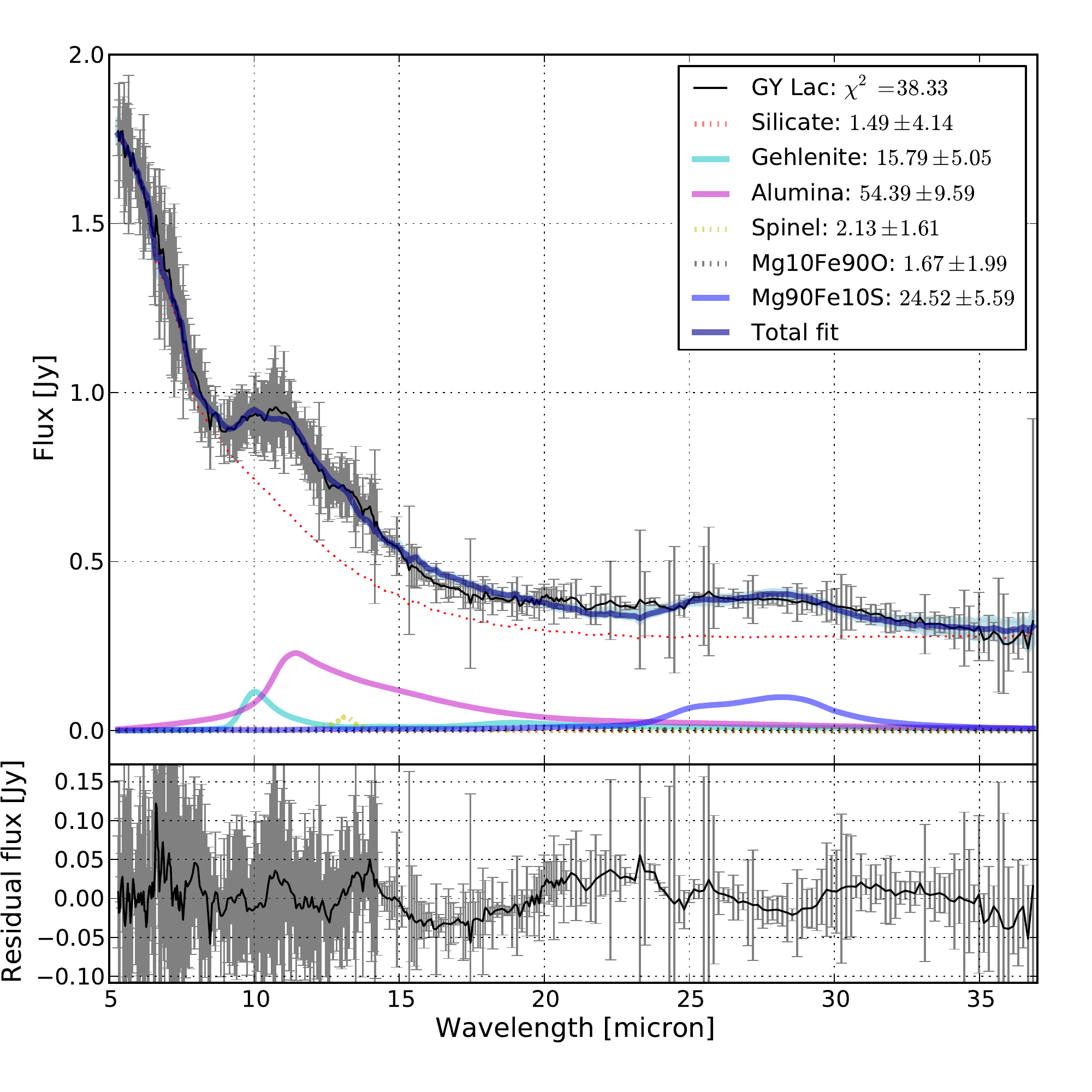}
\label{fig:subfigure_appendix6d}
}
\subfigure{
\includegraphics[width=8.0cm]{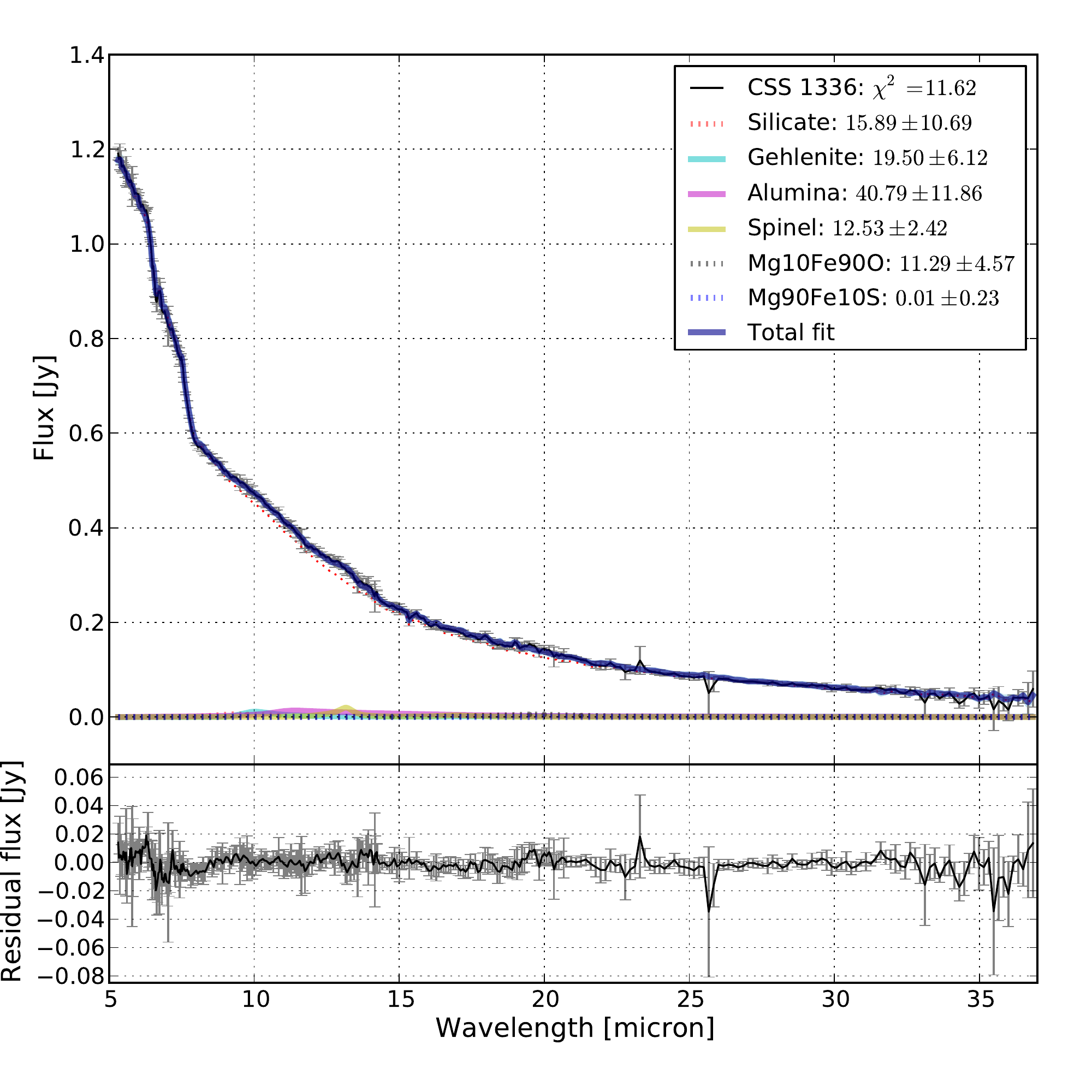}
\label{fig:subfigure_appendix6e}
}
\caption[]{\label{fig:subfigure_appendix6}
Continued}
\end{figure*}

\end{appendix}
\end{document}